\documentclass[letterpaper,12pt,unknownkeysallowed]{article} 
\usepackage[utf8]{inputenc}
\usepackage{verbatim}
\usepackage[english]{babel}
\usepackage{amsfonts}
\usepackage{amssymb}
\usepackage{amsmath}
\usepackage[flushleft]{threeparttable}
\usepackage[round]{natbib}
\usepackage{enumitem}
\usepackage{bbm}
\usepackage{mathtools}
\usepackage{subcaption}
\usepackage{booktabs}
\usepackage{setspace}
\usepackage{graphicx}
\usepackage{lscape}
\usepackage{rotating}
\usepackage{fullpage}
\usepackage[table]{xcolor}
\usepackage{multirow}

\usepackage[capposition=top]{floatrow} 
	\usepackage[utf8]{inputenc}
	\usepackage[english]{babel}
	\usepackage{amsfonts}
	\usepackage{amssymb}
	\usepackage{enumitem}
	\usepackage{bbm}
	\usepackage{mathtools}
    \usepackage{subcaption}
	\usepackage{booktabs}
	\usepackage{setspace}
	\usepackage{graphicx}
	\usepackage{lscape}
	\usepackage{rotating}
	\usepackage{fullpage}
    \usepackage[capposition=top]{floatrow} 
    \usepackage{array}
    \newcolumntype{H}{>{\setbox0=\hbox\bgroup}c<{\egroup}@{}}

	\usepackage{natbib}
	\usepackage[framemethod=tikz]{mdframed}
	\usepackage{todonotes}
	\presetkeys{todonotes}{color=purple!10,inline}{} 

	\newcommand{\rf}[1]{\comment{Reference: \url{#1}}}

	\DeclareMathOperator*{\argmin}{arg\,min}
	
\newtheorem{lemma}{Lemma}

\newtheorem{theorem}{Theorem}
\newtheorem{corollary}{Corollary}[theorem]

\renewcommand{\appendix}{\small\parindent 0cm\parskip 5pt\setcounter{equation}{0}
\setcounter{section}{0}
\renewcommand{\thesection}{A.\arabic{section}}
\renewcommand{\theequation}{A.\arabic{equation}}
\setcounter{lemma}{0}\renewcommand{\thelemma}{A.\arabic{lemma}}
\setcounter{theorem}{0}\renewcommand{\thetheorem}{A.\arabic{theorem}}
}

\newtheorem{assump}{Assumption}

  \makeatletter
\newcommand{\oset}[3][0ex]{%
  \mathrel{\mathop{#3}\limits^{
    \vbox to#1{\kern-2\ex@
    \hbox{$\scriptstyle#2$}\vss}}}}
\makeatother

\makeatletter
\def\blfootnote{\xdef\@thefnmark{}\@footnotetext}
\makeatother

\makeatletter
\def\@sect#1#2#3#4#5#6[#7]#8{\ifnum #2>\c@secnumdepth
     \let\@svsec\@empty\else
     \refstepcounter{#1}\edef\@svsec{\csname the#1\endcsname. \hskip 0.4em}\fi
     \@tempskipa #5\relax
      \ifdim \@tempskipa>\z@
        \begingroup #6\relax
          \@hangfrom{\hskip #3\relax\@svsec}{\interlinepenalty \@M #8\par}%
        \endgroup
       \csname #1mark\endcsname{#7}\addcontentsline
         {toc}{#1}{\ifnum #2>\c@secnumdepth \else
                      \protect\numberline{\csname the#1\endcsname}\fi
                    #7}\else
        \def\@svsechd{#6\hskip #3\relax  
                   \@svsec #8\csname #1mark\endcsname
                      {#7}\addcontentsline
                           {toc}{#1}{\ifnum #2>\c@secnumdepth \else
                             \protect\numberline{\csname the#1\endcsname}\fi
                       #7}}\fi
     \@xsect{#5}}
\makeatother

\makeatletter
\renewcommand{\section}{\@startsection{section}{1}{0mm}{-\baselineskip}{0.25\baselineskip}{\centering\normalfont\normalsize\bf}}
\renewcommand{\subsection}{\@startsection{subsection}{2}{0mm}{-\baselineskip}{0.25\baselineskip}{\raggedright\normalfont\normalsize\bf}}
\renewcommand{\subsubsection}{\@startsection{subsubsection}{3}{0mm}{-\baselineskip}{0.25\baselineskip}{\raggedright\normalfont\small}}
\def\@begintheorem#1#2{\trivlist \item[\hskip \labelsep{\bf #1\ #2}]\it}
\makeatother

\renewcommand{\thesection}{\arabic{section}}

\renewenvironment{abstract}
 {\begin{center}\normalsize\bf\text{Abstract}
 \end{center}\begin{quote}\normalsize}
 {\end{quote}}

\allowdisplaybreaks

\begin{document}

\thispagestyle{empty}
\vskip 20pt
\centerline{\Large\bf Bayesian and Frequentist Inference for Synthetic Controls\footnote{Department of Economics, Massachusetts Institute of Technology. Email: vives@mit.edu. We thank Alberto Abadie, Isaiah Andrews, Dmitry Arkhangelsky, Guido Imbens, Anna Mikusheva, Alex Parret, Hal Varian, and participants in EEA-ESEM (2023), ACIC (2022), the Data X: Synthetic Control Workshop (2022), the Google statistics seminar and the MIT Econometrics lunch for useful comments and discussion. Jaume Vives-i-Bastida aknowledges funding from a La Caixa post-graduate fellowship and a Meta Research PhD fellowship. This research project was started during an internship Jaume Vives-i-Bastida did at Google.}}
  \begin{center}%
    \vskip 10pt
    {\large
     \lineskip .5em%
      \begin{tabular}[t]{ccc}%
       Ignacio Martinez&&Jaume Vives-i-Bastida\\[.2ex]
       Google&&MIT\\
      \end{tabular}
      \par}%
      \vskip 1em%
      {\large \today } \par
       \vskip 1em%
  \end{center}\par
  

\bigskip
\begin{abstract}
\noindent The synthetic control method has become a widely popular tool to estimate causal effects with observational data. Despite this, inference for synthetic control methods remains challenging. Often, inferential results rely on linear factor model data generating processes. In this paper, we characterize the conditions on the factor model primitives (the factor loadings) for which the statistical risk minimizers are \textit{synthetic controls} (in the simplex). Then, we propose a Bayesian alternative to the synthetic control method that preserves the main features of the standard method and provides a new way of doing valid inference. We explore a Bernstein-von Mises style result to link our Bayesian inference to the frequentist inference. For linear factor model frameworks we show that a maximum likelihood estimator (MLE) of the synthetic control weights can consistently estimate the predictive function of the potential outcomes for the treated unit and that our Bayes estimator is asymptotically close to the MLE in the total variation sense. Through simulations, we show that there is convergence between the Bayesian and frequentist approach even in sparse settings. Finally, we apply the method to re-visit the study of the economic costs of the German re-unification and the Catalan secession movement. The Bayesian synthetic control method is available in the bsynth R-package.
\end{abstract}

\section{Introduction} \label{sec:introduction}
\medskip

Synthetic control methods (\citet{abadie2003}, \citet{AbaDiaHai2010}) are often used to estimate treatment effects of aggregate policy interventions. The method has been described as ``arguably the most important innovation in the policy evaluation literature in the last 15 years" (\citet{athey2021matrix}). Despite this, statistical inference for synthetic controls is often non-trivial. To highlight the nature of the problem, consider the synthetic control problem. Suppose that we observe data for a unit that is affected by an intervention of interest and for a donor pool of untreated units. We are interested in estimating the treatment effect on the treated after the intervention. Synthetic controls match the treated unit to the weighted average of the donor units that most closely resembles the characteristics of the treated unit before the intervention. The treatment effect on the treated is then estimated by taking the differences in outcomes between the treated unit and the weighted average. Given that only one treated unit is available and that the synthetic control estimator depends on a weighted average of the donor units outcomes, it is not straightforward to provide asymptotic guarantees for inference on the treatment effect. 

With this in mind, inference for synthetic controls has been approached from various angles. First, a body of literature has focused on permutation based inference relative to a benchmark assignment process, as pioneered in \citet{AbaDiaHai2010} and more recently in \citet{Firpo2018sensitivity} and \citet{abadie2021penalized} among others. Second, large sample properties of synthetic control estimators have been studied in linear factor model settings. Starting with \citet{AbaDiaHai2010} who provide a bound for the bias of synthetic controls and \citet{ferman2021properties} that derives asymptotic results under the assumption that the treated unit factor loading is in the convex hull of the donor units factor loadings as the number of pre-treatment periods and donor units grows. More recently, \citet{imbens2023identificationinferencesyntheticcontrols} and \citet{arkhangelsky2023largesamplepropertiessyntheticcontrol} consider the large sample properties of synthetic control estimators in settings with unmeasured confounding and selection into treatment. Other related research has focused on projection theory results for average treatment effects estimated by synthetic controls (\citet{li2020}) and panel data approaches that relax the simplex restriction (\citet{hsiao2012panel}). Third, conformal inference procedures have also been proposed. For example, in \citet{chernozhukov2021conformal} under the assumption that the synthetic control estimator is as good as a true synthetic control that recovers the treated unit. Finally, Bayesian inference methods have also been studied. For instance, \citet{pang2022} and \citet{pinkney2021} for linear factor model structures, Bayesian regression methods in \citet{kim2020}, Bayesian structural time series in \citet{brodersen2015} and \citet{scott2014predicting}, or empirical Bayes approaches in \citet{amjad2018robust}.

A common thread in the literature is that either the synthetic control restriction that the weights are in the simplex is dropped, or the assumption is made that there exist a true synthetic control that is able to perfectly recover the treated unit. Motivated by this, in this paper we ask: \textit{in linear factor models when is the minimizer of the statistical risk a synthetic control}? In a simple model, we derive conditions on the primitives of the factor structure (the factor loadings) that make the target parameter be in the simplex. Then, given this target parameter, we focus on frequentist and Bayesian inferential procedures. We show under which conditions the MLE can uniformly estimate the predictive part of the factor structure for the treated unit in post-treatment periods (without the error shock) and provide conditions for a Gaussian asymptotic approximation. Then, we propose a Bayesian synthetic control that preserves the main features of the standard model and derive a Bernstein-von Mises (BvM) style result to link the frequentist and Bayesian inference.

The BvM result may be of particular interest to applied researchers as it provides a new way to perform frequentist inference using synthetic controls that is computationally appealing. While the conditions for the result require a large number of pre-treatment periods relative to the number of donor units, we show through simulations that BvM convergence can be achieved for medium size panels even in sparse settings. 

This paper contributes to the synthetic control literature in two ways. First, it provides a new characterization result for synthetic controls under linear factor models. Given that these models are often used to motivate synthetic controls, we see our result as relevant to applied researchers; we provide conditions under which factor models are well suited to study synthetic controls. Our main results reinforce the rule of thumb in the literature that without good pre-treatment fit synthetic control estimates may be misleading (\citet{abadie2021using}, \citet{abadievives21}). However, we also  expand the scope of the class of linear factor models that motivate synthetic controls. We show that the condition in \citet{ferman2021properties} that the treated unit factor loadings fall in the convex full of the donor unit factor loadings, while sufficient, might not be necessary; other factor structures may also motivate the use of synthetic controls. We see this a positive result, supporting design assumptions in synthetic control papers requiring the existence of weights that recover the true treated unit factor structure, as required for example in \citet{chernozhukov2021conformal} or as the limited confoundedness over units assumption enforces in \citet{imbens2023identificationinferencesyntheticcontrols}.

Second, this paper contributes to the literature by providing a new inferential procedure. While other Bayesian synthetic control estimators have been proposed, implementations often drop the simplex assumption. We propose and justify theoretically a Bayesian synthetic control that preserves the simplex assumption. This feature is important for interpretability and to limit extrapolation. Importantly, it provides an easy way of evaluating if synthetic controls should be used to estimate a causal effect: by checking whether the synthetic treated unit can replicate the pre-treatment outcome of interest, the researcher is able to evaluate whether the synthetic control estimator is likely be biased, and whether their Bayesian model may be miss-specified. We implement our proposed Bayesian synthetic control in the \textit{bsynth} R-package and provide additional features to help researchers understand the posterior treatement effect distribution and implicit weight estimates. 

Finally, we apply the Bayesian procedure to a classic synthetic control application studied by \citet{AbaDiaHai2015}, the German re-unification, and a more recent intervention; the attempted unilateral declaration of independence (UDI) of Catalonia in 2017. The Bayesian synthetic control yields a similar estimate to the frequentist one for the effect of the German re-unification on the GDP of West Germany. We find that the German re-unification lead to a 7.5\% decrease of the GDP per capita. The effect of the UDI on Catalan GDP is smaller, but nevertheless, we find that the UDI lead to a 1\% decrease in GDP. We posit, as \citet{catalansc} note, that this decrease is due to capital reallocation caused by the increased political instability. 

The paper proceeds as follows. Section 2 describes the standard frequentist synthetic control and conditions under which the statistical risk minimizer is a synthetic control and can be estimated consistently by MLE. Section 3 describes the Bayesian synthetic control, the Bayesian inference procedure, presents the Bernstein-von Mises result, the connection with frequentist inference through simulations and describes the \textit{bsynth} R-package. Finally, Section 4 discusses the empirical application to the German re-unification and the Catalan secession movement.

\section{The Frequentist Synthetic Control}
\label{section:penalized}
\medskip

\subsection{Standard Synthetic Control for a single unit}
Consider a setting in which we observe $J+1$ aggregate units for $T$ periods. The outcome of interest is denoted by $Y_{it}$ and only unit 1 is exposed to the intervention during periods $T_0+1, \dots, T$. We are interested in estimating the treatment effect $\tau_{1t} = Y_{1t}^I - Y_{1t}^N$ for $t>T_0$, where $Y_{1t}^I$ and $Y_{1t}^N$ denote the outcomes under the intervention and in absence of the intervention respectively. Since we do not observe $Y_{1t}^N$ for $t>T_0$ we estimate $\tau_{1t}$ by building a counterfactual $\hat{Y}_{1t}^N$ of the treated unit's outcome in absence of the intervention.

As in the standard synthetic control our counterfactual outcome will be given by a weighted average of the donor units' outcomes, that is  $\hat{Y}_{1t}^N = \sum_{j=2}^{J+1} w_jY_{j t}$ for a set of weights $\mathbf{w} = (w_2, \dots, w_{J+1})'$. To choose the weight vector $\mathbf{w}$ we use observed characteristics of the units and pre-intervention measures of the outcome of interest. Formally, we let the $K \times 1$ design matrix for the treated unit be $\mathbf{X}_1 = (Z_1, \bar{Y}_1^{\mathbf{K}_1}, \dots, \bar{Y}_1^{\mathbf{K}_M})'$, where  $\{\bar{Y}_1^{\mathbf{K}_i}\}_1^M$ represent $M$ linear combination of the outcome of interest for the pre-intervention period. Similarly, for the donor units, $\mathbf{X}_0$ is a $K\times J$ matrix constructed such that its $j$th column is given by  $(Z_j, \bar{Y}_j^{\mathbf{K}_1}, \dots, \bar{Y}_j^{\mathbf{K}_M})'$. We call the $K$ rows of the design matrices $\mathbf{X}_0$ and $\mathbf{X}_1$ the \textit{predictors} of the outcome of interest. This can include, for example, lags of the outcome variable and important context dependent characteristics of the aggregate units averaged over the pre-treatment period. 

\citet{AbaDiaHai2010} propose estimating the $\boldsymbol{w}$ by solving the following program:
	$$
	\min_{\boldsymbol w \in \Delta^J} \| \mathbf{X}_1 - \mathbf{X}_0\mathbf{w} \|_V = \left(\sum_{h=1}^k v_h (X_{h1} - W_2 X_{h2} - \dots - W_{J+1}X_{hJ+1})^2 \right)^{1/2},
	$$
\noindent where $\Delta^J$ denotes the $J-$dimensional simplex and the researcher can choose the predictor weighting matrix $V = \text{diag} (v_1, \dots, v_k)$ using his domain knowledge or using a data-driven procedure to optimize pre-treatment fit.

Given our synthetic control $\hat{\mathbf{w}}$ we can estimate our treatment effect on the treated for $t>T_0$ by:

$$
	\hat{\tau}_{1t} = Y_{1t} - \sum_{j=2}^{J+1} \hat{w}_j Y_{jt}.
$$

\noindent In the following section we motivate synthetic controls when the potential outcomes are given by linear factor models. We derive conditions under which the target parameter will be in the simplex for the case in which the design matrix includes only the pre-treatment periods and show how a MLE can estimate the treated factor structure as J and $T_0$ grow. 

\vskip 10pt

\subsection{Linear factor model}

We start by carefully examining identification and inference of the predictive part of the treated outcome in a simple factor model. Following \citet{ferman2021properties}, \citet{ferman2021synthetic} and \citet{hsiao2012panel} we consider the following linear factor model for the potential outcomes:
\begin{align*}
    Y_{it}(0) &= \boldsymbol{\lambda}_i' \mathbf{F}_t + \epsilon_{it}, \\
    Y_{it}(1) &= \tau_{it} + Y_{it}(0).
\end{align*}

\noindent The observed data $y_{it}$ is given by
$$
y_{it} = d_{it}Y_{it}(1) + (1- d_{it})Y_{it}(0),
$$

\noindent and only the first unit is treated, so $d_{it} = 1$ for $i=1$ and $t>T_0$ and $d_{it}=0$ otherwise. We make the following simplifying assumptions:
\begin{align*}
    \textbf{(A1)}& - \text{\textit{factors}} \\  
    &\text{ (a) } \text{we have only one factor such that } \lambda_i, F_t \in \mathbb{R},\\   
    &\text{ (b) } F_t \sim_{i.i.d} N(0, \sigma^2), \text{ where } 0 <\sigma < \infty, \\ 
    \textbf{(A2)}& - \text{\textit{idiosyncratic shocks}} \\
    &\text{ (a) } \epsilon_{it} \sim_{i.i.d} N(0, \sigma^2_{\epsilon}), \text{ where } 0 <\sigma_{\epsilon} < \infty.
\end{align*}

\noindent Under \textbf{A1-A2} it is shown in the appendix that the conditional distribution of $Y_{1t}$ given a realization of $\boldsymbol{Y}_{Jt} = (Y_{2t}, \dots, Y_{J+1t})$, which we denote by the lowercase $\boldsymbol{y}_{Jt}$, is

$$
Y_{1t} | \boldsymbol{y}_{Jt} \sim N\left(\tilde{\mu}, \tilde{\Sigma}\right),
$$
\noindent where
\begin{align*}
    \tilde{\mu} &= \sum_{j=2}^{J+1} w_j(\boldsymbol{\lambda}, \sigma)y_{jt}, \\
    \tilde{\Sigma} &= 1 + \lambda_1\sigma^2(1 - \sum_{j=2}^{J+1}w_j(\boldsymbol{\lambda}, \sigma)\lambda_j), \text{ and }\\
    w_j(&\boldsymbol{\lambda}, \sigma) = \frac{\sigma^2 \lambda_1 \lambda_j}{\sigma^2_{\epsilon} + \sum_{j=2}^{J+1} \lambda_j^2 \sigma^2}.
\end{align*}

Hence, conditional on the realization of the outcomes for the donor units, the distribution of the treated unit depends only on the weights $w_j(\boldsymbol{\lambda}, \sigma)$. We denote the $J\times 1$ vector of such weights by $\tilde{\boldsymbol{w}}$. While the conditions \textbf{A1-A2} seem restrictive, the main results in the paper can be extended to include settings with multiple factor loadings and a non-trivial time series component.

\subsection{Identification and characterization of synthetic controls}
In the following proposition we show that $\tilde{\boldsymbol{w}}$ is a minimizer of the statistical risk for the square loss amongst predictors that are linear combinations of the observed outcomes of the donor units.

\begin{theorem}[Linear Predictors]\label{thm_predictor}
Let $\boldsymbol{Y}_1(0)$ denote the $T_0\times 1$ vector of outcomes for the treated unit and $\boldsymbol{y}_J$ the $T_0\times J$ matrix of outcome realizations of the donor units for time periods $1,\dots, T_0$. Under assumptions \textbf{A1-A2} it follows that

$$
\tilde{\boldsymbol{w}} \in \argmin_{\boldsymbol{w}}\frac{1}{T_0} \mathbb{E}\left[ (\boldsymbol{Y}_1(0) - \boldsymbol{y}_{J}'\boldsymbol{w})'V(\boldsymbol{Y}_1(0) - \boldsymbol{y}_{J}'\boldsymbol{w})\right],
$$
for any positive semi-definite matrix $V$.
\end{theorem}

Theorem \ref{thm_predictor} is simply stating the well known fact that under the square loss the conditional expectation is the best linear predictor. In our case under the linear factor model, the conditional expectation is parametrized by the $\tilde{\boldsymbol{w}}$ weights. While $\tilde{\boldsymbol{w}}$ is a minimizer of the statistical risk amongst linear predictors, it is not immediate whether it can recover the predictive part of treated outcome in future periods, $\lambda_1 F_{t}$ for $t> T_0$, when the observations $\boldsymbol{y}_{JT}$ are viewed as random under the linear factor model. The following proposition fleshes out the conditions under which $\boldsymbol{y}_{JT_0+1}'\tilde{\boldsymbol{w}}$ converges to $\lambda_1 F_{T_0+1}$.
\begin{theorem}[Predictor convergence]\label{thm_convergence}
Given $\textbf{A1-A2}$ the following hold:

\begin{enumerate}
    \item There exist no values of $\lambda_j$ that allow $\boldsymbol{y}_{JT_0+1}'\tilde{\boldsymbol{w}} \overset{m.s.}{\to} \lambda_1 F_{T_0+1}$ as $J \to \infty$.
    \item If $\frac{1}{\|\boldsymbol{\lambda}_J\|^2_2} \sum_j |\lambda_j| \to 0$ as $J\to \infty$, then $\boldsymbol{y}_{JT_0+1}'\tilde{\boldsymbol{w}} \overset{p}{\to} \lambda_1 F_{T_0+1}$.
\end{enumerate}
\end{theorem}

Theorem 2 may look surprising as it seems to provide a negative result, but it is in line with the synthetic controls literature. Statement (1) speaks to the fact that identification and asymptotic results for synthetic control estimators often require conditioning on perfect (or good) pre-treatment fit (Abadie et al. 2010). The result can be overturned in the case in which $\sigma^2_{\epsilon}/\sigma = 0$, that is, when the noise term is completely dominated by the signal (factor structure). Statement (2) provides conditions for convergence in probability. The condition implies that as $J\to \infty$, $\|\boldsymbol{\lambda}_J\|^2_2 \to \infty$ which implies the condition in Ferman 2021 that $\|\boldsymbol{w}_J\|^2_2 \to 0$ given the analytic form of $\tilde{\boldsymbol{w}}_J$. Furthermore, as $\|\boldsymbol{\lambda}_J\|^2_2 \to \infty$ for $J \to \infty$ we also recover the treated unit factor loading:
$$
\sum_{j=2}^J \tilde{w}_j\lambda_j = \frac{\sigma^2 \lambda_1 \|\boldsymbol{\lambda}_J\|^2_2}{1 + \sigma^2 \|\boldsymbol{\lambda}_J\|^2_2} \to \lambda_1.
$$

Intuitively, unless we can distribute the error terms over all units as the donor pool grows, we are not able to get consistency. This intuition is similar to the requirements in Ferman 2021 and can be thought as justifying their results in our conditional normal setting.

Next, we consider the question of whether $\tilde{\boldsymbol{w}}$ is a \textit{synthetic control}. By this we mean whether the sum to one and non-negative constraints can be justified under our factor model. In general, if $\lambda_1$ is fixed and does not depend on other factor loadings then the result will be negative. However, if we allow $\lambda_1$ to depend on the $\boldsymbol{\lambda}_J$ then the consistency conditions and synthetic control constraints can be reconciled.

\begin{theorem}[Synthetic Control Characterization]\label{thm_sc}
For fixed $J$ under \textbf{A1}-\textbf{A2}, $\tilde{\boldsymbol{w}} \in \Delta^J$ iff the following conditions hold
\begin{enumerate}
    \item $sign(\lambda_1) = sign(\lambda_j)$ for all $j$,
    \item $\sum_j \lambda_j^2 - \lambda_1 \sum_j \lambda_j + \frac{\sigma^2_{\epsilon}}{\sigma^2} = 0.$
\end{enumerate}

\noindent Furthermore, the following statements follow
\begin{enumerate}
    \item For a fixed $\lambda_1$, a sufficient condition for the existence of sequences $\{\lambda_j\}$ such that (1) and (2) hold is that $\lambda_1^2 \geq \frac{4\sigma^2_{\epsilon}}{J\sigma^2}$.
    \item For a fixed $\lambda_1$, as $J\to \infty$ if $\frac{1}{\|\boldsymbol{\lambda}_J\|^2_2} \sum_j |\lambda_j| \to 0$ then there exist no sequences  $\{\lambda_j\}$ for which (2) and (1) hold simultaneously.
    \item Suppose condition (1) holds, let $\lambda_1 = h(\boldsymbol{\lambda}_J)$ for a component-wise weakly increasing odd function $h: \mathbb{R}^J \to \mathbb{R}$, then if as $J\to \infty$, $\| \boldsymbol{\lambda}_J\|_2^2\to \infty$, a sufficient condition for $\tilde{\boldsymbol{w}} \in \Delta^J$ is $|h(\boldsymbol{\lambda}_J)|\frac{\| \boldsymbol{\lambda}_J\|_1}{\| \boldsymbol{\lambda}_J\|_2^2} \to 1$.
    \item For any $\boldsymbol{\lambda}_J$, if $\frac{\| \boldsymbol{\lambda}_J\|_1}{\| \boldsymbol{\lambda}_J\|_2^2}\in \Delta(\boldsymbol{\lambda}_J)$ then any function $h$ such that $h(\boldsymbol{\lambda}_J) = \boldsymbol{\lambda}_J'\boldsymbol{w}$ for $\boldsymbol{w}\in \Delta_J$ satisfies the condition in (3).
\end{enumerate}
\end{theorem}

Theorem \ref{thm_sc} clarifies the conditions on the factor loadings under which the target parameter $\tilde{\boldsymbol{w}}$ is a synthetic control. The main result is that a sufficient condition for $\tilde{\boldsymbol{w}}$ to be a synthetic control is 
$$
|h(\boldsymbol{\lambda}_J)|\frac{\| \boldsymbol{\lambda}_J\|_1}{\| \boldsymbol{\lambda}_J\|_2^2} \to 1,
$$
\noindent when $\lambda_1 = h(\boldsymbol{\lambda}_J)$. Together with our sufficient conditions from Theorem \ref{thm_convergence}, this implies that $\lambda_1$ has to grow with the factor loadings of the donor units. In particular, this rules out the possibility that $\lambda_1$ is a fixed constant. This however does not imply that synthetic controls are not possible, in fact, this condition is more general than the one required by Ferman 2021. When $\frac{\| \boldsymbol{\lambda}_J\|_1}{\| \boldsymbol{\lambda}_J\|_2^2}\in \Delta(\boldsymbol{\lambda}_J)$, it follows that there exists a $\boldsymbol{w}^*$ such that $\lambda_1 = \boldsymbol{\lambda}_J'\boldsymbol{w}^*$ that satisfies the condition and, therefore, implies that as $J\to \infty$, $\boldsymbol{\tilde{w}}\in \Delta^J$. Hence, in settings in which the treated unit factor loading is in the convex hull of the donor units factor loadings the target parameter \textit{is} a synthetic control.

\subsection{Inference}
In this section we consider how to estimate $\tilde{\boldsymbol{w}}_J$ using a data set of pre-treatment outcomes $\{y_{1t}(0), \boldsymbol{y}_{Jt}(0)\}_{t=1}^{T_0}$. Given that we are interested in comparing frequentist and Bayesian procedures we focus on the maximum likelihood estimator. We do not directly observe the factor loadings, but we can estimate the $\tilde{\boldsymbol{w}}$ weights by maximizing the following \textit{pseudo} log-likelihood for parameter $\boldsymbol{\theta} = (\boldsymbol{w}, \Sigma)$:
$$
l_{T_0}(\boldsymbol{\theta}) = -\frac{1}{2}\log(2\pi\Sigma) - \frac{1}{T_0}\sum_{t=1}^{T_0}\frac{1}{2\Sigma}\left(y_{1t} - \sum_{j=2}^{J+1} w_j y_{jt}\right)^2.
$$

We derive our theoretical results for the MLE in two parts. First, in Theorem \ref{thm_mle_fixed} we show that for fixed $J$, the size of the donor pool, the MLE can recover the predictive part of the treated unit factor model and has the standard Gaussian approximation as $T_0 \to \infty$. Recall, however, that our characterization and identification results in the previous section required $J\to\infty$. Therefore, we also derive conditions under which as $J$ and $T_0$ go to $\infty$ the MLE uniformly converges to the predictive part of the treated unit factor model. The second set of results Theorem \ref{thm_mle} and Corollary \ref{cor_mle} extend results in the semi-parametric estimation literature to the synthetic control framework.

\begin{theorem}[MLE for fixed J]\label{thm_mle_fixed}
Let $\hat{\boldsymbol{\theta}}_{MLE} \in \text{argmax}_{\boldsymbol{\theta}} l_{T_0}(\boldsymbol{\theta} \in \Theta)$ for a compact parameter space $\Theta$, then under $\textbf{A1-A2}$:

\begin{enumerate}
    \item $\hat{\boldsymbol{w}}_{MLE} \overset{p}{\to} \tilde{\boldsymbol{w}}$ as $T_0 \to \infty$ for fixed $J$.
    \item  $\sqrt{T_0}(\hat{\boldsymbol{w}}_{MLE} - \tilde{\boldsymbol{w}}) \overset{a}{\sim} N(0, V_{T_0})$ as $T_0 \to \infty$ for fixed $J$, for $V_{T_0}= \frac{1}{T_0} \mathbb{E}[(\nabla_{\boldsymbol{w}} l_{T_0}(\tilde{\boldsymbol{\theta}}) \nabla_{\boldsymbol{w}} l_{T_0}(\tilde{\boldsymbol{\theta}})')^{-1}]$, where $\tilde{\boldsymbol{\theta}} = (\tilde{\boldsymbol{w}}, \tilde{\Sigma})$.
\end{enumerate}
\end{theorem}

For fixed $J$, the MLE consistency and asymptotic normality result is straightforward. Theorem \ref{thm_mle_fixed} shows that the target parameter $\tilde{\boldsymbol{w}}$ can be consistently estimated by the MLE. However, as discussed, our identification result requires that the donor pool grows with the sample size, $J\to \infty$. This means that the parameter space is growing with the sample size $T_0$. In order to account for this, Theorem \ref{thm_mle} provides conditions for uniform consistency and asymptotic normality to the target parameter as $J,T_0 \to \infty$.

\begin{theorem}[MLE with growing J]\label{thm_mle}
Let $\hat{\boldsymbol{\theta}}_{MLE} \in \text{argmax}_{\boldsymbol{\theta}} l_{T_0}(\boldsymbol{\theta} \in \Theta)$ for a compact parameter space $\Theta$, then under $\textbf{A1-A2}$ and $\lambda_j$ are uniformly bounded:

\begin{enumerate}
    \item $\frac{1}{T_0}\sum_t \boldsymbol{y}_{Jt}\boldsymbol{y}_{Jt}' = D_{T_0}$ where $0<\text{lim inf}_{T_0} \sigma_{min}(D_{T_0}) \leq \text{lim sup}_{T_0} \sigma_{max}(D_{T_0}) < \infty$,
    \item $\text{max}_{t\leq T_0} \|\boldsymbol{y}_{Jt}\|^2_2 = O_p(J)$,
    \item $\sup_{\boldsymbol{\beta}, \boldsymbol{\gamma} \in \mathcal{S}_J(1)} \sum_t |\boldsymbol{y}_{Jt}'\boldsymbol{\beta}|^2 |\boldsymbol{y}_{Jt}'\boldsymbol{\gamma}|^2 = O_p(T_0)$.
\end{enumerate}
Then, it follows that if $o(T_0) = J (\log J)^3$
$$
\| \hat{\boldsymbol{w}}_{MLE} - \tilde{\boldsymbol{w}}\|_2^2 = O_p(J/T_0).
$$

\noindent If $o(T_0) = J^2 \log(J)$ then
$$
\sqrt{T_0} \boldsymbol{\alpha}'(\hat{\boldsymbol{w}}_{MLE} - \tilde{\boldsymbol{w}})/\sigma_{\alpha} \overset{d}{\to} N(0,1),
$$
for any $\boldsymbol{\alpha} \in S_J(1)$ and 
$$
\sigma^2_{\alpha} = (\mathbb{E}[\epsilon^2_{Jt}]) \boldsymbol{\alpha}'D_{T_0}^{-1}\boldsymbol{\alpha},
$$
where $S_J(1)$ denotes the Euclidean ball of radius 1 in $\mathbb{R}^J$.
\end{theorem}

Theorem \ref{thm_mle} shows that $L^2$ norm convergence and uniform Gaussian approximation is possible when $T_0$ grows at rate faster $J$. In particular, we require that $T_0$ grows faster than $J$ for the consistency result and that $T_0$ grows faster than $J^2$ for the asymptotic normality result. While these rates might be impractical in settings with small $T_0$, they speak to the discussion in Abadie et al. 2010 and Abadie and Vives-i-Bastida 2021 that large donor pools may increase the bias in synthetic control estimators. Intuitively, larger donor pools imply more parameters to estimate which might increase the finite sample bias of the estimator. Next, we apply this result to our specific setting when we use the data $\boldsymbol{y}_{JT_0+1}$ to predict the treated unit outcome in absence of the intervention.

\begin{corollary}\label{cor_mle}
Under the conditions of Theorem \ref{thm_mle}, as $J,T_0\to \infty$:

\begin{enumerate}
    \item If $o(T_0) = J (\log J)^3$ and $\frac{1}{\|\boldsymbol{\lambda}_J\|^2_2} \sum_j |\lambda_j| \to 0$, then $$\boldsymbol{y}_{JT_0+1}'\hat{\boldsymbol{w}}_{MLE} \overset{p}{\to} \lambda_1 F_{T_0+1}.$$
    \item If $o(T_0) = J^2 \log(J)$ and $\frac{1}{\|\boldsymbol{\lambda}_J\|^2_2} \sum_j |\lambda_j| \to 0$, then 
$$
\sqrt{T_0} (\boldsymbol{y}_{JT_0+1}'\hat{\boldsymbol{w}}_{MLE} - \lambda_1F_{T_0+1})/\sigma_{y_{JT_0+1}} \overset{d}{\to} N(0,1).
$$
\end{enumerate}
\end{corollary}

Corollary \ref{cor_mle} provides conditions for valid frequentist inference to the predictive part of the treated unit factor model as $T_0,J \to \infty$. Similar semi-parametric results have also been derived in \citet{ferman2021properties}. Our results are different in that they apply to a wider class of models under our characterization conditions and provide explicit rate conditions for $J$ and $T_0$. Indeed, all conditions are imposed directly on the factor structure.

\section{The Bayesian Synthetic Control}

We propose the Bayesian equivalent of the program described in section 2 to generate synthetic controls. In particular, our Bayesian formulation includes two key aspects of synthetic controls: (1) the synthetic treated unit is a convex combination of the donor units, that is, we do not want to extrapolate outside the convex hull of the donor units and (2) we construct the synthetic control by matching the predictors of treated unit and donor units for the outcome of interest. An advantage of the Bayesian approach is that we can directly quantify the uncertainty in our estimates. In this section, first we discuss a valid Bayesian inference procedure and then the Bayesian synthetic control model.

\subsection{Bayesian Inference}

We derive conditions for valid Bayesian inference in a general setting that includes our set up. Following \citet{imbensrubin} and \citet{pang2022} we consider the Bayesian inference problem as a missing data problem. Let the adoption process of the intervention for one treated unit be encoded by a treatment random variable $\mathbf{d}_1 = (d_{11}, \dots, d_{1T})$ that is fully determined by an adoption time random variable $a_1$. Similarly, we can define these variables for the donor pool as $\mathbf{D}$ and $\mathbf{a}$.

The potential outcome function of our outcome of interest $Y_{it}$ under SUTVA is given by $Y_{it}(\mathbf{d}_i(a_i)) = Y_{it}(a_i)$, where no anticipation implies that $Y_{it}(a_i) = Y_{it}(c)$ for $t<a_i$ and $c$ denoting the counterfactual state in which the unit is never treated.

\textbf{Assumptions}:
\begin{enumerate}
    \item \textbf{(B1) - (B2)}: SUTVA and no anticipation. 
    \item \textbf{(B3)} \textbf{Latent Ignorability}: there exist latent variables $\mathbf{U}_i = (u_{i1}, \dots,u_{iT})$ such that:
    $$
    \mathbf{y}_i(0) \perp \mathbf{D}_i | \mathbf{U}_i.
    $$
    \item \textbf{(B4)} \textbf{Exchangeability}: given $\mathbf{U}$, permutations of the indices $it$ of the sequence $\{(Y_{it}(0)\}_{i\in [N], t \in [T}$ do not alter the joint distribution.
\end{enumerate}

The above assumptions allow us to get an expression for the posterior on the "missing" data, the outcome counterfactual for treated units after treatment assignment if they had not been treated. The predictive inference consists in sampling from the posterior distribution of the missing data conditional on the observed data and a model indexed by parameter vector $\boldsymbol{\theta}$ with prior distribution $\pi(\mathbf{\theta})$. Let the missing and observed data be denoted by $\mathbf{Y}^{mis} = \{Y_{1t}\}_{t>T_0}$ and $\mathbf{Y}^{obs} = (\{Y_{it}\}_{i\in [J+1], t\leq T_0}$, $\{ Y_{jt}\}_{j\in [J], t>T_0}, \textbf{D})$. Then, under \textbf{B1-B3} we can factor out the assignment:
\begin{align*}
    P(\mathbf{Y}^{mis} | \mathbf{Y}(0)^{obs}, \mathbf{D}, \mathbf{\theta}) &\propto P(\mathbf{Y}(0)^{mis}, \mathbf{Y}(0)^{obs}, \mathbf{D}, \mathbf{\theta}) \\
    &\propto P(\mathbf{Y}(0), \mathbf{U}) P(\mathbf{D} | \mathbf{Y}(0), \mathbf{\theta}) \\ 
    &\propto P(\mathbf{Y}(0), \mathbf{\theta}) P(\mathbf{D}| \mathbf{\theta}) \\
    &\propto P(\mathbf{Y}(0), \theta).
\end{align*}

\noindent Then under \textbf{B4} by de Finetti's Theorem we can separate out the posterior predictive distribution and the likelihood
\small
    \begin{align*}
        P(\mathbf{Y}(0)^{mis} | \mathbf{Y}(0)^{obs}, \mathbf{D}, \theta) &\propto P(\mathbf{Y}(0), \theta) \\
        &\propto \int \Pi_{it} f(Y_{it}(0) | \theta ) \pi(\theta) d\theta\\
        &\propto \int \underbrace{\left(\Pi_{it \in \text{ mis}} f(Y_{it}(0)^{mis} | \theta)\right)}_{\text{Posterior Predictive Distribution}}  \underbrace{\left(\Pi_{it \in \text{ obs}} f(Y_{it}(0)^{obs} | \theta)\right)}_{Likelihood} \pi(\theta) d\theta
    \end{align*}
\noindent where $f$ is the marginal density of the potentially observable data.

The assumptions imposed by our sampling model \textbf{A1}-\textbf{A2} in which the data is i.i.d and the potential outcomes are given by a latent factor model implicitly satisfy \textbf{B1}-\textbf{B4}. Therefore, under our sampling we will be able to sample from the Bayes posterior distribution to recover the distribution of the missing data and recover the distribution of the target treatment effect. In order to estimate the model, we use a HMC sampler (NUTS) to get $M$ draws $\{\theta^m\}$ and then our predictive posterior is given by the realizations of
$$
\tau_{1t}^m = Y_{1t}(1) - y^m_{1t}(0),
$$
In Section 3.5 we describe the Bayesian estimation procedure we implement in the \textit{bsynth} package in more detail. In the next section, we consider under which conditions on the prior model can we use a Bayes estimator to approximate the frequentist inference.

\subsection{Bayesian Model}

A Bayesian model imposes a functional form for the prior distribution and data density $f$, by modelling explicitly the potential outcome in absence of treatment. In the previous section we showed under which conditions can we perform valid Bayesian inference. In this section, we impose restrictions on the prior structure. We start by considering a Gaussian prior model
\begin{align*}
y_{1t} | \boldsymbol{y}_{Jt}, \boldsymbol{w}, \sigma_y &\sim N( \boldsymbol{y}_{Jt}'\boldsymbol{w}, \sigma_y^2), \\
w_j | \boldsymbol{y}_{Jt} &\sim N(\mu_j, \tau^2_j).
\end{align*}

\noindent Given that the Gaussian conjugate prior is Gaussian, it can be shown that the 
\textbf{Bayes estimator} for the implicit weights is given by 
$$\hat{w}^B_j = \mathbb{E}_B[w_j | \boldsymbol{y}_t] = \int w_j p(w_j | \boldsymbol{y}_t) dw_j.$$

\noindent Furthermore, in this case the predictive posterior distribution is normally distributed and

\begin{align*}
    \hat{Y}_{1t}^B  =  \boldsymbol{y}_{Jt}'\mathbb{E}_B[w_j | \boldsymbol{y}_t] &= \frac{\sigma_y^2}{\sigma_y^2 + \sum_j \tau_j^2}\boldsymbol{y}_{Jt}'\mu_J + \frac{\sum_j \tau_j^2}{\sigma^2_y + \sum_j \tau_j^2}y_{1t}, \\
    \mathbb{V}_B(\boldsymbol{y}_{Jt}'\boldsymbol{w} | \boldsymbol{y}_{t}) &= \frac{\sigma_y^2 \sum_j\tau_j^2 }{\sigma_y^2 + \sum_j\tau_j^2}.
\end{align*}

Motivated by the characterization of synthetic controls in Theorem \ref{thm_sc} we consider Bayesian models in which the parameters $\mu_j$ are in the simplex. This suggests adding the following restriction:

$$
\mu_j \sim Dir(1),
$$

\noindent where $Dir(1)$ denotes the Dirichlet distribution with scale one. Intuitively, this restriction forces the means of the target weights to be in the simplex.

In Section 4 we describe alternative Bayesian models implemented in the \textit{bsynth} package that allow for additional covariates and a Gaussian process term. The Bayesian synthetic control with additional covariates, which has not been implemented in the literature yet, and the use of Gaussian processes, also suggested by \citet{arbour2021}. Our theoretical results could potentially be extended to apply to both cases with additional regularity conditions. In the case of the Gaussian process note that in a general setting a BvM result for Gaussian processes was derived in \citet{Ray_2020}. In the next section, we link the Bayesian and frequentist inference for the baseline Bayesian model.

\subsection{Bernstein-von Mises Result}

In this section we consider how the Bayesian inference can be used to approximate the frequentist inference. We derive a Bernstein-von Mises style result in which under the correct prior specification the Bayesian posterior predictive distribution converges in the total variation sense to the MLE sample distribution as $T_0,J\to \infty$. Intuitively, our result states that if we assume that the factor loading of the treated unit can be recovered by a convex combination of the treated units then under the same assumptions that yield a valid MLE estimator, the Bayes estimator is able to consistently estimate the predictive term $\lambda_1 F_{T_0+1}$. If additionally, the uncertainty in our predictions converges to the frequentist sampling variance, then the two estimators distributions are close in the total variation sense.

\begin{theorem}[BvM]\label{thm_bvm}
    Under \textbf{A1-A2}, the assumptions of Corollary \ref{cor_mle} and
    \begin{enumerate}
        \item \textbf{Prior conditions}: $\|\mu_J\|_2^2 \to 0$, $\{\tau_j\}$ such that $\sum_j \tau_j^2 = O(J^{\alpha})$, for $0<\alpha<1$, as $J\to \infty$, and $\sigma_y \to \sigma_{\epsilon}$. 
        \item \textbf{Convex recovery}: $\| \lambda_1 - \boldsymbol{\lambda_J}'\boldsymbol{\mu}_J\|_2 \to 0$ as $J\to \infty$.
    \end{enumerate}
    Then, as $T_0,J \to \infty$ at rate $o(T_0) = J^2 \log(J)$,
    $$
    \boldsymbol{y}_{JT_0+1}'\mathbb{E}_B[\boldsymbol{w} | \boldsymbol{y}_{T_0}] \overset{p} \to \lambda_1F_{T_0+1},
    $$
    and 
    $$
    \| \Phi^{MLE}_{T_0,J} - Q_{T_0,J}\|_{TV} \to 0,
    $$
    where $\Phi^{MLE}_{T_0+1,J}$ denotes the MLE finite sample distribution and $Q_{T_0+1,J}$ the Bayes posterior predictive distribution.
\end{theorem}
    
Theorem \ref{thm_bvm} imposes strong conditions on the class of priors necessary to approximate the MLE distribution. In particular, it is key that we choose the $\mu_j$ in the simplex and in a way that recovers the treated unit factor loading. The requirement that such a sequence of priors exists is motivated by our characterization conditions in Theorem \ref{thm_sc}. In the same spirit as in the frequentist synthetic control, if the Bayesian synthetic control is unable to find implicit weights that replicate the treated unit outcomes in the pre-treatment period, then it is likely the Bayesian synthetic control will be biased as the model is miss-specified. 

The other requirement in the proof of Theorem \ref{thm_bvm} is that the Bayesian posterior is Gaussian. While this requirement is important for the proof method, which relies on the analytical form of the KL divergence between Gaussian distributions, it is not a necessary requirement. A more general result could be derived by imposing weaker restrictions on the functional form and the second moments of the posterior distribution. 

The BvM result in Theorem \ref{thm_bvm} is important because it gives conditions under which researchers might be able to interpret their Bayesian credible intervals as valid confidence intervals. As $T_0, J\to \infty$ under the conditions of Theorem \ref{thm_bvm} the $1-\alpha$ credible interval defined by the limit Bayesian posterior predictive distribution and the $1-\alpha$ confidence interval defined by the MLE estimator will coincide. Hence, using Bayesian synthetic controls offers a new way of performing valid asymptotic inference for synthetic controls without the need of exact inference or permutation tests.

\subsection{Simulation Evidence}

In this section we compare the standard synthetic control and the Bayesian synthetic control for a grouped linear factor model data generating process. Similar data generating processes have been used to study the properties of synthetic controls estimators, for example in \citet{Firpo2018sensitivity} or in \citet{abadievives21}. In particular, we let the potential outcome in absence of intervention be given by
\begin{align}\label{grouped_fm}
    Y_{it}(0) = \lambda_{f(i)t} + \epsilon_{it}.
\end{align}
\noindent where the $\lambda_{ft}$ follow an $AR(1)$ with $\rho=0.5$, standard Gaussian innovations and noise given by $\epsilon_{it} \sim N(0,\sigma^2)$. Without loss of generality we assume that only unit 1 is treated and that the treatment effect is zero. We consider two designs: a \textit{dense} design in which units are grouped in 2 groups of 10 units, such that $f(1)=f(2)=\dots=f(10)$, and a \textit{sparse} design in which units are paired in 10 groups, such that $f(1)=f(2)$. In the \textit{dense} case the optimal synthetic control assigns equal weight to units 2 to 10, whereas in the \textit{sparse} case the optimal synthetic control puts all the weight on unit 2.

The simulation settings satisfy the convex recovery assumptions of Theorem \ref{thm_sc} and Theorem $\ref{thm_bvm}$ as there exists a sequence of weights that perfectly recreates the factor loading of the treated unit. In the sparse case, for example,  $w_2 =1$ and $w_j=0$ for all $j>2$. However, each setting does not necessarily satisfy the \textit{density} condition, or the prior condition, meaning that the target parameter does not satisfy that $\| \tilde{\boldsymbol{w}}\| \to 0$ as $J\to \infty$. In fact, to check the relevance of the proposed Bayesian method and the theoretical results we consider a realistic case with a fixed number of units $J = 20$. We will see however, that in this realistic setting, there will still be BvM convergence in both the \textit{dense} and \textit{sparse} designs as $T_0\to \infty$.

\begin{figure}[H]
\centering
  \begin{subfigure}[b]{0.5\textwidth}
  \centering
  	\includegraphics[width=1\linewidth]{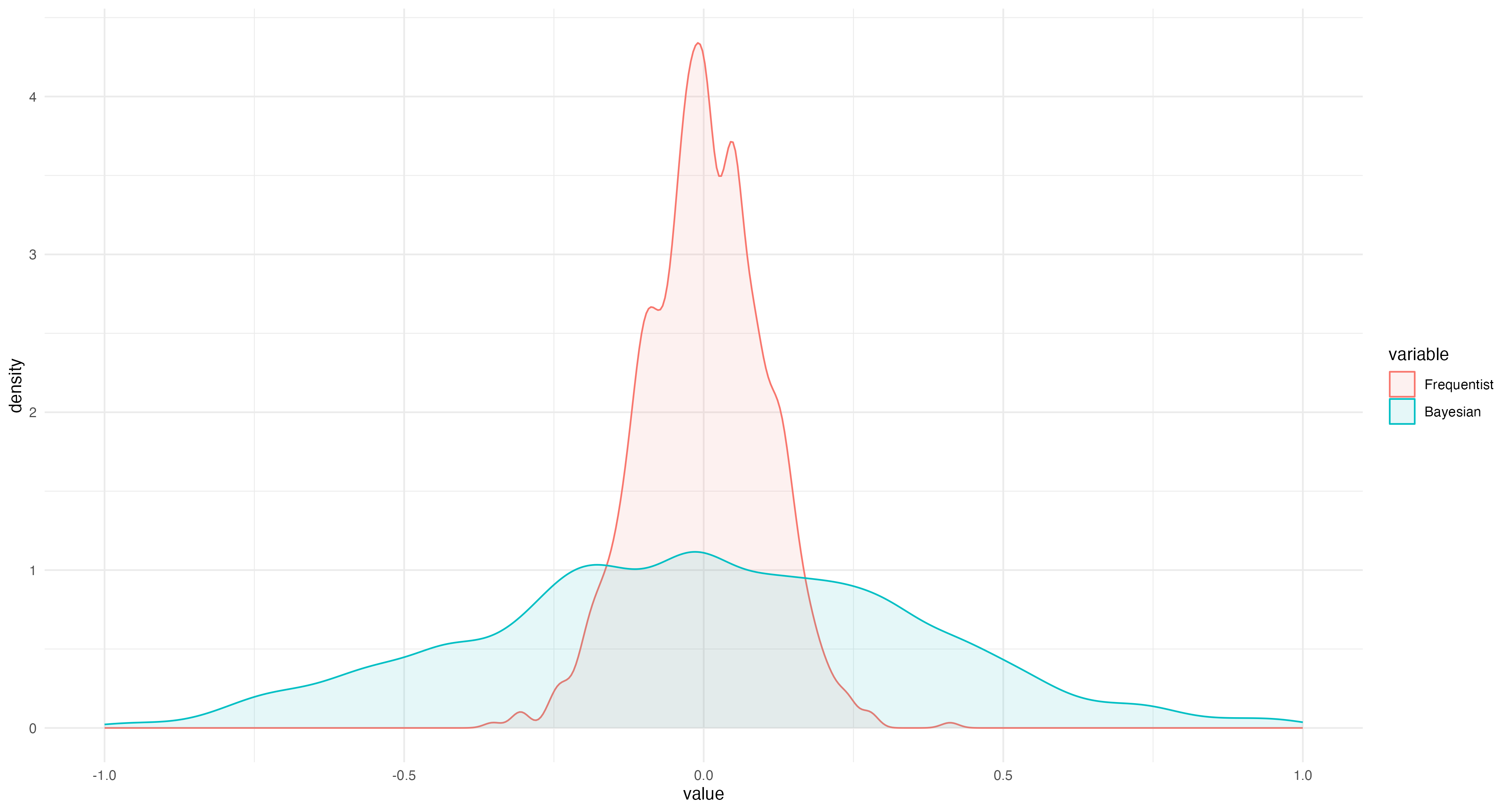}
  	\caption{$T_0 = 10$}
  \end{subfigure}%
  \begin{subfigure}[b]{0.5\textwidth}
  \centering
  	\includegraphics[width=1\linewidth]{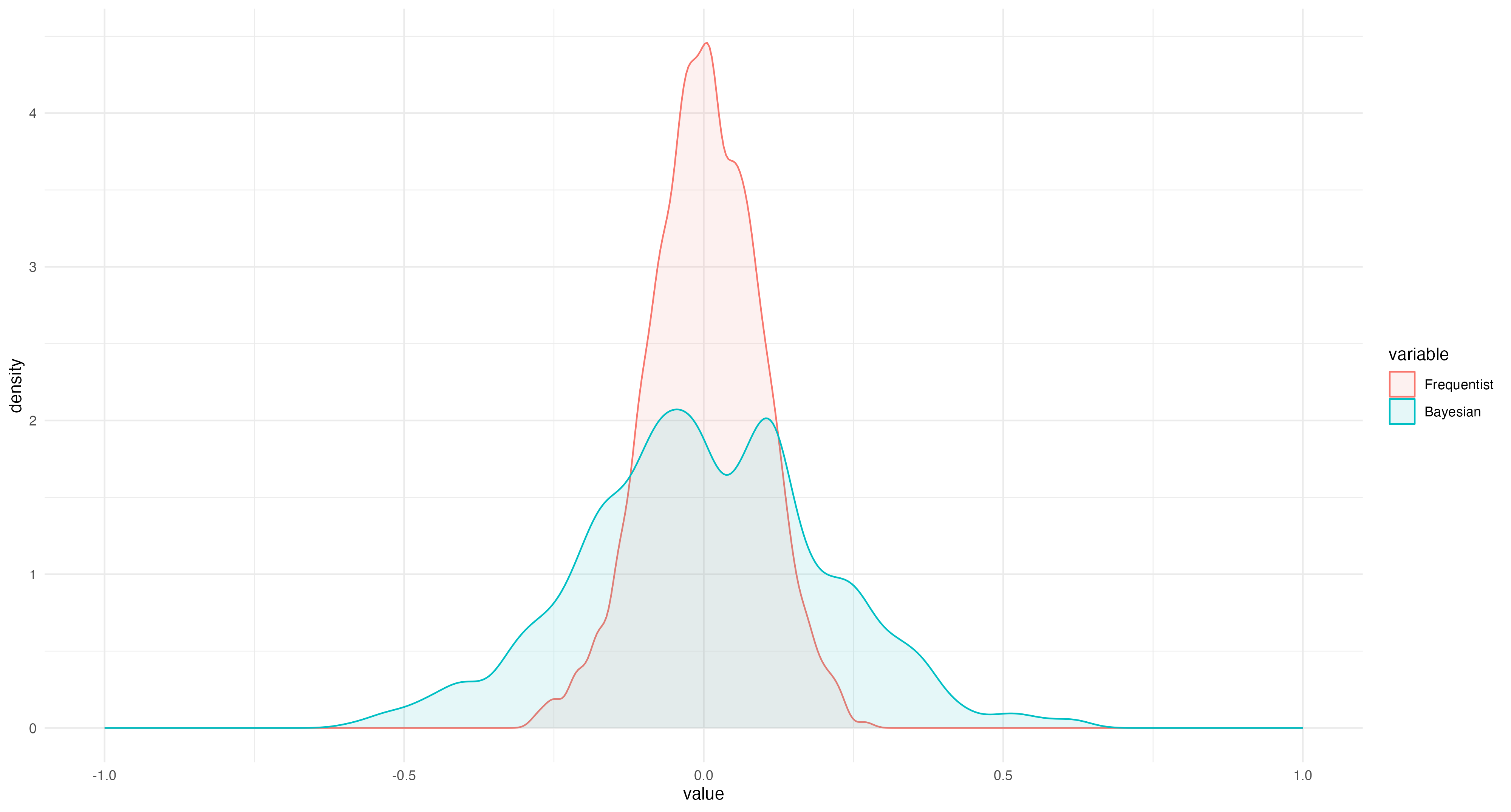}
  	\caption{$T_0 = 20$}
  \end{subfigure}%
  
   \begin{subfigure}[b]{0.5\textwidth}
  \centering
  	\includegraphics[width=1\linewidth]{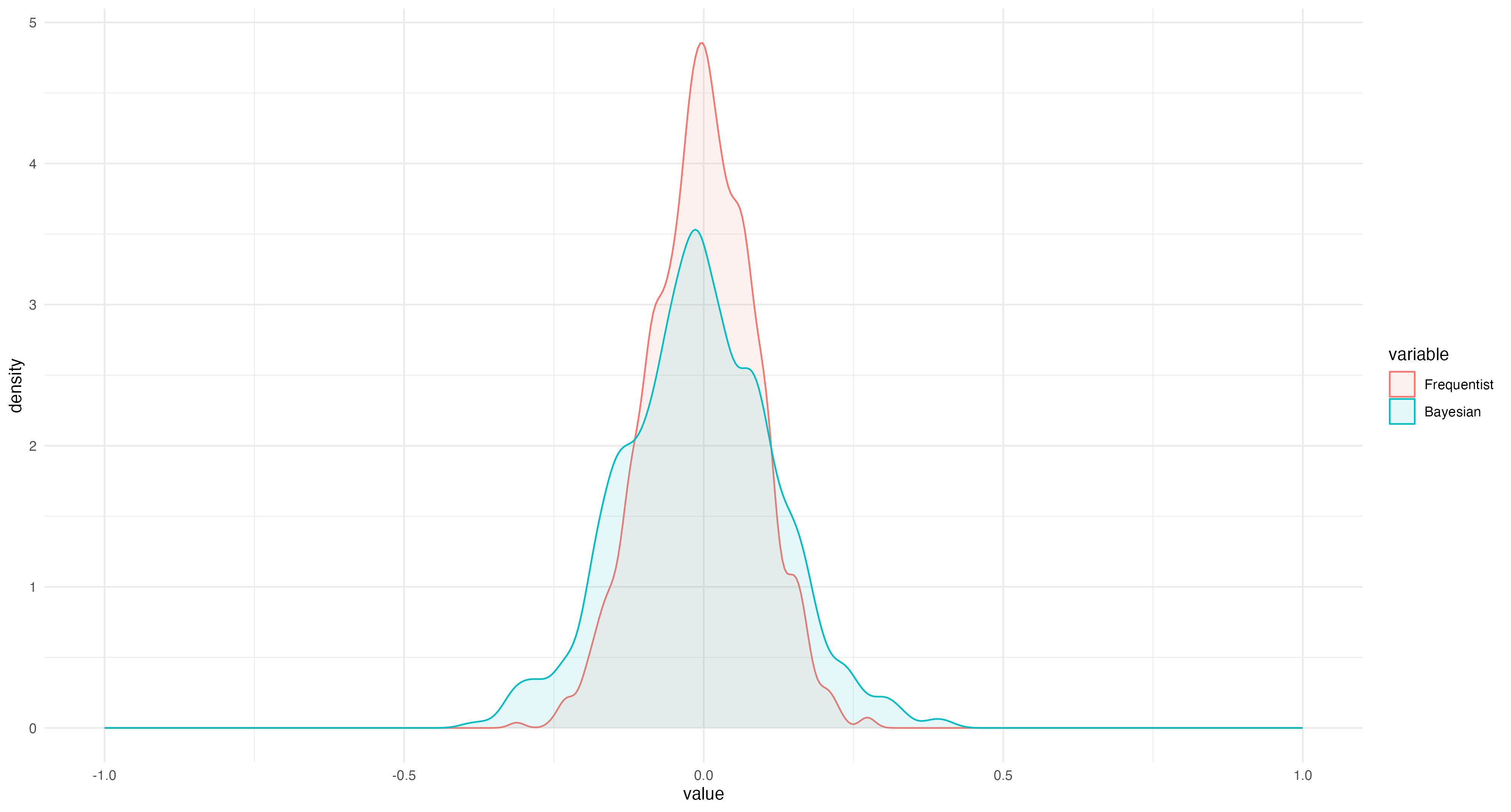}
  	\caption{$T_0 = 40$}
  \end{subfigure}%
  \begin{subfigure}[b]{0.5\textwidth}
  \centering
  	\includegraphics[width=1\linewidth]{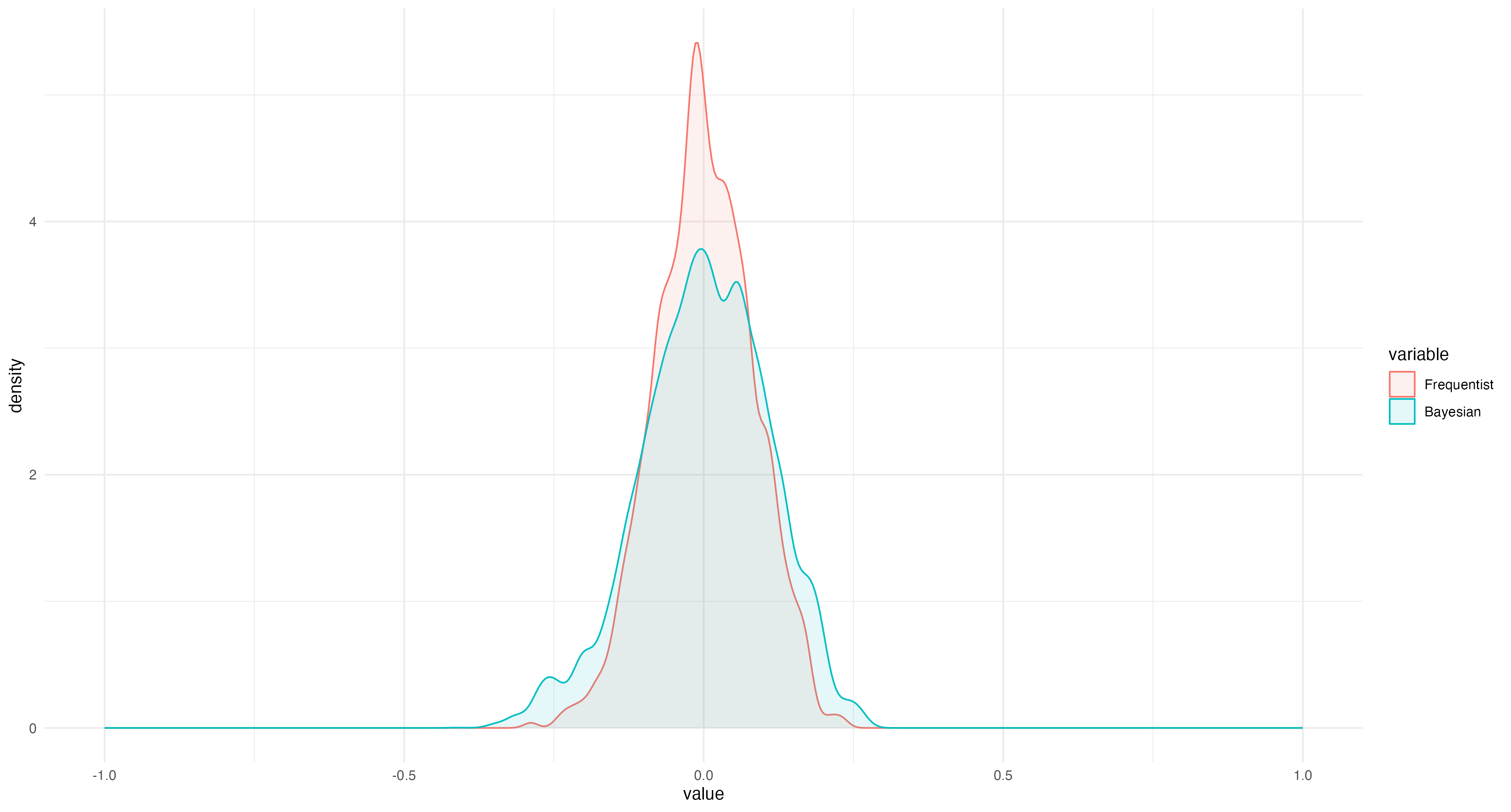}
  	\caption{$T_0 = 60$}
  \end{subfigure}%

     \begin{subfigure}[b]{0.5\textwidth}
  \centering
  	\includegraphics[width=1\linewidth]{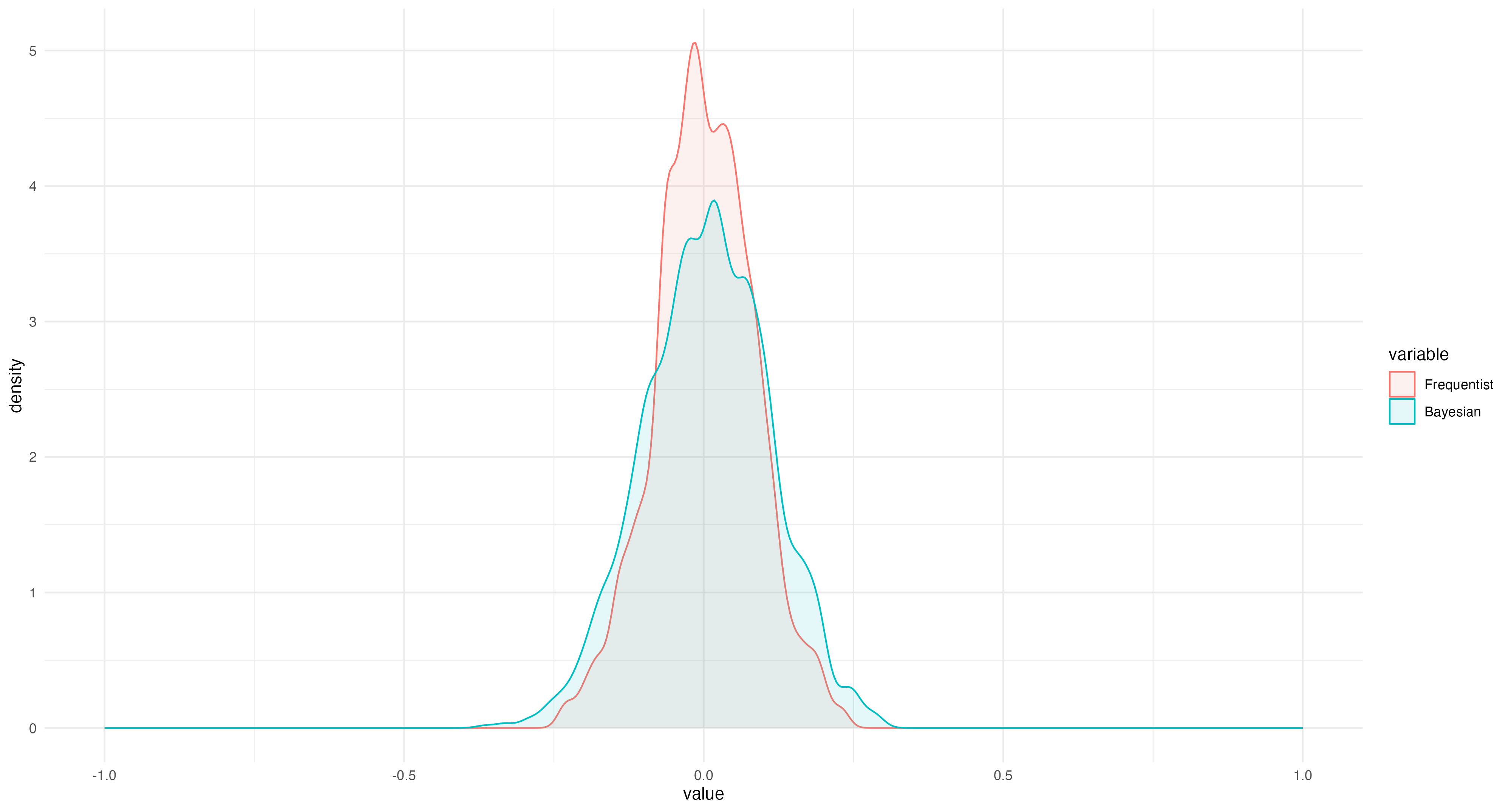}
  	\caption{$T_0 = 80$}
  \end{subfigure}%
  \begin{subfigure}[b]{0.5\textwidth}
  \centering
  	\includegraphics[width=1\linewidth]{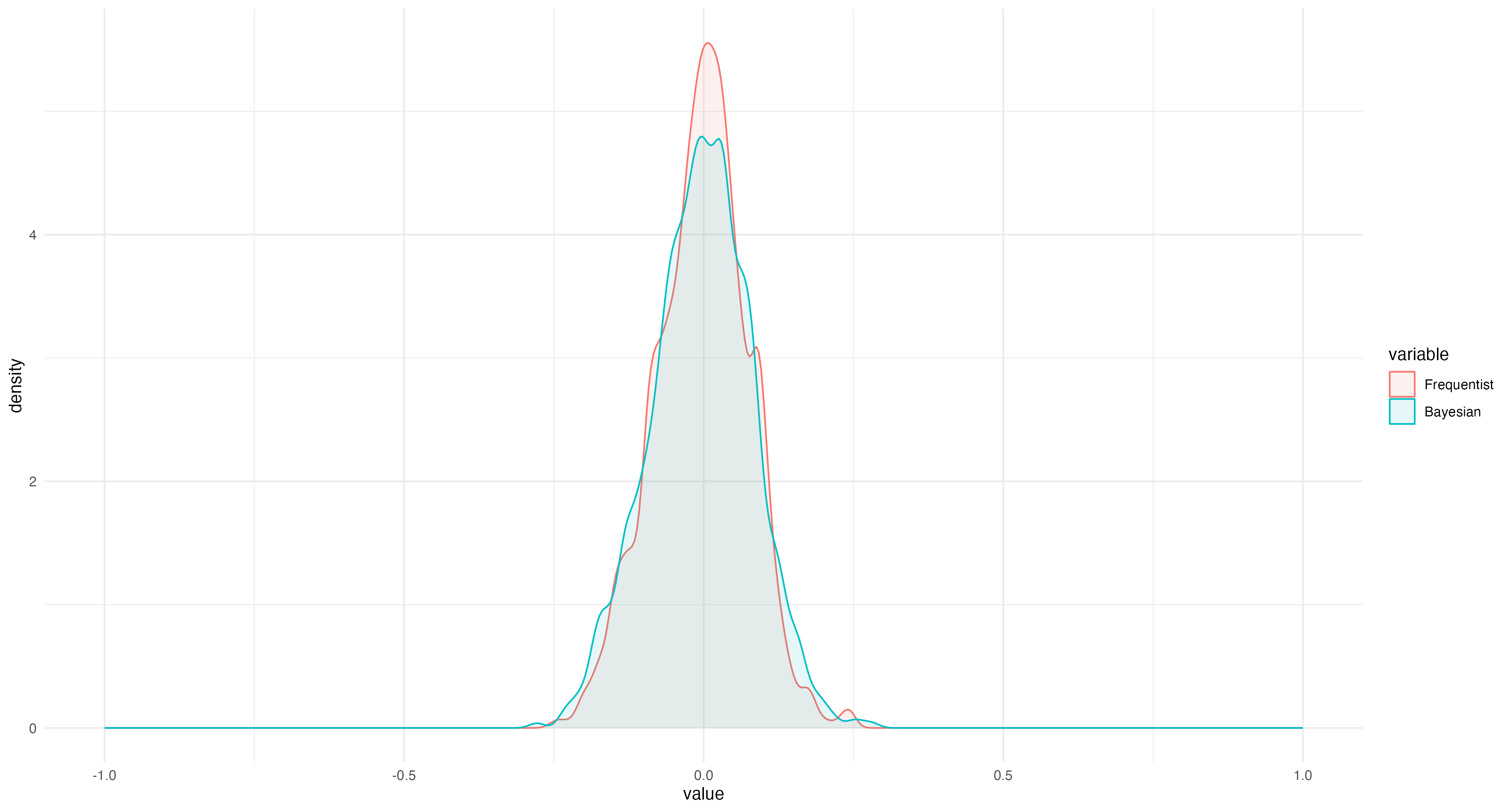}
  	\caption{$T_0 = 200$}
  \end{subfigure}%
  \caption{Convergence of frequentist and Bayesian coverage as $T\to\infty$ for the dense case.}
    \begin{tablenotes}
		\small\item\textbf{Notes}: Kernel densities of the frequentist empirical distribution of the estimated treatment effect (in red) and the Bayesian posterior mean (in blue) over 10000 draws for different values of $T_0$. The potential outcomes are generated by the grouped factor model (\ref{grouped_fm}) with $\sigma = 0.25$ in a dense design with factor loadings grouped in two groups of 10 units.
   \end{tablenotes}
  \label{fig_bvm_dense}
\end{figure}

In our simulation analysis, for each $T_0$ we estimate the standard synthetic control and report our estimated treatment effect on the treated for 10000 draws. We compute the treatment effect over 10 post-treatment periods $\hat{\tau}_1 = \frac{1}{T-T_0}\sum_t \hat{\tau}_{1t}$, which allows us to average over the additional noise terms $\epsilon_{it}$ for $t>T_0$. We compare this empirical distribution to the mean posterior predictive distribution of our Bayesian model for the 10000 draws. Figure \ref{fig_bvm_dense} shows the empirical distribution for each estimator in the dense setting for different values of $T_0$.

\begin{figure}[H]
\centering  
   \begin{subfigure}[b]{0.5\textwidth}
  \centering
  	\includegraphics[width=1\linewidth]{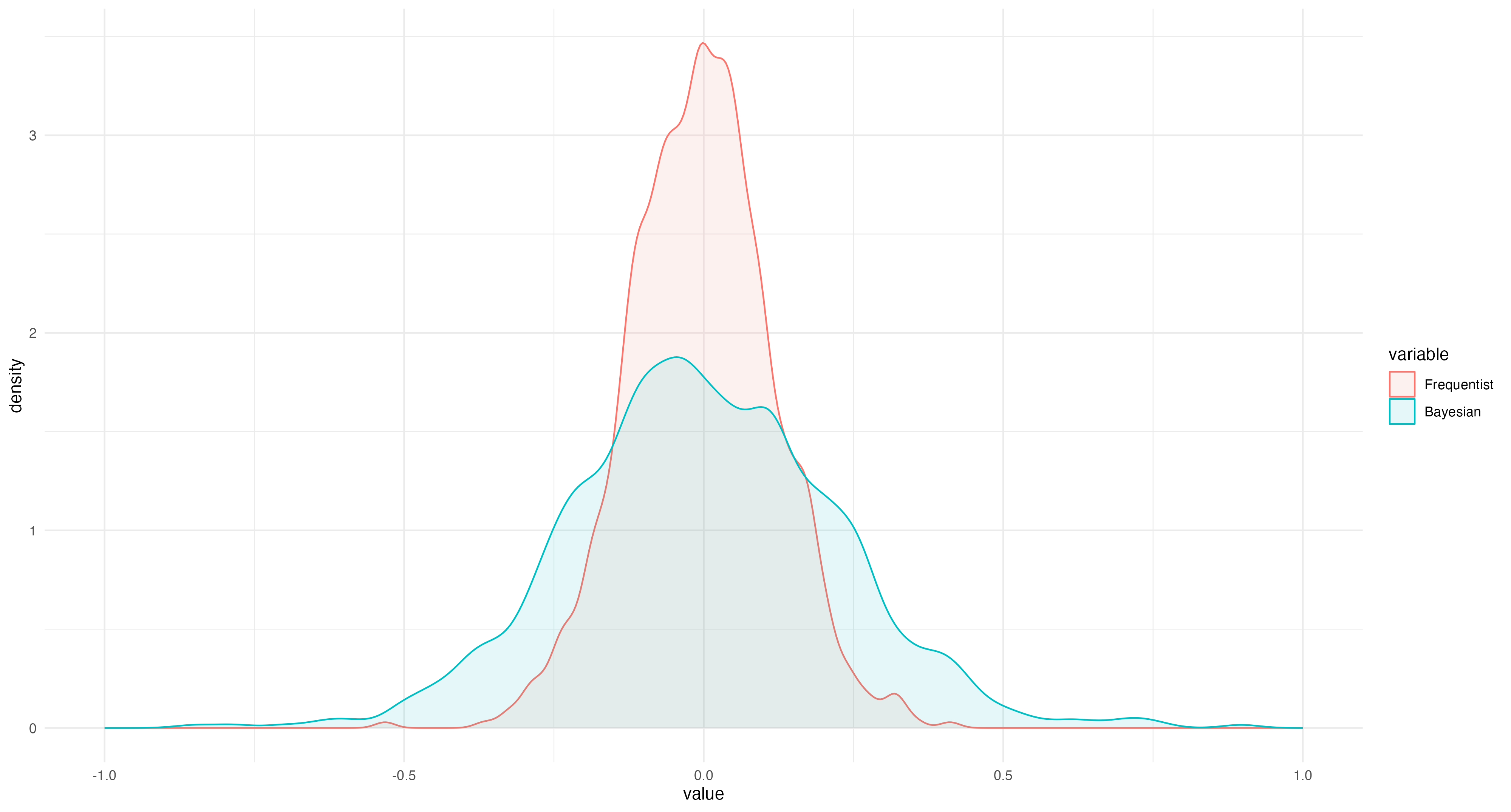}
  	\caption{$T_0 = 40$}
  \end{subfigure}%
  \begin{subfigure}[b]{0.5\textwidth}
  \centering
  	\includegraphics[width=1\linewidth]{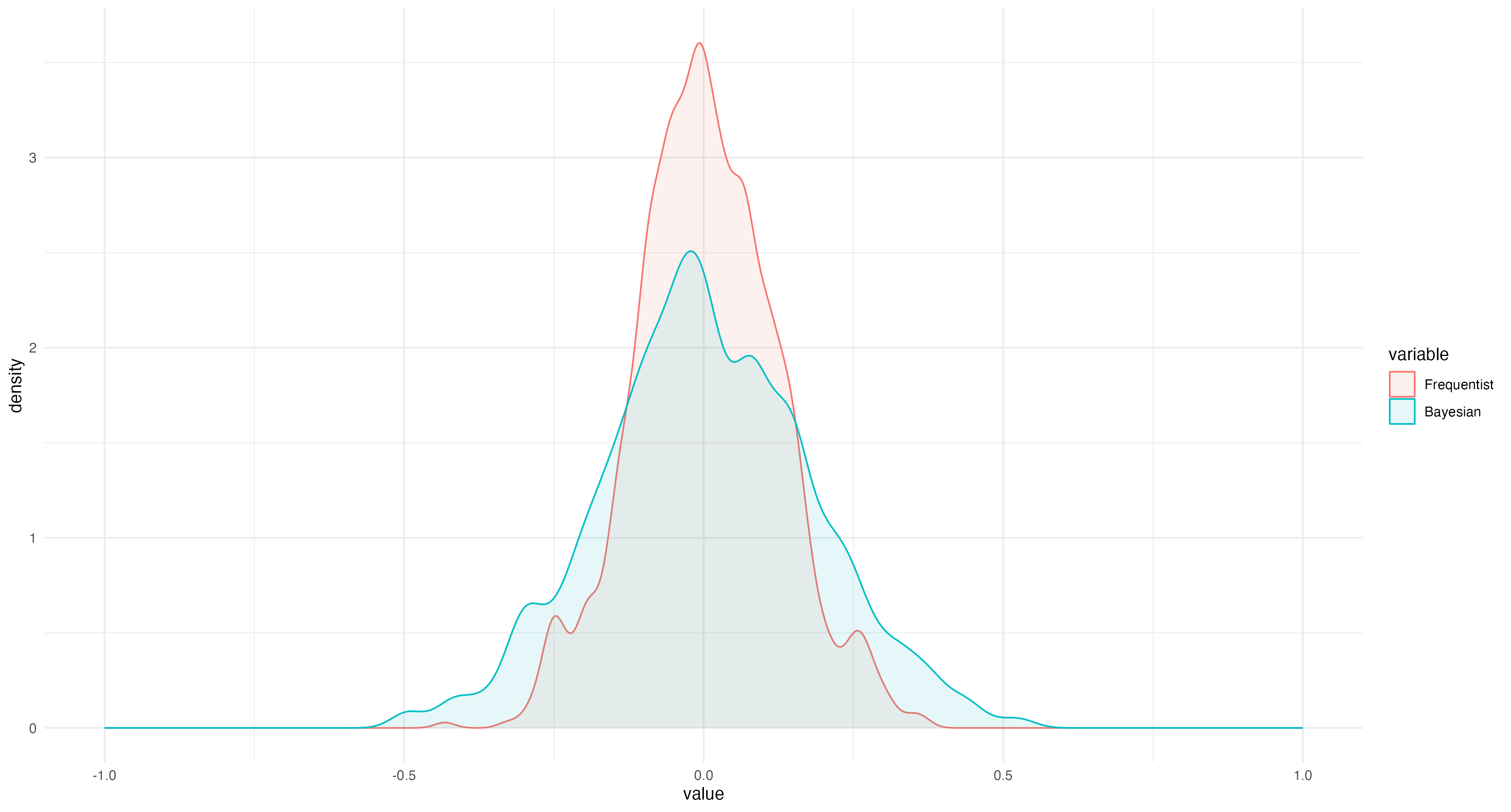}
  	\caption{$T_0 = 60$}
  \end{subfigure}%

     \begin{subfigure}[b]{0.5\textwidth}
  \centering
  	\includegraphics[width=1\linewidth]{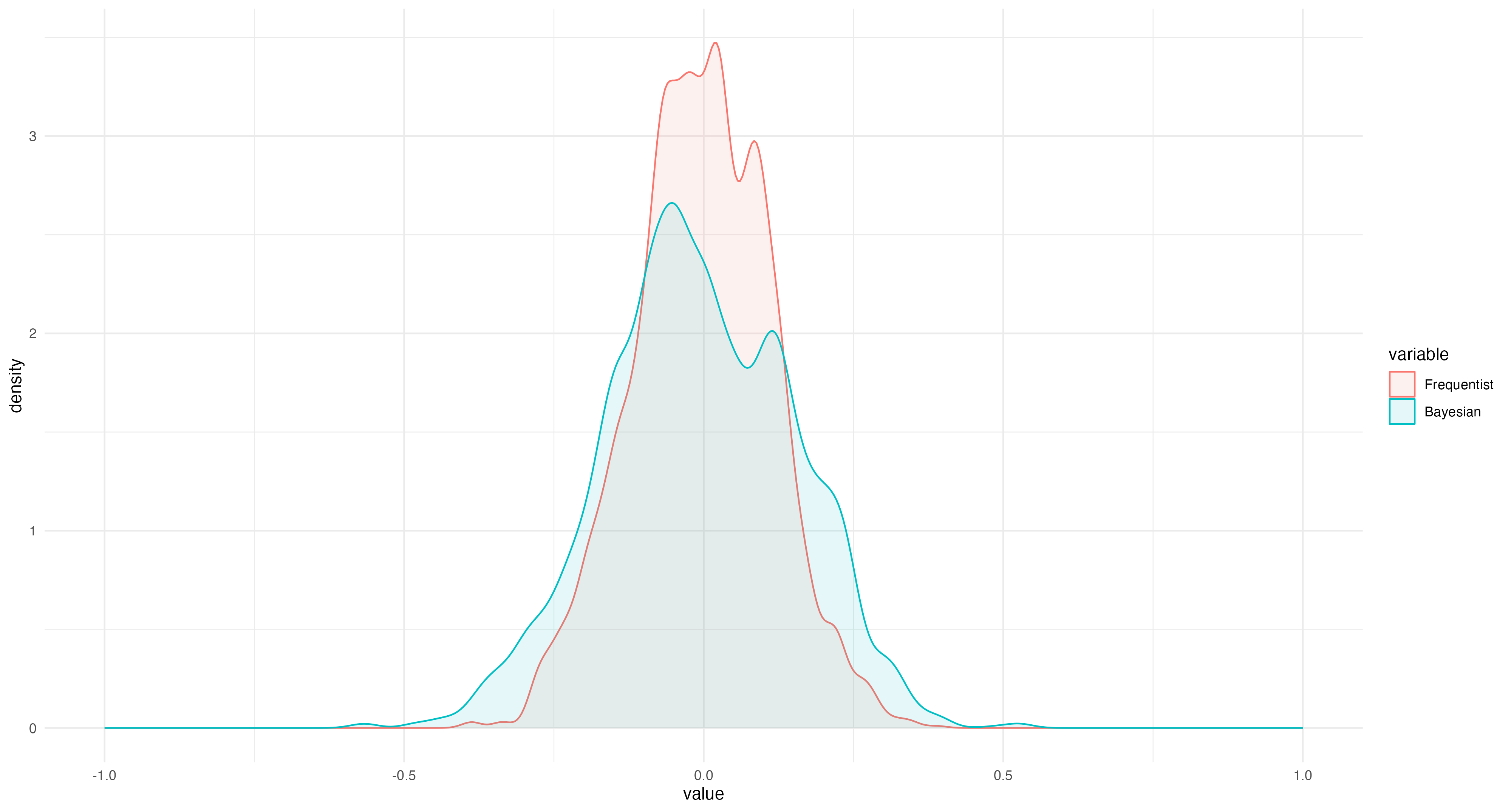}
  	\caption{$T_0 = 80$}
  \end{subfigure}%
  \begin{subfigure}[b]{0.5\textwidth}
  \centering
  	\includegraphics[width=1\linewidth]{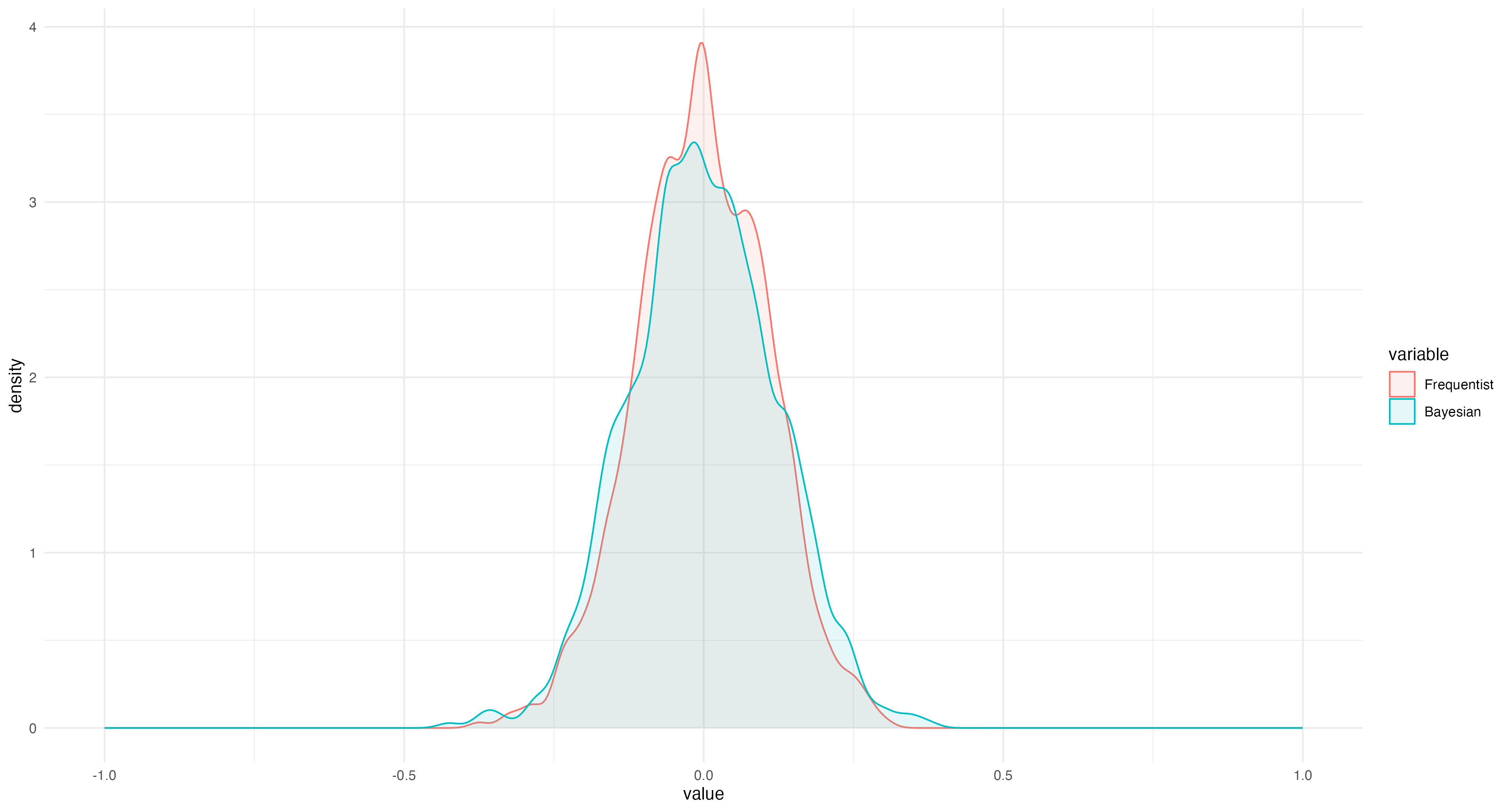}
  	\caption{$T_0 = 200$}
  \end{subfigure}%
  
  \centering
    \begin{subfigure}[b]{0.5\textwidth}
  \centering
  	\includegraphics[width=1\linewidth]{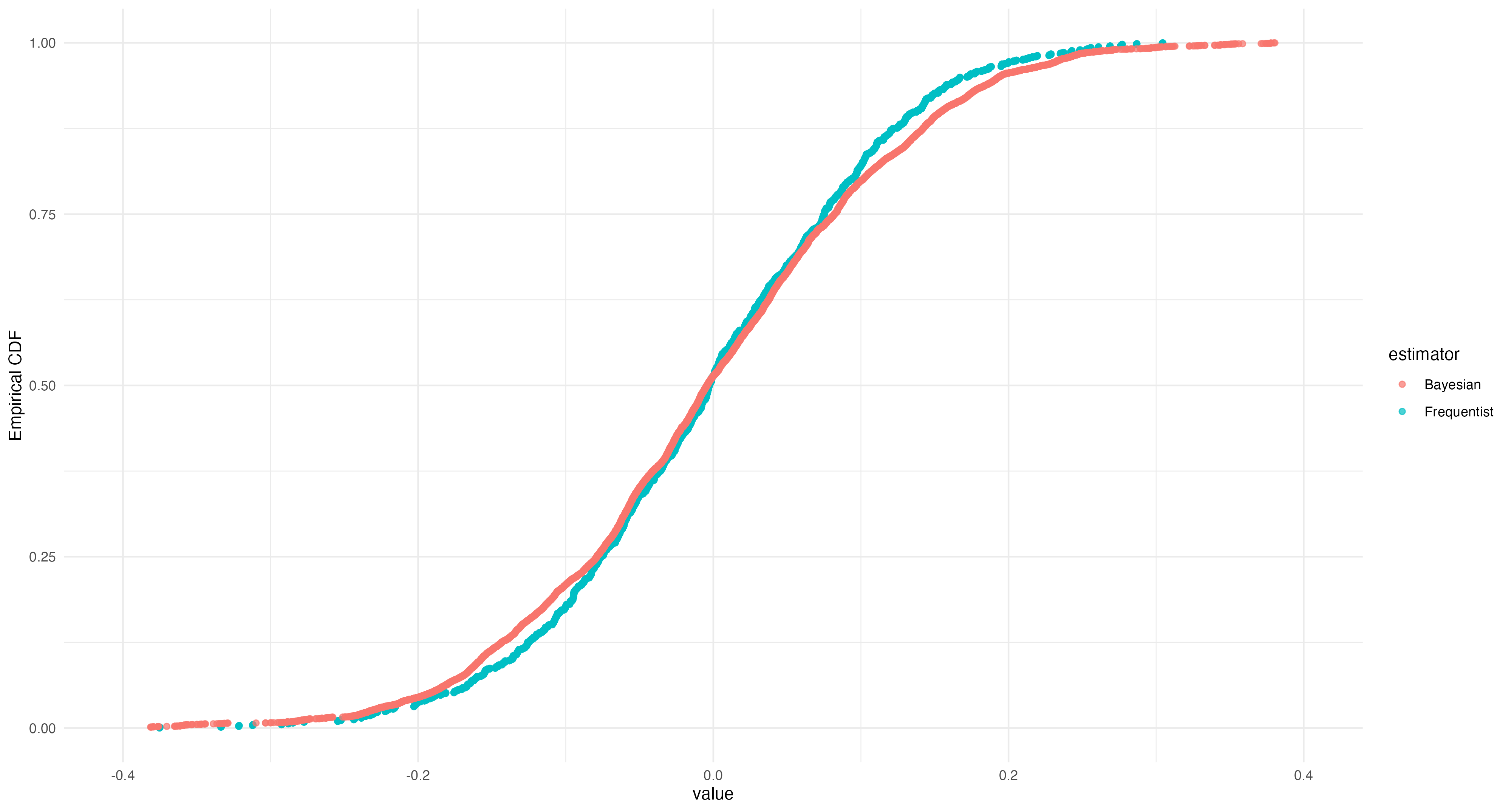}
  	\caption{CDF for $T_0 = 200$}
  \end{subfigure}%
  \begin{subfigure}[b]{0.5\textwidth}
  \centering
  	\includegraphics[width=1\linewidth]{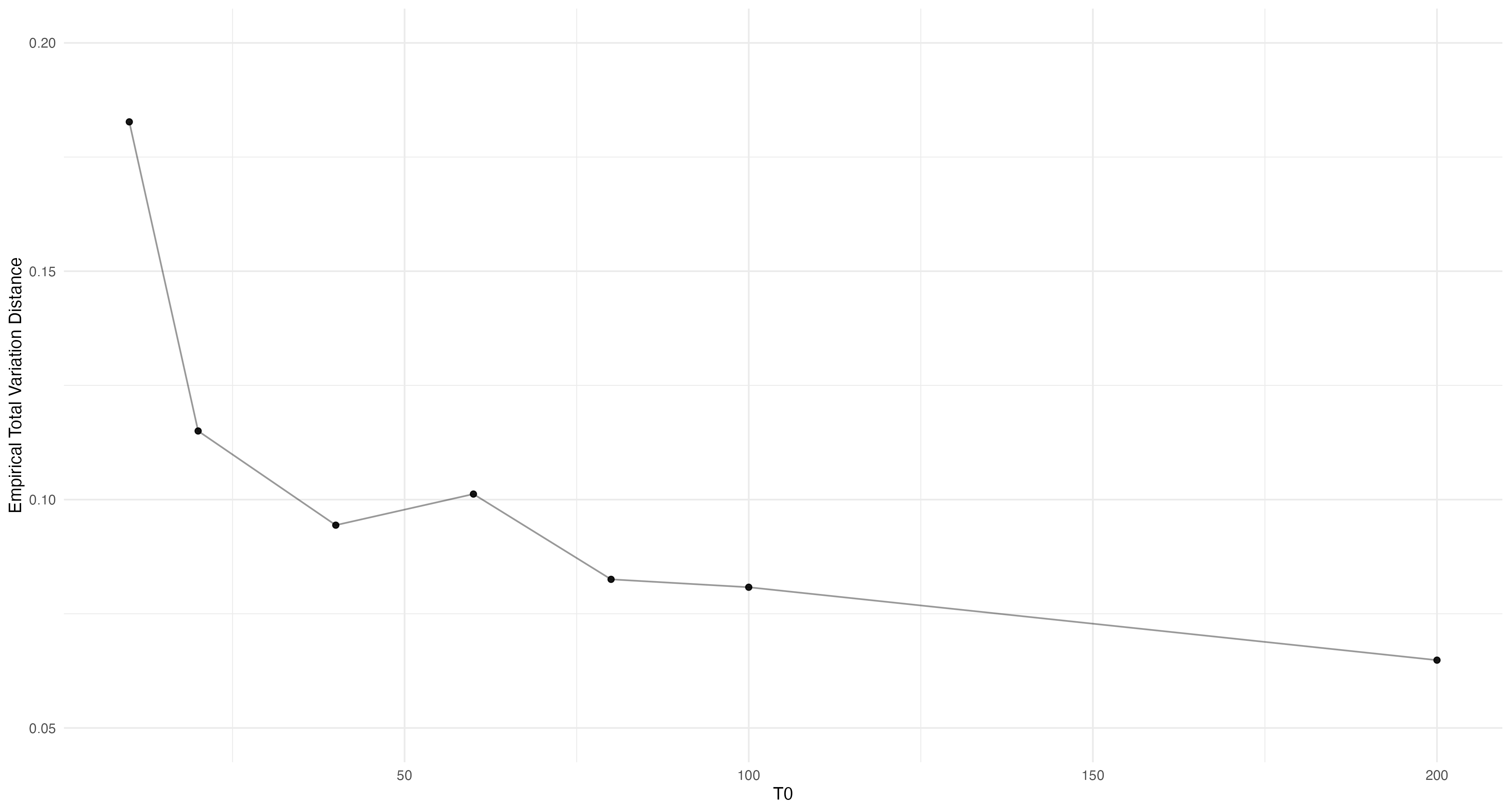}
  	\caption{Total variation distance}
  \end{subfigure}%
  
  \caption{Convergence of frequentist and Bayesian coverage as $T\to\infty$ for the sparse case.}
    \begin{tablenotes}
		\small\item\textbf{Notes}: Kernel densities of the frequentist empirical distribution of the estimated treatment effect (in red) and the Bayesian posterior mean over 10000 draws (in blue) for different values of $T_0$. Panel (e) shows the empirical CDF for each case. Panel (f) shows the empirical total variation distance between both distributions for different values of $T_0$. The potential outcomes are generated by the grouped factor model (\ref{grouped_fm}) with $\sigma = 0.25$ in a sparse design with factor loadings grouped in two groups of 10 units.
   \end{tablenotes}
  \label{fig_bvm_sparse}
\end{figure}

As can be seen in Figure \ref{fig_bvm_dense} the Bayesian estimator exhibits over coverage when $T_0$ is small. However, for moderate values of $T_0$, such as 40 or 60 time periods, the coverage is close to that of the frequentist estimator. As $T_0$ grows, the prior becomes dominated, and the Bayesian posterior distribution converges to the finite sample frequentist distribution, as can be seen in panels (e) and (f) in Figure \ref{fig_bvm_dense}. A similar pattern can be observed in Figure \ref{fig_bvm_sparse} for the \textit{sparse} design, albeit, as expected, the convergence rate is slower than in the dense case. Panel (e) in Figure \ref{fig_bvm_sparse} shows that for large values of $T_0$ the empirical CDFs of the two estimators coincide, and panel (f) shows that the empirical total variation distance between the two distributions decreases as $T_0$ grows. Overall, the main takeaway is that in both sparse and dense settings for medium sized panels the Bayesian posterior distribution can be used to approximate the frequentist finite sample distribution.

Consequently, for applied researchers with medium $T_0$ and $J$ it may be possible to use Bayesian synthetic control methods and interpret the Bayesian credible intervals for the estimated treatment effects as valid confidence intervals, providing a new way of doing inference in the framework of synthetic controls. 

\begin{figure}[H]
\centering
  \begin{subfigure}[b]{0.5\textwidth}
  \centering
  	\includegraphics[width=1\linewidth]{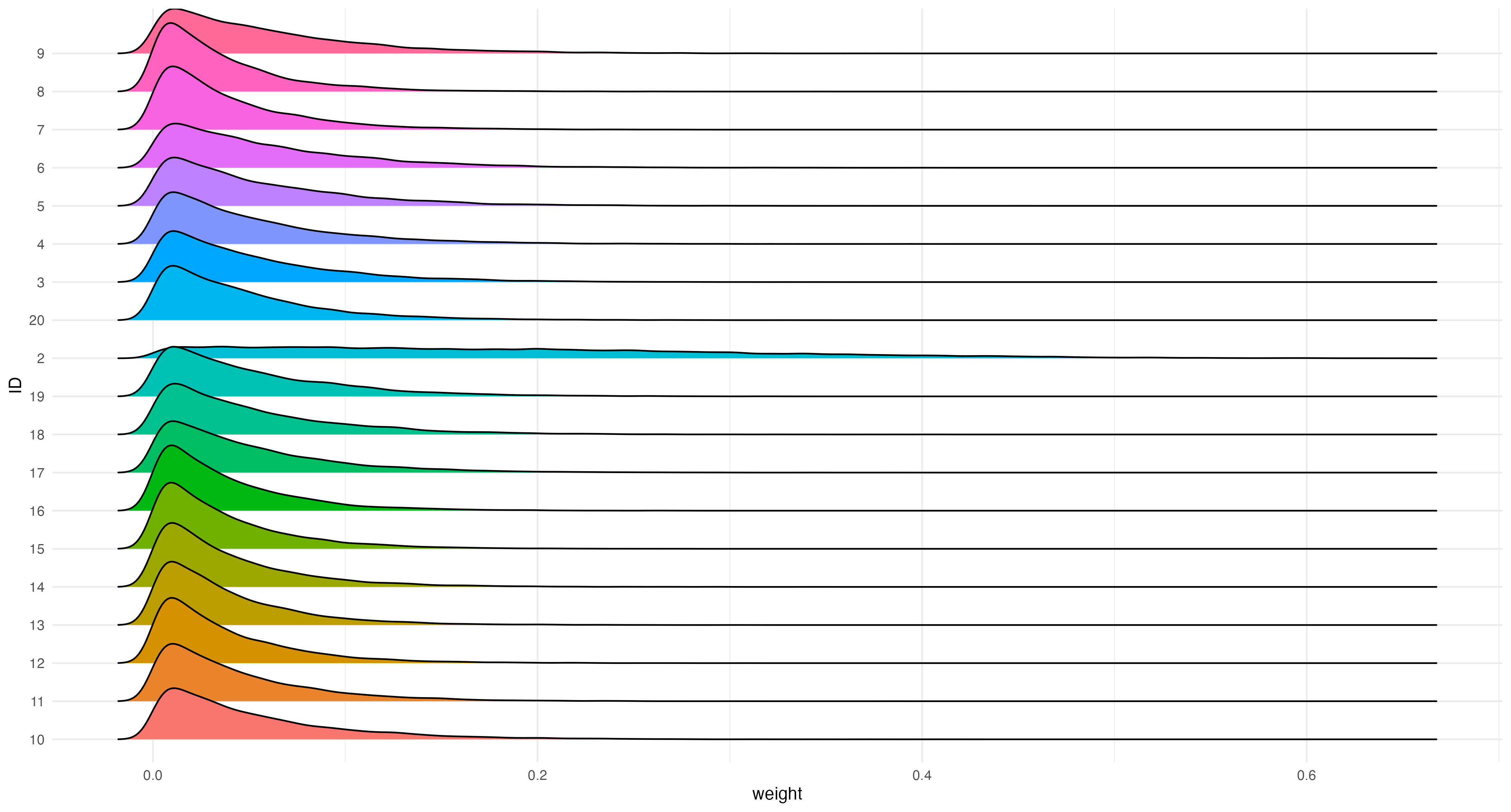}
  	\caption{$T_0 = 20$}
  \end{subfigure}%
  \begin{subfigure}[b]{0.5\textwidth}
  \centering
  	\includegraphics[width=1\linewidth]{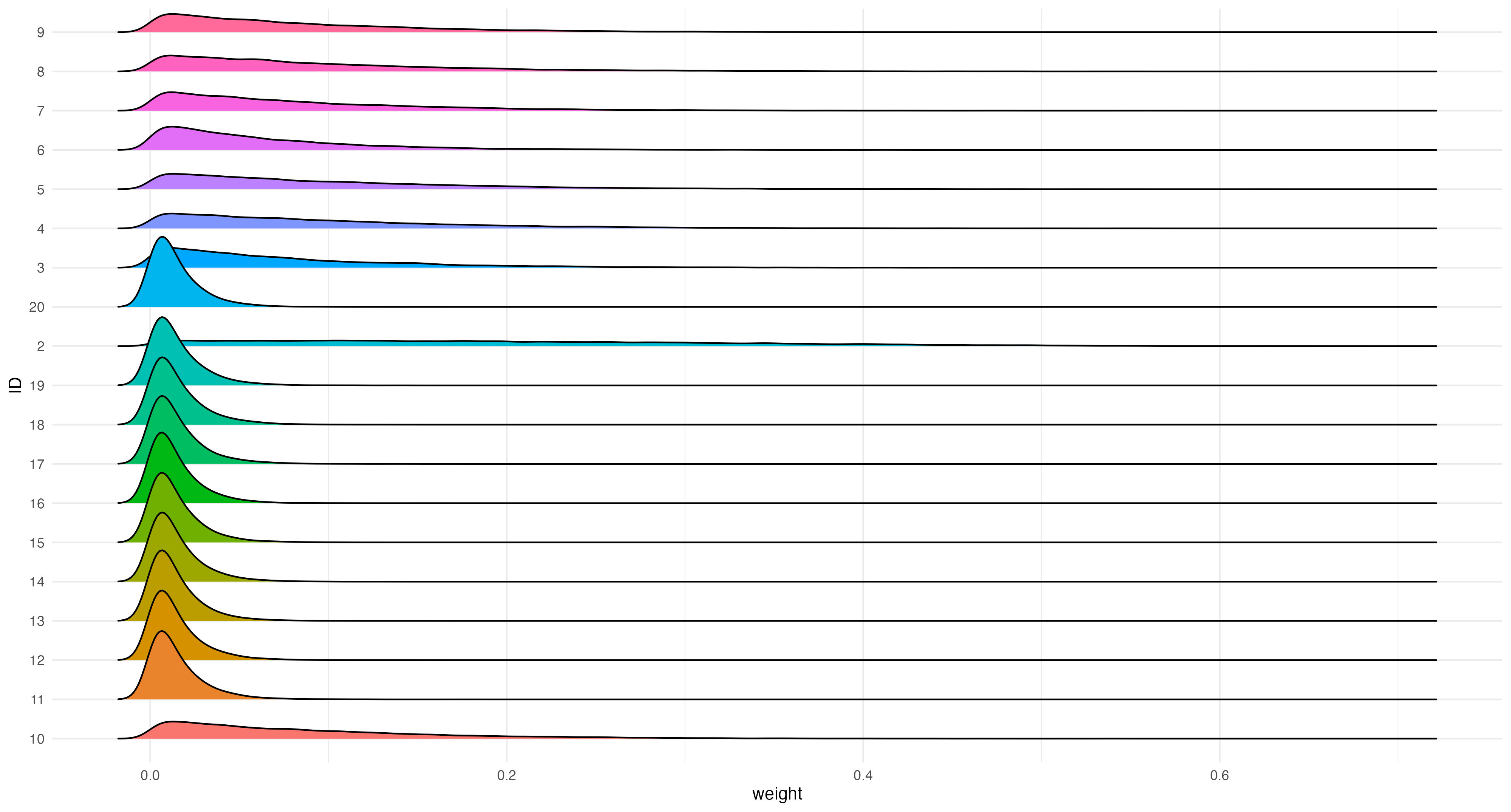}
  	\caption{$T_0 = 20$}
  \end{subfigure}%

  \centering
    \begin{subfigure}[b]{0.5\textwidth}
  \centering
  	\includegraphics[width=1\linewidth]{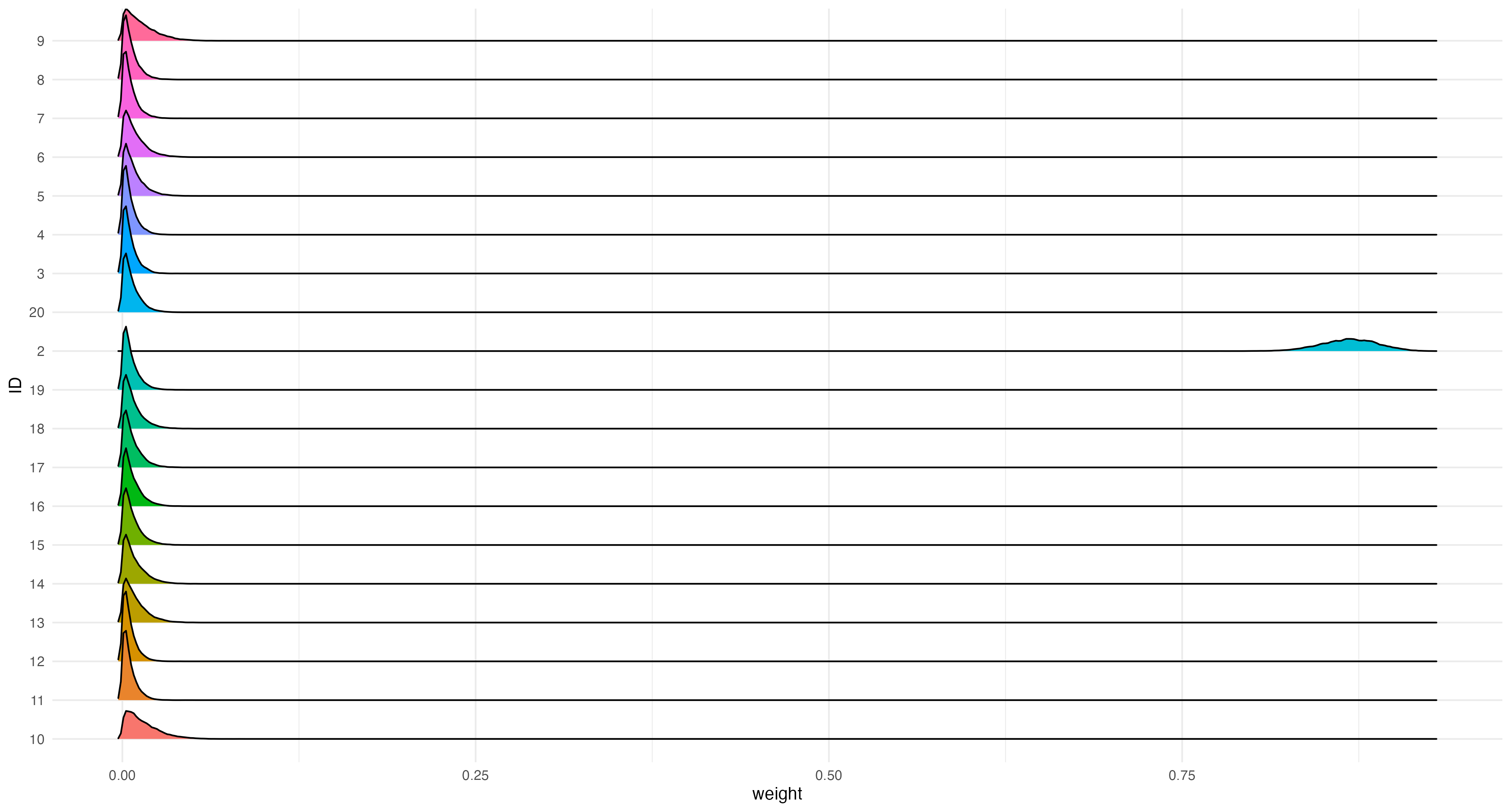}
  	\caption{$T_0 = 200$}
  \end{subfigure}%
  \begin{subfigure}[b]{0.5\textwidth}
  \centering
  	\includegraphics[width=1\linewidth]{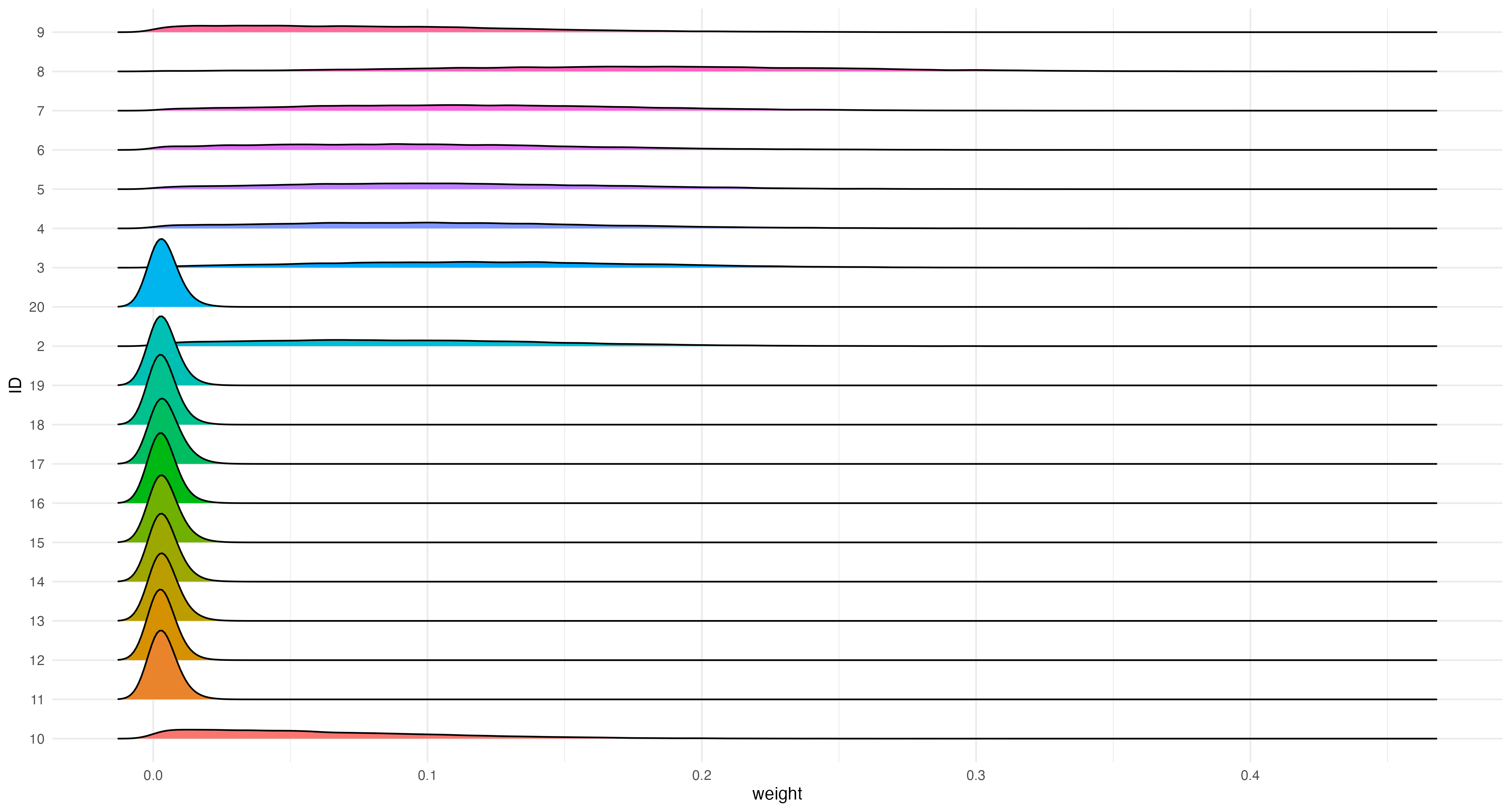}
  	\caption{$T_0 = 200$}
  \end{subfigure}%
  
  \caption{Implicit weights as $T_0\to\infty$.}
    \begin{tablenotes}
		\small\item\textbf{Notes}: Kernel densities of the Bayesian implicit weights for different values of $T_0$. The potential outcomes are generated by the grouped factor model (\ref{grouped_fm}) with $\sigma = 0.25$. Panel (a) and (c) refer to the \textit{sparse} design and panels (b) and (d) refer to the \textit{dense} design.
   \end{tablenotes}
  \label{fig_weights}
\end{figure}

Another point highlighted by the simulations is the use of Bayesian synthetic controls in consistently estimating the synthetic control weights. Figure \ref{fig_weights} shows the implicit weights posterior distribution of the Bayesian estimator for different values of $T_0$. As can be seen in panels (a) and (c) for the sparse design and (b) and (d) for the dense design, as $T_0\to \infty$, the mean of the implicit weight distribution is centered around the optimal value of the synthetic controls ($\tilde{w}$s), which in the sparse case is $\tilde{w}_2 = 1$ and in the dense case is $\tilde{w}_2 = \dots = \tilde{w}_{10} = 1/9$. This is evidence of unit wise convergence which implies the $L^2$ uniform convergence result in Theorem \ref{thm_mle}. Overall, this adds to the previous discussion that Bayesian synthetic control methods may be useful both for performing inference for the estimated treatment effect and in interpreting the synthetic control weights. 

In the appendix, we also consider simulations in which we compare the 10000 frequentist draws to the posterior distribution of the Bayesian model for 1 draw and show that even in this case as $T_0\to \infty$ we have convergence between the two distributions.

\subsection{The \textit{bsynth} package}

We have implemented the Bayesian synthetic control model we propose in this paper in the publicly available \textit{bsynth} R-package.\footnote{The package can be accessed at https://github.com/google/bsynth.} The \textit{bsynth} R-package extends the Bayesian model we propose to include more complex models. In particular, it allows for additional features such as modelling the time series component with a Gaussian process and adding additional covariates. The \textit{bsynth} package allows for Bayesian models with the following form:
\begin{align*}
X_1 | w, \sigma &\sim N( X_0 w + f_{1t}, \sigma^2\text{diag}(\Gamma)^{-2}), \\
w &\sim Dir(1), \\
f_{1t} &\sim \mathcal{GP}, \\
\sigma &\sim N(0,1)^+, \\
\Gamma &\sim Dir(v_1, \dots, v_K), \quad \text{ } v_k \in \Delta^k,
\end{align*}
where here $X_1$ and $X_0$ denote the design matrices for the treated and donor units which may include the outcome variables as well as additional predictors. The $f_{1t}$ term is modelled through a Gaussian process and the weight of the predictors are modelled by $\Gamma$. To preserve the main features of synthetic controls both the $w$ and the $v$ weights are assumed to be in the simplex.

The \textit{bsynth} package offers the possibility to compute different statistics of the posterior distribution. Of special interest is an upper bound on the frequentist bias given by the Bayesian model. This bound is motivated by the finite sample bound first developed in \citet{AbaDiaHai2010} and expanded in \citet{vives2021} to include additional predictors. Given the BvM style result from Theorem \ref{thm_bvm} the bound can be used to check the likelihood that the Bayesian synthetic control is badly biased due to model mispecification. Intuitively, if the Bayesian synthetic control can not replicate in the pre-treatement period the outcomes of the treated unit then our frequentist interpretation of the method will be biased in the same way standard synthetic controls are biased when perfect pre-treatment fit can not be achieved. In the following section we use the Bayesian synthetic control to study two political economy questions.

\section{Empirical applications}

One of the most salient applications of synthetic controls is the study of the impact of the German re-unification in 1990 to the GDP of West Germany. In this paper, we replicate this finding using the Bayesian synthetic control and we highlight the usefulness of the Bayesian inference procedure.

\subsection{Re-visiting the German re-unification}

\begin{figure}[H]
\centering
  \begin{subfigure}[b]{0.53\textwidth}
  \centering
  	\includegraphics[width=1\linewidth]{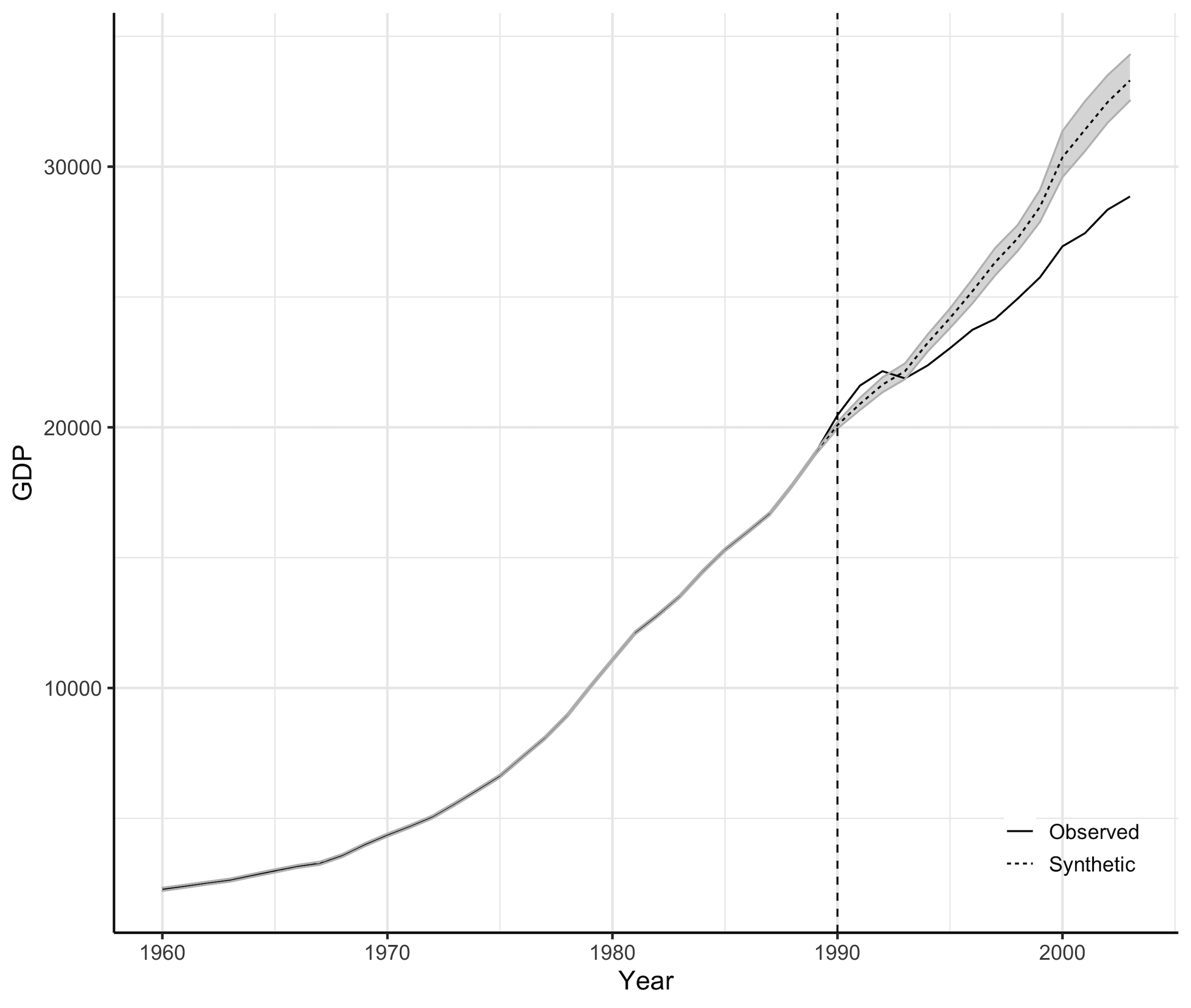}
  	\caption{Treatment effect}
  \end{subfigure}%
  \begin{subfigure}[b]{0.47\textwidth}
  \centering
  	\includegraphics[width=1\linewidth]{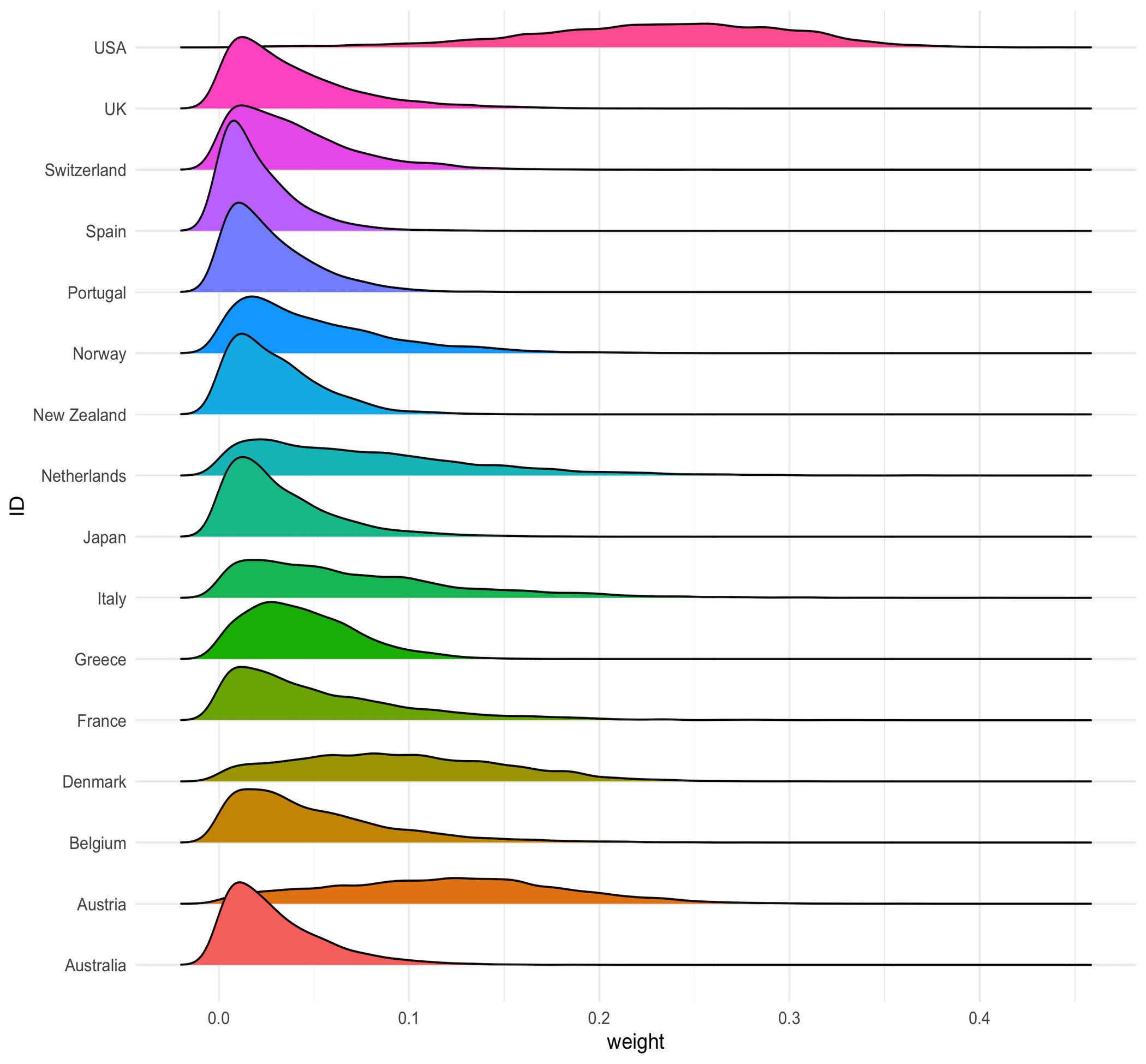}
  	\caption{Implicit weight marginals}
  \end{subfigure}%
  \caption{Bayesian synthetic control for West Germany.}
    \begin{tablenotes}
		\small\item\textbf{Notes}: Panel (a) shows the West Germany GDP per capita and Bayesian synthetic control estimates from 1960 to 2003 with credible intervals shaded in grey. Panel (b) shows the marginal distributions of the implicit weights of the Bayesian synthetic control. 
   \end{tablenotes}
   \label{fig_germany}
\end{figure}

In 1989, after the fall of the Berlin wall, the process of re-unifying West Germany and East Germany started. \citet{AbaDiaHai2015} found that in absence of the re-unification, West Germany's GDP would have been 8\% higher in 2003, 13 years later. Using the same data and specification as \citet{AbaDiaHai2015} in Figure \ref{fig_germany} we display the Bayesian synthetic control for West Germany over and the marginal distribution of implicit weights of the donor units. 

The Bayesian synthetic control in Figure \ref{fig_germany} is similar to the standard synthetic control in \citet{AbaDiaHai2015}. It shows an overall increasing trend in absence of the intervention with a slight reversal in the first few years after the intervention. Similarly, in panel (b) the implicit weights of the Bayesian synthetic control also indicate that West Germany is best replicated by a combination the United States and Austria. In the appendix, we also report the correlations between the implicit weights, a statistic that could be of interest to applied researchers seeking to understand the robustness of their synthetic control estimate. The correlation matrix shows that the implicit weight distributions of the United Sates and Austria are positively correlated, as expected, but it also shows that an alternative synthetic control could have included Denmark, Spain, Portugal and Australia which are also correlated with each other. 

\begin{figure}[H]
\centering
\centering
  	\includegraphics[width=0.7\linewidth]{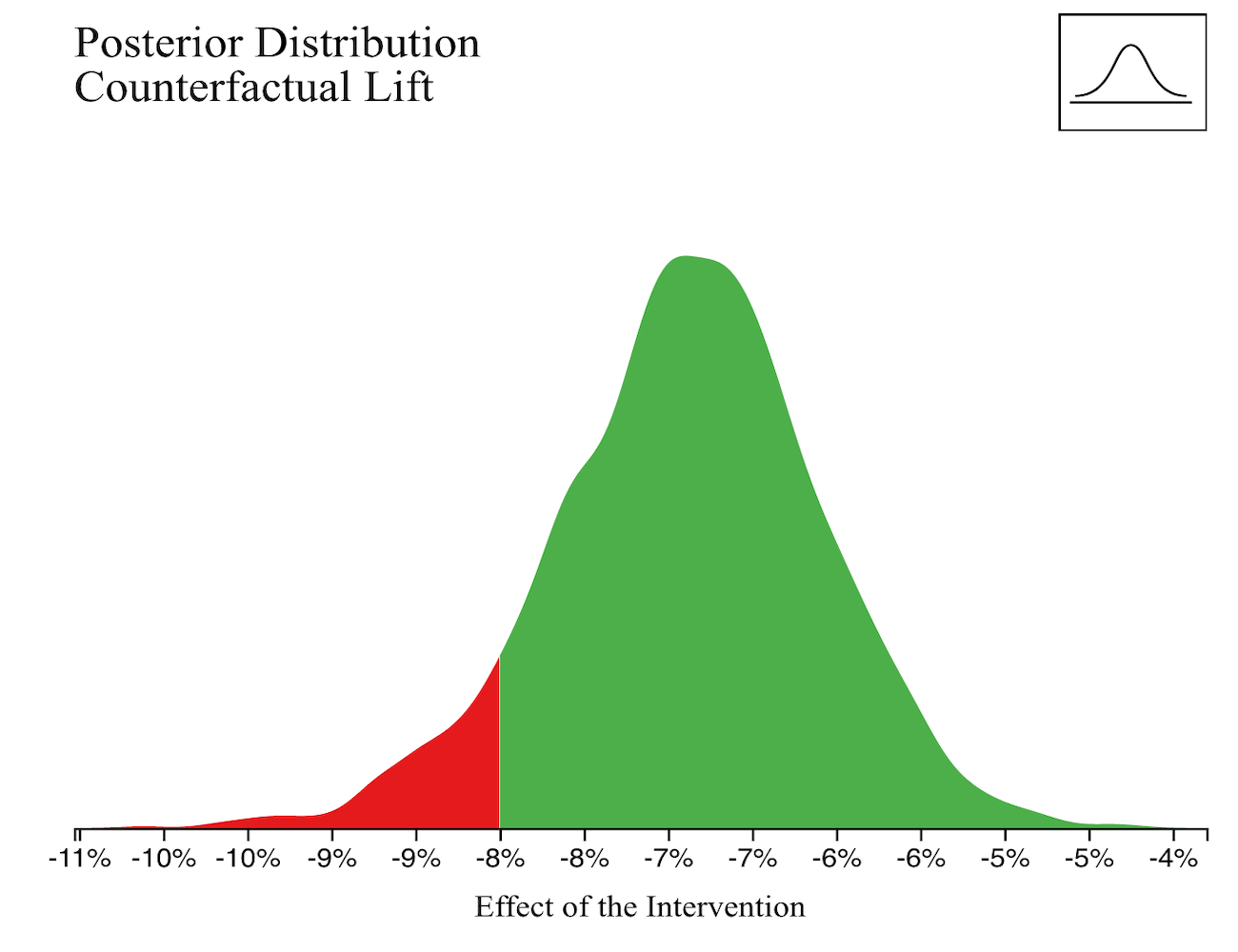}
  \caption{Treatment effect posterior distribution.}
    \begin{tablenotes}
		\small\item\textbf{Notes}: Posterior distribution of the average treatment effect in the post-treatment period, the change of color indicates the frequentist estimated effect of 8\%. This figure is generated directly from the output of the \textit{bsynth} R-package.
   \end{tablenotes}
   \label{fig_posterior}
\end{figure}

An advantage of the Bayesian synthetic control is that we estimate a full posterior distribution. Figure \ref{fig_posterior} shows the treatment posterior distribution for the post-treatment period relative to the baseline year. The mean of the posterior distribution is $7.5\%$ which is slightly lower than the frequentist estimate of the ATET of $8\%$. The 95\% Bayesian credible interval spans $[-9\%, -6\%]$ and contains the frequentist estimate. Given that in this setting the number of units is small ($J=16$) and the number of pre-treatment periods is moderate ($T_0 = 30$), we expect the Bayesian credible interval to be a good approximation of the confidence interval. Furthermore, in the appendix, we show the posterior distribution of the frequentist bias bound that depends on the MAD in the pre-treatment period. The bound indicates that the likelihood that the sign of the result could be overturned due to the bias induced by model misspecification (bad pre-treatment fit) is small. 

\subsection{The impact of the Catalan secession movement}

The pro-independence demands of the Catalan electorate, commonly referred as the Catalan secession process, have been the focus of Catalan politics, and Spanish politics, since 2012. While the movement started with a plea for higher fiscal autonomy (the ability of the Catalan government to collect and administer its own taxes), quickly it shifted towards independence through a series of plebiscitary elections that were won by pro-independence parties. This transition towards a pro-independence mandate culminated in 2017 with the celebration of a non-binding independence referendum on the 1st of October which received wide international media coverage. Following the referendum, on the 27th of October a unilateral declaration of independence (UDI) was voted and approved by the Catalan parliament. However, the UDI was deemed unconstitutional by the Spanish constitutional court and the Spanish government suspended the Catalan government (activating the \textit{article 155}). More so, the members of the Catalan parliament and the government behind the UDI were arrested while several high ranking members of the government, including the president Carles Puigdemont, sought asylum in other countries. 

Overall the UDI and the events that ensued after the Spanish government suspended the Catalan government had a great impact on the Catalan economy. An extensive analysis of different outcomes affected by the UDI has been done by \citet{catalansc}, also using a synthetic control methodology. The main finding in \citet{catalansc} is that the UDI lead to a significant drop in short term bank deposits (capital out-flow), accounting to up to 2.5bn euros. This result is in line with reported evidence in the media\footnote{See for example the following Financial Times article: https://www.ft.com/content/18f0ca8c-607b-4633-a83f-e99f71a046e8.} that the UDI and political instability lead to many firms and individuals to reallocate outside of Catalonia. In this paper, we focus on quarterly GDP as our main outcome of interest. To estimate the Bayesian synthetic control we consider all quarters between 2012 and the UDI in 2017 as the pre-treatment period and all quarters from the UDI to 2020 as the post-treatment period. We choose this time frame to avoid the effects of the 2008 financial crisis and the 2010 European debt crisis. Given that Catalonia has a different economic structure than other regions in Spain the effect of the two crises was different and therefore matching outcomes in that time period is potentially misleading. We also restrict the post period to avoid the covid-19 shock in 2020.

\begin{figure}[H]
\centering
  \begin{subfigure}[b]{0.5\textwidth}
  \centering
  	\includegraphics[width=1\linewidth]{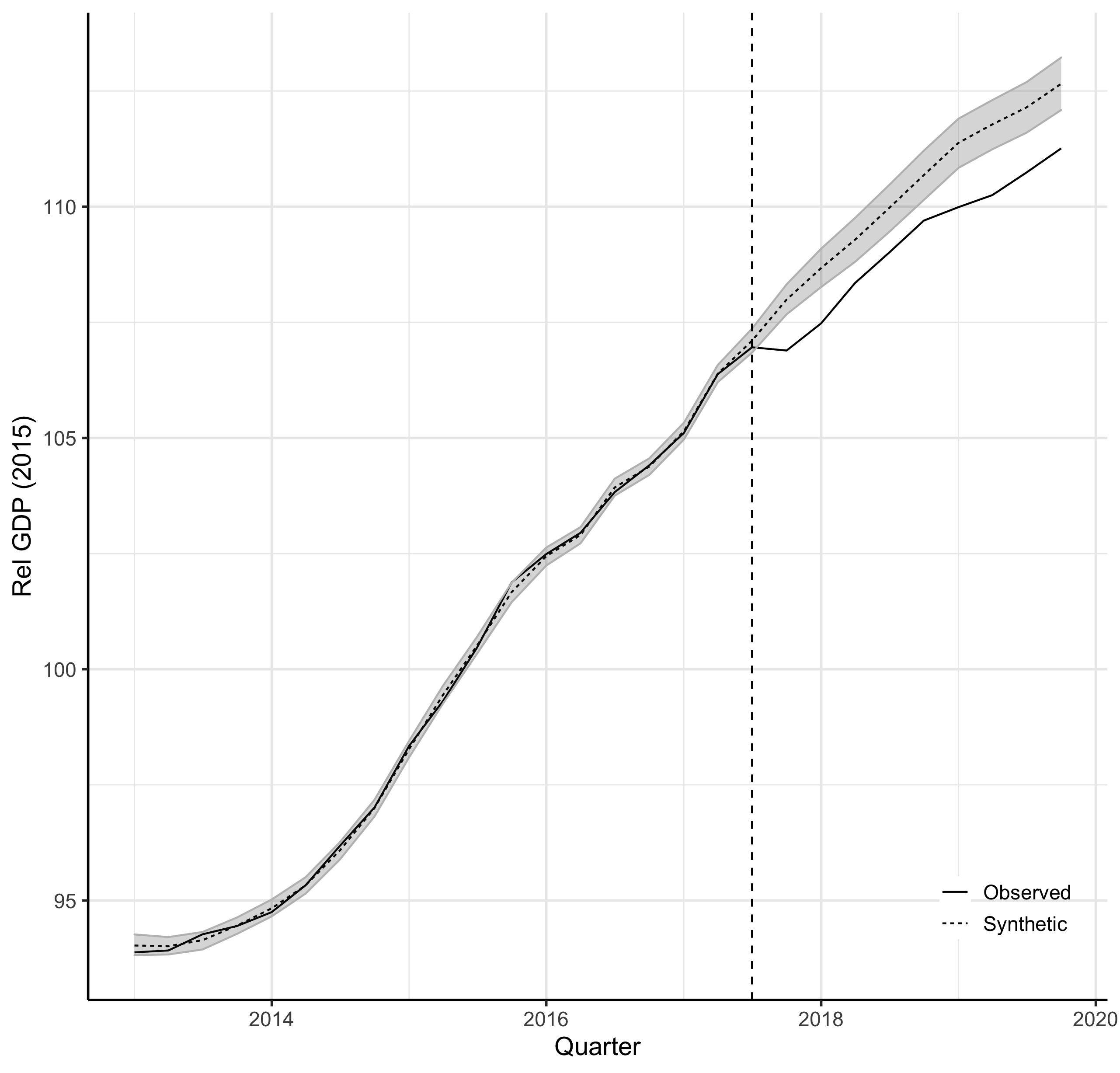}
  	\caption{Treatment effect}
  \end{subfigure}%
  \begin{subfigure}[b]{0.5\textwidth}
  \centering
  	\includegraphics[width=1\linewidth]{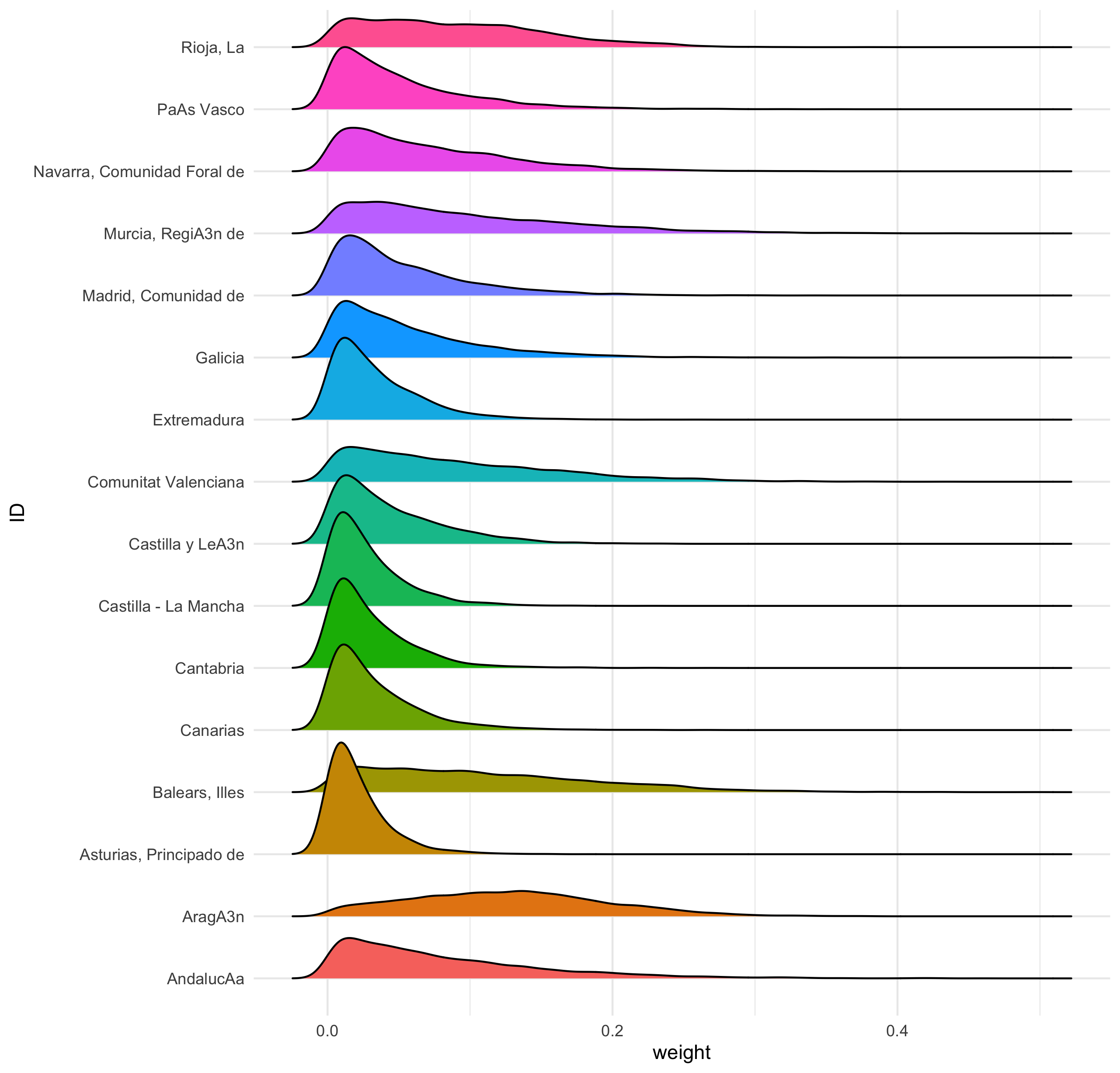}
  	\caption{Implicit weight marginals}
  \end{subfigure}%
  \caption{Bayesian synthetic control for Catalonia.}
    \begin{tablenotes}
		\small\item\textbf{Notes}: Panel (a) shows the Catalan quarterly GDP (relative to 2015) and Bayesian synthetic control estimates from 2013 to 2020 with 75\% credible intervals shaded in grey. Panel (b) shows the marginal distributions of the implicit weights of the Bayesian synthetic control. 
   \end{tablenotes}
   \label{fig_cat}
\end{figure}

Figure \ref{fig_cat} shows the Bayesian synthetic control and implicit weight marginal distributions for Catalonia. The main finding is that the UDI lead to a quarterly GDP decrease of 1\% over 2017-2020 relative to the third quarter of 2017. The Bayesian synthetic control allows us to quantify uncertainty with the 75\% credible interval being $[-0.6\%, -1.3\%]$ and the 95\% credible interval slightly larger at $[-0.3\%, -1.6\%]$. Overall this is indicative evidence that the capital flight described in the press and \citet{catalansc} may have had a significant impact on the Catalan GDP. In panel (b) of Figure \ref{fig_cat} we can see that the Bayesian synthetic Catalonia is mostly composed of neighboring regions: Aragon, the Balearic islands and Valencia. A potential concern is that the UDI also affected positively the regions that received the out-flowing capital, biasing the estimated treatment effect upwards. In particular, as detailed in media coverage, many Catalan firms migrated their fiscal head quarters to Madrid or the neighboring Valencia.\footnote{See the article in footnote 2 for reference.} In light of this, in the appendix, we re-do the exercise removing Madrid and Valencia from the donor pool and find similar treatment effects.  

\section{Conclusion}

This paper contributes to the synthetic control literature in two ways. First, we characterize the conditions on the primitives of factor models (the factor loadings) that generate target parameters (minimizers of the statistical risk) that are \textit{synthetic controls}. This result complements the existing literature on the asymptotic properties of synthetic controls under linear factor models by providing guidance on the set of data generating processes for which synthetic control estimators are best suited. We show that the target parameters can be estimated by MLE and derive rate conditions for uniform consistency as the number of time periods and donor units grow.

Second, we propose the Bayesian synthetic control as an alternative to perform inference for synthetic control methods. We derive a Bernstein-von Mises style result that states conditions under which the Bayesian synthetic control and the MLE estimator converge in the total variation sense. This result can be used to approximate the frequentist inference using the Bayesian synthetic control in large samples. Through simulations we show that this result might also be useful in settings in which the theoretical assumptions don't hold, highlighting the potential for theoretical extensions in future work.

Finally we apply the Bayesian synthetic control to study two important political economy interventions: that of the re-unification of a country and that of secession. We replicate using the Bayesian synthetic control the findings in \citet{AbaDiaHai2015} that the German re-unification lead to a significant decrease in West Germany's GDP and show that the Catalan unilateral declaration of independence of 2017 may have lead to a 1\% decrease in GDP. The Bayesian synthetic control allows us to quantify uncertainty and in both applications we can conclude that the effects are unlikely to be zero. 

Future work may aim to generalize the results in the paper to a larger class of models and extend the functionality of the \textit{bsynth} R-package. 

\newpage
\bibliography{Bibliography}

\newpage
\centerline{\Large\bf Appendix}
\medskip

\appendix

\section{Conditional Distribution Model Derivations}
Suppose we have the following simple model independent of time:
$$
Y_{i} = \lambda_iF_t + \epsilon_i,
$$
\noindent where $F_t\sim N(0, \sigma^2)$ and $\epsilon_i \sim N(0,\sigma^2_{\epsilon})$. Then it follows that $\mathbb{E}[Y_1] = 0$, $\mathbb{E}[\boldsymbol{Y}_J] = \boldsymbol{0}$, $cov(Y_i, Y_j) = \lambda_i \lambda_j \sigma^2$ and $var(Y_j) = \sigma^2_{\epsilon} + \lambda_i^2\sigma^2$. Then, the joint distribution of ys is normal. We are interested in the conditional distribution of $Y_1$ given $\boldsymbol{y}_{J} = (y_2, \dots, y_{J+1})$:

$$
Y_1 | \boldsymbol{y}_J \sim N\left(\tilde{\mu}, \tilde{\Sigma}\right),
$$
\noindent where
\begin{align}
    \tilde{\mu} &= cov(Y_1, \boldsymbol{Y}_J)\Sigma_{(2,J+1)}^{-1}\boldsymbol{y}_J = \sum_{j=2}^{J+1} w_j(\boldsymbol{\lambda}, \sigma)y_j \\
    \tilde{\Sigma} &= var(Y_1) - cov(Y_1, \boldsymbol{Y}_J) \Sigma_{(2,J+1)}^{-1}cov(\boldsymbol{Y}_J, Y_1).
\end{align}

In this set up the data are repeated observations over time. For a pre-treatment period of length $T_0$ then we will be interested in estimators $\hat{w}_j$ such that our estimated counterfactual is given by:
$$
\hat{Y}_{1t}(0)  = \sum_{j=2}^{J+1} \hat{w}_j y_{jt}.
$$

Start by noting that in this setting $\Sigma_{(2,J+1)}$ is positive definite and invertible. Hence, by the spectral theorem we can express it as a linear combination of eigenvalues and eigenvectors:
$$
\Sigma_{(2,J+1)} = \begin{pmatrix}
\sigma^2_{\epsilon} + \lambda_2^2 \sigma^2 & \lambda_2\lambda_3\sigma^2 & \cdots & \lambda_2\lambda_{J+1}\sigma^2 \\
\lambda_2\lambda_3\sigma^2 & \sigma^2_{\epsilon} + \lambda_3^2\sigma^2 & \cdots & \lambda_3\lambda_{J+1}\sigma^2 \\
\vdots  & \vdots  & \ddots & \vdots  \\
\lambda_2\lambda_{J+1}\sigma^2 & \lambda_3\lambda_{J+1}\sigma^2 & \cdots & \sigma^2_{\epsilon} + \lambda_{J+1}^2\sigma^2
\end{pmatrix} = \sum_{j=2}^{J+1} s_j \boldsymbol{u}_j \boldsymbol{u}_j^T,
$$

where $s_j$ is the eigenvalue associated with the $\boldsymbol{u}_j$ eigenvector. Observe that the eigenvalues are given by $s_2 = \dots = s_{J} = 1$ and $s_{J+1} = \sigma^2_{\epsilon} + \sum_{j=2}^{J+1} \lambda_j^2 \sigma^2$. Therefore, we can express $\tilde{\mu}$ as a linear combination of the data, $\boldsymbol{\lambda}$ and $\sigma$: 

\begin{align*}
    \tilde{\mu} &= \sigma^2 \lambda_1 \sum_{j=2}^{J+1} \sum_{i=2}^{J+1} \lambda_i [\Sigma_{(2,J+1)}^{-1}]_{ji} y_j \\
    &= \sum_{j=2}^{J+1} \sigma^2 \lambda_1 \sum_{i=2}^{J+1} \lambda_i \sum_{k=2}^{J+1} \frac{1}{s_k} [\boldsymbol{u}_k \boldsymbol{u}_k^T]_{ji} y_j.
\end{align*}

By substituting the eigenvectors and eigenvalues we can find the closed form for the weights:
\begin{align*}
    w_j(\boldsymbol{\lambda}, \sigma) &= \sigma^2 \lambda_1 \sum_{i=2}^{J+1} \lambda_i \sum_{k=2}^{J+1} \frac{1}{s_k} [\boldsymbol{u}_k \boldsymbol{u}_k^T]_{ji} \\
    &= \frac{\sigma^2 \lambda_1 \lambda_j}{\sigma^2_{\epsilon} + \sum_{j=2}^{J+1} \lambda_j^2 \sigma^2}.
\end{align*}

Similarly, we can derive a closed form for the variance:
\begin{align*}
    \tilde{\Sigma} &= \sigma^2_{\epsilon} + \lambda_1\sigma^2 - \frac{\sigma^4\lambda_1^2\sum_{j=2}^{J+1} \lambda_j^2}{\sigma^2_{\epsilon} + \sigma^2 \sum_{j=2}^{J+1} \lambda_j^2} \\
    &= \sigma^2_{\epsilon} + \lambda_1\sigma^2(\sigma^2_{\epsilon} - \sum_{j=2}^{J+1}w_j\lambda_j) \\
    &= \sigma^2_{\epsilon} + \lambda_1\sigma^2 - (\sigma^2_{\epsilon} + \sigma^2 \sum_{j=2}^{J+1} \lambda_j^2)\sum_{j+2}^{J+1}w_j^2.
\end{align*}

\section{Proof of Theorem \ref{thm_predictor}}
Observe that without loss of generality we can consider $V=I$. For notational convenience we drop the potential outcome subscript in what follows such that $\boldsymbol{Y}_1 \equiv \boldsymbol{Y}_1(0)$. Observe that this proof is equivalent to showing that the conditional expectation is the best linear predictor under the square loss. Consider the objective function:
\begin{align*}
    \mathbb{E}\left[ (\boldsymbol{Y}_1 - \boldsymbol{y}_{J}'\boldsymbol{w})'(\boldsymbol{Y}_1 - \boldsymbol{y}_{J}'\boldsymbol{w})\right] &= \mathbb{E}\left[ (\boldsymbol{Y}_1'\boldsymbol{Y}_1 - 2 \boldsymbol{w}'\boldsymbol{y}_{J}\boldsymbol{Y}_1 + \boldsymbol{w}'\boldsymbol{y}_{J}\boldsymbol{y}_{J}'\boldsymbol{w}\right] \\
    &= \mathbb{E}\left[ \mathbb{E}[\boldsymbol{Y}_1'\boldsymbol{Y}_1 |\boldsymbol{Y}_{J}]  - 2 \boldsymbol{w}'\boldsymbol{y}_{J}\mathbb{E}[\boldsymbol{Y}_1|\boldsymbol{Y}_{J}] + \boldsymbol{w}'\boldsymbol{y}_{J}\boldsymbol{y}_{J}'\boldsymbol{w}\right]\\
    &= \mathbb{E}\left[ \tilde{\boldsymbol{w}}'\boldsymbol{Y}_{J}\boldsymbol{Y}_{J}'\tilde{\boldsymbol{w}} + T_0\tilde{\Sigma}^2  - 2 \boldsymbol{w}'\boldsymbol{y}_{J}\boldsymbol{Y}_{J}'\tilde{\boldsymbol{w}} + \boldsymbol{w}'\boldsymbol{y}_{J}\boldsymbol{y}_{J}'\boldsymbol{w}\right]
\end{align*}


\noindent where the steps follow from the derivations for the conditional multivariate normal, the law of iterated expectations and the definition of conditional variance, $V(Y|X) + \mathbb{E}[Y|X]^2 = E[Y^2|X]$. Given that the $w$ parameters are only on the observed $\boldsymbol{y}$, we only can restrict attention to the event $\{\boldsymbol{Y}_J = \boldsymbol{y_j}\}$. Conditional on this event, it is straightforward to see that setting $\boldsymbol{w} = \tilde{\boldsymbol{w}}$ minimizes the expression. Therefore, it follows that conditional on the data,  $\tilde{\boldsymbol{w}}$ is a global minimizer of the expression.

\section{ Proof of Theorem \ref{thm_convergence}}
\textbf{Statement (1)}: we want to show that as $J \to \infty$
$$
\mathbb{E}\left[ (\boldsymbol{y}_{JT_0+1}'\tilde{\boldsymbol{w}} - \lambda_1 F_{T_0+1})^2 \right] \to 0.
$$

Expanding the expression conditional on $F_{T_0+1}$ and given the assumptions and derivations for $\tilde{\boldsymbol{w}}$: 
\begin{align*}
    \mathbb{E}\left[ (\boldsymbol{y}_{JT_0+1}'\tilde{\boldsymbol{w}} - \lambda_1 F_{T_0+1})^2 | F_{T_0+1} \right] &= \frac{\lambda_1^2 \sigma^4}{(\sigma^2_{\epsilon} + \|\boldsymbol{\lambda}_J\|^2_2 \sigma^2)^2}\mathbb{E}\left[\left(F_{T_0+1} \|\boldsymbol{\lambda}_J\|^2_2 + \sum_j \lambda_j^2 \epsilon_{jT_0+1}\right)^2 | F_{T_0+1} \right] \\
    &+ \lambda_1^2F_{T_0+1}^2\left( \frac{\sigma^2_{\epsilon} - \sigma^2 \|\boldsymbol{\lambda}_J\|^2_2}{\sigma^2_{\epsilon} + \sigma^2 \|\boldsymbol{\lambda}_J\|^2_2}\right) \\
    &- \frac{2\sigma^2\lambda_1^2}{\sigma^2_{\epsilon} + \sigma^2 \|\boldsymbol{\lambda}_J\|^2_2}\mathbb{E}\left[ \sum_j\lambda_j \epsilon_{jT_0+1} | F_{T_0+1} \right] \\
    &= \frac{\lambda_1^2 \sigma^4}{(\sigma^2_{\epsilon} + \|\boldsymbol{\lambda}_J\|^2_2 \sigma^2)^2}\left(F_{T_0+1}^2 \|\boldsymbol{\lambda}_J\|^4_2 + \|\boldsymbol{\lambda}_J^2\|^2_2\right) \\
    & + \lambda_1^2F_{T_0+1}^2\left( \frac{\sigma^2_{\epsilon} - \sigma^2 \|\boldsymbol{\lambda}_J\|^2_2}{\sigma^2_{\epsilon} + \sigma^2 \|\boldsymbol{\lambda}_J\|^2_2}\right)
\end{align*}

Observe that there exists no value of $\|\boldsymbol{\lambda}_J\|^2_2$ for which the above expression is zero. By the law of iterated expectations this implies that convergence in mean squared can not be achieved unless $\sigma^2_{\epsilon}=0$.

\textbf{Statement (2)}: By Markov's inequality we have that a for $t>0$

$$
\mathbb{P}(| \boldsymbol{y}_{JT_0+1}'\tilde{\boldsymbol{w}} - \lambda_1 F_{T_0+1} | \geq t ) \leq \frac{\mathbb{E}[| \boldsymbol{y}_{JT_0+1}'\tilde{\boldsymbol{w}} - \lambda_1 F_{T_0+1} |]}{t}.
$$
\noindent Under the assumption that independently $\epsilon_{jt} \sim N(0, \sigma^2_{\epsilon})$:
\begin{align*}
   \mathbb{E}\left[\left| \boldsymbol{y}_{JT_0+1}'\tilde{\boldsymbol{w}} - \lambda_1 F_{T_0+1} \right| | F_{T_0+1}\right] &= \mathbb{E}\left[\left|\frac{\sigma^2 \lambda_1}{\sigma^2_{\epsilon} + \|\boldsymbol{\lambda}_J\|^2_2 \sigma^2}\left(\|\boldsymbol{\lambda}_J\|^2_2F_{T_0+1} + \sum_j \lambda_j \epsilon_{jT_0+1}\right) - \lambda_1 F_{T_0+1} \right| | F_{T_0+1}\right] \\ 
   &= \mathbb{E}\left[\left|\frac{-F_{T_0+1}\sigma^2\lambda_1}{\sigma^2_{\epsilon} + \|\boldsymbol{\lambda}_J\|^2_2 \sigma^2}  + \frac{\sigma^2 \lambda_1}{\sigma^2_{\epsilon} + \|\boldsymbol{\lambda}_J\|^2_2 \sigma^2}\sum_j \lambda_j \epsilon_{jT_0+1}  \right| | F_{T_0+1}\right] \\
   &\leq \left|\frac{\sigma^2 \lambda_1 F_{T_0+1}}{\sigma^2_{\epsilon} + \|\boldsymbol{\lambda}_J\|^2_2 \sigma^2}\right| + \left|\frac{\sigma^2 \lambda_1}{\sigma^2_{\epsilon} + \|\boldsymbol{\lambda}_J\|^2_2 \sigma^2} \right|\mathbb{E}\left[\left|\sum_j \lambda_j \epsilon_{jT_0+1}  \right| | F_{T_0+1}\right] \\
   &\leq \left|\frac{\sigma^2 \lambda_1 F_{T_0+1}}{\sigma^2_{\epsilon} + \|\boldsymbol{\lambda}_J\|^2_2 \sigma^2}\right| + \left|\frac{\sigma^2 \lambda_1}{\sigma^2_{\epsilon} + \|\boldsymbol{\lambda}_J\|^2_2 \sigma^2} \right|\sum_j |\lambda_j|\frac{\sqrt{2}\sigma_{\epsilon}}{\sqrt{\pi}}.
\end{align*}

\noindent The last step uses the mean of the half-normal distribution. Hence, by the law of iterated expectations convergence in probability holds if as $J\to \infty$, $\frac{1}{\|\boldsymbol{\lambda}_J\|^2_2} \sum_j |\lambda_j| \to 0$. Observe, that by Statement 1 we know that this proof would not work with the application of Chebyshev's inequality. 

\section{Proof of Theorem \ref{thm_sc}}
The first part of the theorem follows directly from the derivation for $\tilde{\boldsymbol{w}}$ given A1-A2, setting

$$
\frac{\sigma^2 \lambda_1 \sum_{j=2}^{J+1} \lambda_j}{\sigma^2_{\epsilon} + \sum_{j=2}^{J+1} \lambda_j^2 \sigma^2} = 1,
$$
and simplifying yields the second condition. For the sign restriction, given that the denominator of $\tilde{w}_j$ is non-negative,
$$
\tilde{w}_j=\frac{\sigma^2 \lambda_1 \lambda_j}{\sigma_{\epsilon}^2 + \|\boldsymbol{\lambda}_J\|^2_2 \sigma^2},
$$
\noindent it follows that $\tilde{w}_j\geq 0$ if and only if all factor loadings $\lambda_j$ have the same sign.

For statement 1, Condition (2) implies that given a $\lambda_1$ the $\lambda_j$ lie in a sphere in $\mathbb{R}^J$. A sufficient condition for existence of such $\lambda_j$ is that the discriminant for the second degree equation for each $\lambda_j$ is positive. We exemplify this argument for $J=2$. Condition (2) then requires

$$
\lambda_2^2 -\lambda_1\lambda_2 + \sigma^2_{\epsilon}/(2\sigma^2) + \lambda_3^2 -\lambda_1\lambda_3 + \sigma^2_{\epsilon}/(2\sigma^2) = 0,
$$
which has real roots when the discriminant of each second degree regression is non-negative. In both cases, the condition required is $\lambda_1^2 - 4\sigma^2_{\epsilon}/(2\sigma^2) \geq 0$. For an arbitrary $J$ it follows that a sufficient condition for real roots is  $\lambda_1^2 \geq 4\sigma^2_{\epsilon}/(J\sigma^2)$.

For statement 2, observe that given that $\|\boldsymbol{\lambda}_J\|_2^2 \geq 0$, we can re-write condition (2) as
$$
1 - \lambda_1 \frac{\sum_j \lambda_j}{\|\boldsymbol{\lambda}_J\|_2^2} + \frac{\sigma^2_{\epsilon}}{\|\boldsymbol{\lambda}_J\|_2^2\sigma^2} = 0.
$$
\noindent Given condition (1) it is without loss to write the second term as
$$
\frac{\sum_j |\lambda_1\lambda_j|}{\|\boldsymbol{\lambda}_J\|_2^2}\leq|\lambda_1|\frac{\sum_j |\lambda_j|}{\|\boldsymbol{\lambda}_J\|_2^2} \to 0 
$$
by the assumption $\frac{1}{\|\boldsymbol{\lambda}_J\|^2_2} \sum_j |\lambda_j| \to 0$. Given that this assumption implies $\|\boldsymbol{\lambda}_J\|^2_2\to \infty$, as $J\to \infty$ condition (2) is violated as $1 \neq 0$.

For statement 3, recall that an odd function $h$ satisfies that $h(-x) = -h(x)$ which we extend for multidimensional functions to mean this definition simultaneously holds for all components $\lambda_j$:  $h(-\lambda_2, \dots, -\lambda_{J+1}) = - h(\lambda_2, \dots, \lambda_{J+1})$. When $h$ is component-wise weakly increasing, this ensures that for all $j$, $sign(\lambda_j) = sign(h(\boldsymbol{\lambda}_J)) = sign(\lambda_1)$, so condition (1) is satisfied. As before, if $\| \boldsymbol{\lambda}_J\|_2^2\to \infty$ condition (2) is satisfied when as $J\to \infty$
$$
|\lambda_1|\frac{\sum_j |\lambda_j|}{\|\boldsymbol{\lambda}_J\|_2^2} \to 1,
$$
\noindent where the $\lambda_1$ can be substituted in to get the desired result.

For statement 4, note that we can write the system of equations 
$$
 \boldsymbol{\lambda}_J'\boldsymbol{w}\| \boldsymbol{\lambda}_J\|_1 = \| \boldsymbol{\lambda}_J\|_2^2, \quad \boldsymbol{w}\in\Delta^J,
$$
\noindent as a linear programming problem. A sufficient condition for this program to have a solution is for $\frac{\| \boldsymbol{\lambda}_J\|_2^2}{\|\boldsymbol{\lambda}_J\|_1}$ to be in the convex hull of the $\boldsymbol{\lambda}_J$. 

\section{Proof of Theorem \ref{thm_mle_fixed}}
Recall that under assumptions $\textbf{A1-A2}$ by the weak law of large numbers we have that as $T_0 \to \infty$, $\frac{1}{T_0}\sum_t F_t^2 \overset{p}{\to} \sigma^2$, $\frac{1}{T_0}\sum_t F_t \overset{p}{\to} 0$, $\frac{1}{T_0}\sum_t \epsilon_{it}^2 \overset{p}{\to} \sigma^2_{\epsilon}$ and $\frac{1}{T_0}\sum_t \epsilon_{it} \overset{p}{\to} 0$. For notational convenience drop in what follows the $MLE$ subscript in the weight estimator. By an application of l'hopital rule for the limit, observe that the sum of squares in the log-likelihood $l_{T_0}$ is proportional to

\begin{align*}
   \frac{1}{T_0}\sum_{t=1}^{T_0}\left(Y_{1t} - \sum_{j=2}^{J+1} \hat{w}_j Y_{jt}\right)^2 &= \frac{1}{T_0}\sum_{t=1}^{T_0} (\lambda_1 -  \sum_{j=2}^{J+1} \hat{w}_j \lambda_{j})^2 F_t^2 \\
   &\quad+ 2(\lambda_1 -  \sum_{j=2}^{J+1} \hat{w}_j \lambda_{j})F_t(\epsilon_{1t} -  \sum_{j=2}^{J+1} \hat{w}_j \epsilon_{jt}) \\
   &\quad+ (\epsilon_{1t} -  \sum_{j=2}^{J+1} \hat{w}_j \epsilon_{jt})^2\\
   &= \frac{1}{T_0}\sum_{t=1}^{T_0} (\lambda_1 -  \sum_{j=2}^{J+1} \hat{w}_j \lambda_{j})^2 \sigma^2 + \sigma_{\epsilon}^2(1 + \sum_{j=2}^{J+1} \hat{w}_j^2) + o_p(1).
\end{align*}

Minimizing the last expression as $T_0\to \infty$ is equivalent to minimizing the Ridge regression problem, and given that the problem is convex, a global minimizer is given by
$$
\hat{w}_j = \frac{\sigma^2 \lambda_1 \lambda_j}{\sigma_{\epsilon}^2 + \|\boldsymbol{\lambda}_J\|^2_2 \sigma^2}.
$$

Hence, as $T_0 \to \infty$ we have that $\hat{\boldsymbol{w}}_{MLE} \overset{p}{\to} \tilde{\boldsymbol{w}}$. The second statement follows from the conditions given in Theorem 2 and the continuous mapping theorem.

For the third statement, note that under the model assumptions, the standard M-estimator regularity conditions (\citet{vaart_1998}) are satisfied, therefore the result follows from known results. A sketch of the arguments is as follows: by part (1) we have that the estimator is consistent, and by the standard CLT, given that $ \mathbb{E}[\nabla_{\boldsymbol{w}}l_{T_0}(\boldsymbol{\theta})|_{\tilde{\boldsymbol{w}}} = 0$, we have that
$$
\sqrt{T_0} \nabla_{\boldsymbol{w}}l_{T_0}(\boldsymbol{\theta})|_{\tilde{\boldsymbol{w}}} \overset{d}{\to} N(0, V),
$$

where $V = V_{T_0}$. The proof then follows from an application of the mean value theorem.

\section{Proof of Theorem \ref{thm_mle}}

This proof follows results in \citet{heshao2000} and \citet{heshao1996}. A more modern treatment of similar results can also be found in \citet{belloni2015}. The proof considers the more general problem for $M-$estimators in linear models. Start by defining the objective function
$$
\rho(\boldsymbol{w}) = \sum_t (y_{1t} - \boldsymbol{y}_{Jt}'\boldsymbol{w})^2,
$$
and the score function
$$
\phi(\boldsymbol{w}) = 2\sum_t (y_{1t} - \boldsymbol{y}_{Jt}'\boldsymbol{w}) \boldsymbol{y}_{Jt}.
$$

Focus on the first assumption:
$$
\frac{1}{T_0}\sum_t \boldsymbol{y}_{Jt}\boldsymbol{y}_{Jt}' = D_{T_0},
$$
where  $0<\text{lim inf}_{T_0} \sigma_{min}(D_{T_0}) \leq \text{lim sup}_{T_0} \sigma_{max}(D_{T_0}) < \infty$. First note that under A1-A2 the matrix $\boldsymbol{y}_{Jt}\boldsymbol{y}_{Jt}
'$ is invertible and has eigenvalues bounded away from zero, so this assumption is satisfied in our setting. Then, under the other assumptions,

\begin{align*}
    \left| \boldsymbol{\alpha}'\left(\sum_t \mathbb{E}((y_{1t} - \boldsymbol{y}_{Jt}'\hat{\boldsymbol{w}}_{MLE}) -(y_{1t} - \boldsymbol{y}_{Jt}'\tilde{\boldsymbol{w}}))\boldsymbol{y}_{Jt}\right) - T_0\boldsymbol{\alpha}'D_{T_0}(\hat{\boldsymbol{w}}_{MLE} - \tilde{\boldsymbol{w}})\right| \\
    \leq c \sum_t |\boldsymbol{y}_{Jt}'\boldsymbol{\alpha}| |\boldsymbol{y}_{Jt}'(\hat{\boldsymbol{w}}_{MLE} - \tilde{\boldsymbol{w}})|^2
\end{align*}
Which implies that there exist a sequence of $J\times J$ matrices $D_{T_0}$ with bounded eigenvalues such that for any $\delta>0$, uniformly in $\boldsymbol{\alpha} \in \mathcal{S}_J(1)$,
\begin{align*}
\text{sup}_{\|w -\tilde{w}\|\leq \delta(J/T_0)^{1/2}} \left|\boldsymbol{\alpha}'\left(\sum_t \mathbb{E}((y_{1t} - \boldsymbol{y}_{Jt}'\boldsymbol{w}) -(y_{1t} - \boldsymbol{y}_{Jt}'\tilde{\boldsymbol{w}}))\boldsymbol{y}_{Jt}\right) - T_0\boldsymbol{\alpha}'D_{T_0}(\hat{\boldsymbol{w}}_{MLE} - \tilde{\boldsymbol{w}})\right| \\
=o((T_0 J)^{1/2}).
\end{align*}

The above expression means that condition (C3) in \citet{heshao2000} is satisfied. Next, we check that condition (C2) is satisfied. Observe that

$$
\|\phi(\tilde{\boldsymbol{w}})\| \leq 2 \sum_t\|  y_{1t} - \boldsymbol{y}_{Jt}'\boldsymbol{\tilde{w}}\| \|\boldsymbol{y}_{Jt}\| = O((T_0J)^{1/2}),
$$
\noindent given assumptions (2) and (3) in the theorem. By a similar argument (C1) is satisfied and (C3) implies that (C4) or (C5) are satisfied. Hence, we can apply Corollary 2.1 in \citet{heshao2000} to get the desired result.

\section{Proof of Corollary \ref{cor_mle}}

To derive the first statement in Corollary 5.1 observe that

$$
\boldsymbol{y}_{Jt}'(\tilde{\boldsymbol{w}} - \boldsymbol{w}) \leq \|\boldsymbol{y}_{Jt}\|^2\|\tilde{\boldsymbol{w}} - \boldsymbol{w}\|^2 = O_p(J^2/T_0),
$$
by condition 2 and result 1 in Theorem \ref{thm_mle}. A similar, approach to \citet{heshao2000} can be followed to improve the rate to $J/T_0$ up to log terms.

To derive the second statement, we show that Theorem \ref{thm_mle} applies when $\alpha$ is itself a stochastic sequence. We will show that in our setting the boundedness assumptions on the DGP allows us to preserve stochastic equi-continuity, and so condition (C2) in \citet{heshao2000} and in the proof of Theorem \ref{thm_mle} applies. Consider the following quantity 
$$
\gamma(\boldsymbol{\alpha}) = T_0\boldsymbol{\alpha}'D_{T_0}(\hat{\boldsymbol{w}}_{MLE} - \tilde{\boldsymbol{w}}).
$$
Observe that when $\boldsymbol{\alpha}\in S^J$, a ball of dimension $J$, by a standard maximal inequality for simplex random variables we get that this object can be bounded by $\sqrt{J}T_0$. It is without loss to substite $\boldsymbol{\alpha}$ with $\boldsymbol{y}_{Jt}$ given assumption 2 in Theorem \ref{thm_mle}. Indeed, in that case we can also upper bound the quantity by the same rate. Overall, this implies that condition (C2) in \citet{heshao2000} is satisfied and so the result follows.

\section{Proof of Theorem \ref{thm_bvm}}

Recall that our Bayes model given an $i.i.d$ sample of size $T_0$ with $J+1$ units means that the posterior predictive distribution is normal with the following parameters:
\begin{enumerate}
    \item Mean: $$\mu^B_{T_0,J} =  \frac{\sigma_y^2}{\sigma_y^2 + T_0\sum_j \tau_j^2}\sum_t \boldsymbol{y}_{Jt}'\mu_J + \frac{\sum_j \tau_j^2}{\sigma^2_y + T_0\sum_j \tau_j^2}\sum_t y_{1t}.$$
    \item Variance: $$
    \Sigma^B_{T_0,J} = \frac{\sigma_y^2 \sum_j\tau_j^2 }{\sigma_y^2 + T_0 \sum_j\tau_j^2}.
    $$
\end{enumerate}

Suppose, as in the assumptions, that $\| \boldsymbol{\mu}_J\|^2 \to 0$ and $\| \lambda_1 - \boldsymbol{\lambda_J}'\boldsymbol{\mu}_J\| \to 0$ as $J\to \infty$. Under the DGP given by \textbf{A1}-\textbf{A2} we have that $y_{jt} = \lambda_j F_t + \epsilon_{jt}$, for $\epsilon_{jt} \sim_{i.i.d} N(0,1)$. It follows that
\begin{align*}
    \boldsymbol{y}_{JT_0+1}'\boldsymbol{\mu}_J 
    &= \sum_j (\lambda_jF_{T_0+1} + \epsilon_{jT_0+1})\mu_j \\
    &= F_{T_0+1}\sum_j \lambda_j\mu_j + \sum_j \epsilon_{jT_0+1}\mu_j.
\end{align*}

Next, consider the following expectation
\begin{align*}
    \mathbb{E}[( \boldsymbol{y}_{JT_0+1}'\boldsymbol{\mu}_J - \lambda_1 F_{T_0+1}) ^2 |F_{T_0+1}] &= \mathbb{E}[(F_{T_0+1}(\sum_j \lambda_j\mu_j - \lambda_1) + \sum_j \epsilon_{jT_0+1}\mu_j)^2|F_{T_0+1}] \\
    &= \mathbb{E}[(F^2_{T_0+1}(\sum_j \lambda_j\mu_j - \lambda_1)^2 + (\sum_j \epsilon_{jT_0+1}\mu_j)^2|F_{T_0+1}] \\
    &\leq \mathbb{E}[(F^2_{T_0+1}(\sum_j \lambda_j\mu_j - \lambda_1)^2|F_{T_0+1}] + \sigma_{\epsilon}^2\sum_j \mu_j^2\\
        &\to 0.
\end{align*}
\noindent where the second and third inequalities follow from $\epsilon_{jt}$ being mean zero and $i.i.d$. Under the assumptions on $\boldsymbol{\mu}_J$ it follows that this inequality goes to zero as $J\to \infty$. Therefore, the convergence in probability follows by Chebyshev's inequality. 

Next, we show that the mean of the posterior predictive distribution $\mu^B_{T_0,J}$ converges to the same mean as the MLE estimator. First, note that under \textbf{A1}-\textbf{A2} and $\sum_j \tau_j^2 = O(T_0^{-\alpha})$ the second term in $\mu^B_{T_0,J}$ is $o_p(1)$ as $\frac{1}{T_0}\sum_t y_{1t} \overset{p}{\to} 0$. The first term then converges to the treated unit factor loading by a similar argument as the first part of this proof given our convex recovery assumption.

To derive the Bernstein-von Mises result we start by noting that due to Pinsker's inequality
$$
\| \Phi_{MLE} - Q \|_{TV} \leq \sqrt{\frac{1}{2} D_{KL}(\Phi_{MLE} || Q)}.
$$

Hence, we procced in bounding the KL divergence. A useful result is \citet{barron1986}, which provide conditions under which the KL divergence and CLT can be related. The following Lemma summarizes the result.

\begin{lemma}[KL Convergence (Barron 1986)]
Let $\Phi_{J,T}$ be the MLE estimator distribution and $Q_{T,J}$ be the smooth, bounded Bayes posterior predictive distribution for fixed $J$ and $T_0$. Suppose that as $J,T\to \infty$,

\begin{enumerate}
    \item $\Phi_{J,T} \to P^*$,
    \item $Q_{T,J} \to Q^*$,
    \item $Q^*$ and $P^*$ have the same mean and have bounded fourth moments.
\end{enumerate}
Then, it follows that
$$
D_{KL}(\Phi_{J,T} || Q_{T,J}) = D_{KL}(\Phi^* || Q^*) + O(1/(T J)).
$$
\end{lemma}

The conditions in Lemma A.1 are satisfied given our assumptions, the previous derivations and the results in Theorem \ref{thm_mle}. Therefore, we need to bound $D_{KL}(\Phi^* || Q^*)$. Given our Gaussian assumption we can bound this quantity by comparing the variances of the two distributions. The following Lemma gives an exact derivation of the KL divergence for Gaussian distributions.

\begin{lemma}[Gaussian KL] Suppose that $Q$ and $P$ are normal random variables with equal means and $k\times k$ covariance matrices $\Sigma_Q$ and $\Sigma_P$. Then, 

$$
D_{KL}(P || Q) = \frac{1}{2}\left(\log\frac{|\Sigma_P|}{|\Sigma_Q|} - k + tr(\Sigma_Q^{-1} \Sigma_P)\right).
$$
\end{lemma}

Observe that in our case the distributions are one dimensional so it is sufficient to show that as $T_0, J \to \infty$
$$
\frac{\mathbb{V}(\Phi^{MLE}_{T_0,J})}{\Sigma_{T_0,J}^B} \to 1.
$$

Starting with the Bayesian model, recall that $\sum_j \tau_j^2 = O(J^{\alpha})$, so for the prediction period $T_0+1$,
$$
\Sigma_{T_0+1,J}^B = \frac{\sigma_y^2}{\sigma_y^2( \sum_j\tau_j^2 )^{-1} + 1} \to \sigma_y^2.
$$
From Theorem \ref{thm_mle} recall that $\sigma^2_{\alpha} = (\mathbb{E}[\epsilon^2_{Jt}])^{-1} \boldsymbol{\alpha}'D_{T_0}^{-1}\boldsymbol{\alpha}$. Following, Corollary \ref{cor_mle} when $\boldsymbol{\alpha}$ is given by $\boldsymbol{y}_{Jt}$ and $\frac{1}{T_0} \sum_t \boldsymbol{y}_{Jt}\boldsymbol{y}_{Jt}' = D_{T_0}$ observe that
\begin{align*}
   \sigma^2_{\alpha} &= (\sigma_{\epsilon}^2) \boldsymbol{\alpha}'D_{T_0}^{-1}\boldsymbol{\alpha}  \\
   &= (\sigma_{\epsilon}^2) T_0 \boldsymbol{y}_{JT_0+1}'(\boldsymbol{y}_{Jt}\boldsymbol{y}_{Jt}')^{-1}\boldsymbol{y}_{JT_0+1} \\
   &=(\sigma_{\epsilon}^2)T_0\text{tr}((\boldsymbol{y}_{Jt}\boldsymbol{y}_{Jt}')^{-1}\boldsymbol{y}_{JT_0+1}\boldsymbol{y}_{JT_0+1}')\\
   &= (\sigma_{\epsilon}^2) O(T_0),
\end{align*}
\noindent where the last step follows from Theorem $\ref{thm_mle}$'s assumption 2 regarding the $l_2$ norm of $\boldsymbol{y}_Jt$. This implies that the sample MLE variance is given by $1/T_0 \sigma_{\alpha}^2 \to \sigma_{\epsilon}^2$. If the time series structure is such that $\text{tr}((\boldsymbol{y}_{Jt}\boldsymbol{y}_{Jt}')^{-1}\boldsymbol{y}_{JT_0+1}\boldsymbol{y}_{JT_0+1}') = 1$ as $T_0, J \to\infty$, then it follows that a sufficient condition for
$$
D_{KL}(\Phi^* || Q^*) \to 0
$$
is that $\sigma_y /\sigma_{\epsilon} \to 1$ as $J,T\to \infty$. Given \textbf{A1} - \textbf{A2}, it follows that a sufficient condition is $\sigma_y \to \sigma_{\epsilon}$. The trace condition is satisfied under assumption 2 of Theorem \ref{thm_mle} and \textbf{A1}-\textbf{A2}, but it would also be satisfied under weaker conditions such as mixing time series regimes.

\section{Additional simulations}

\begin{figure}[ht!]
\centering
  \begin{subfigure}[b]{0.5\textwidth}
  \centering
  	\includegraphics[width=1\linewidth]{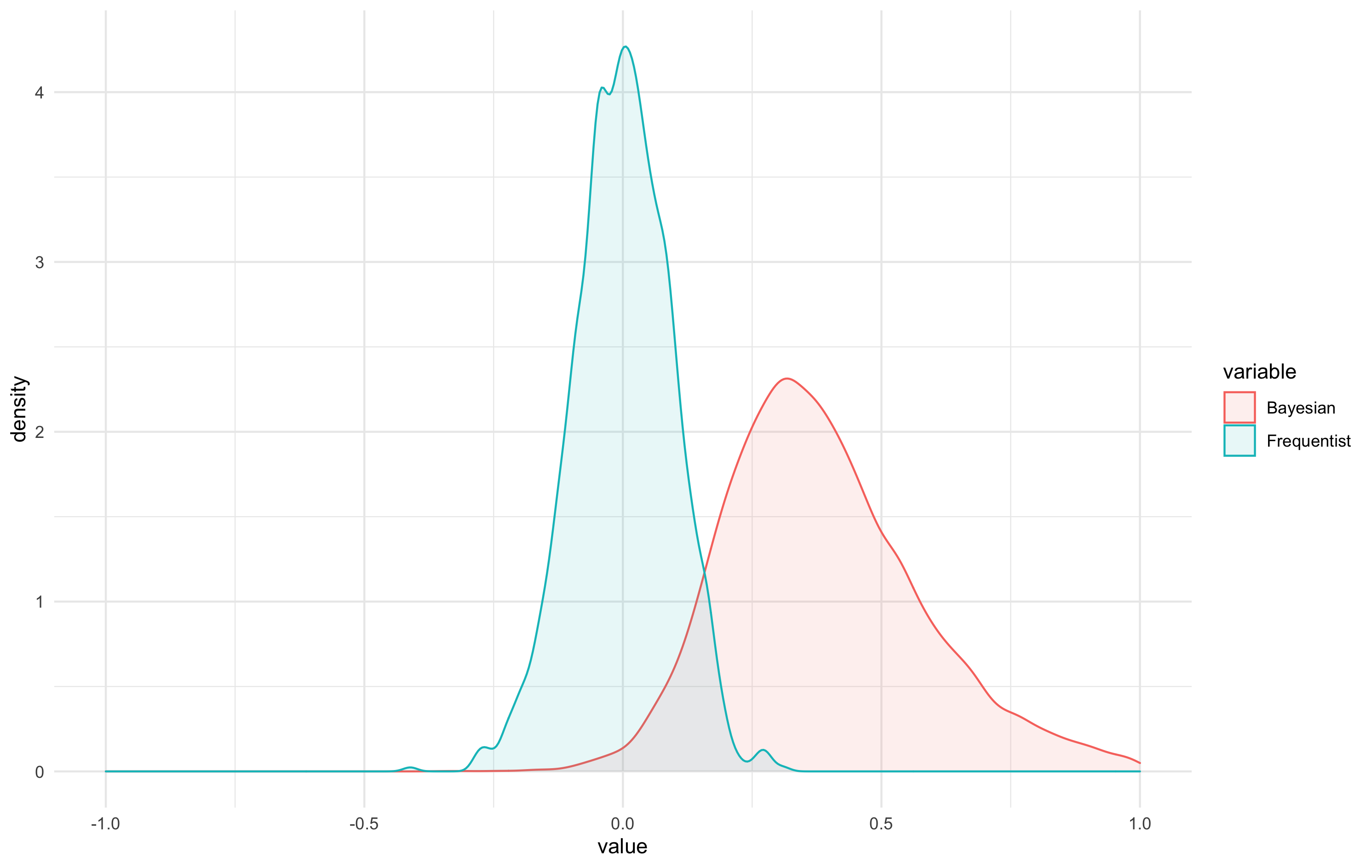}
  	\caption{$T = 30$}
  \end{subfigure}%
  \begin{subfigure}[b]{0.5\textwidth}
  \centering
  	\includegraphics[width=1\linewidth]{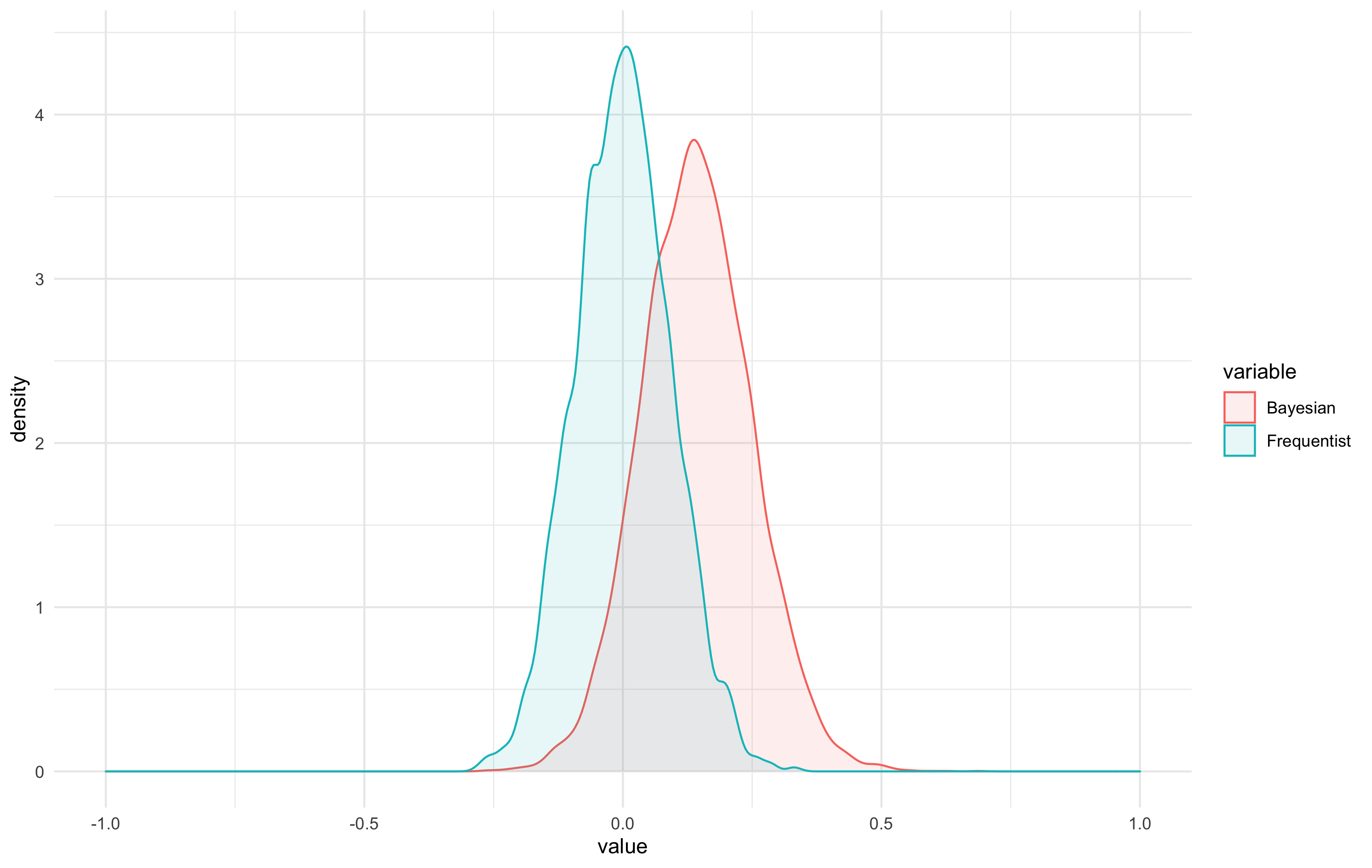}
  	\caption{$T = 50$}
  \end{subfigure}%
  
   \begin{subfigure}[b]{0.5\textwidth}
  \centering
  	\includegraphics[width=1\linewidth]{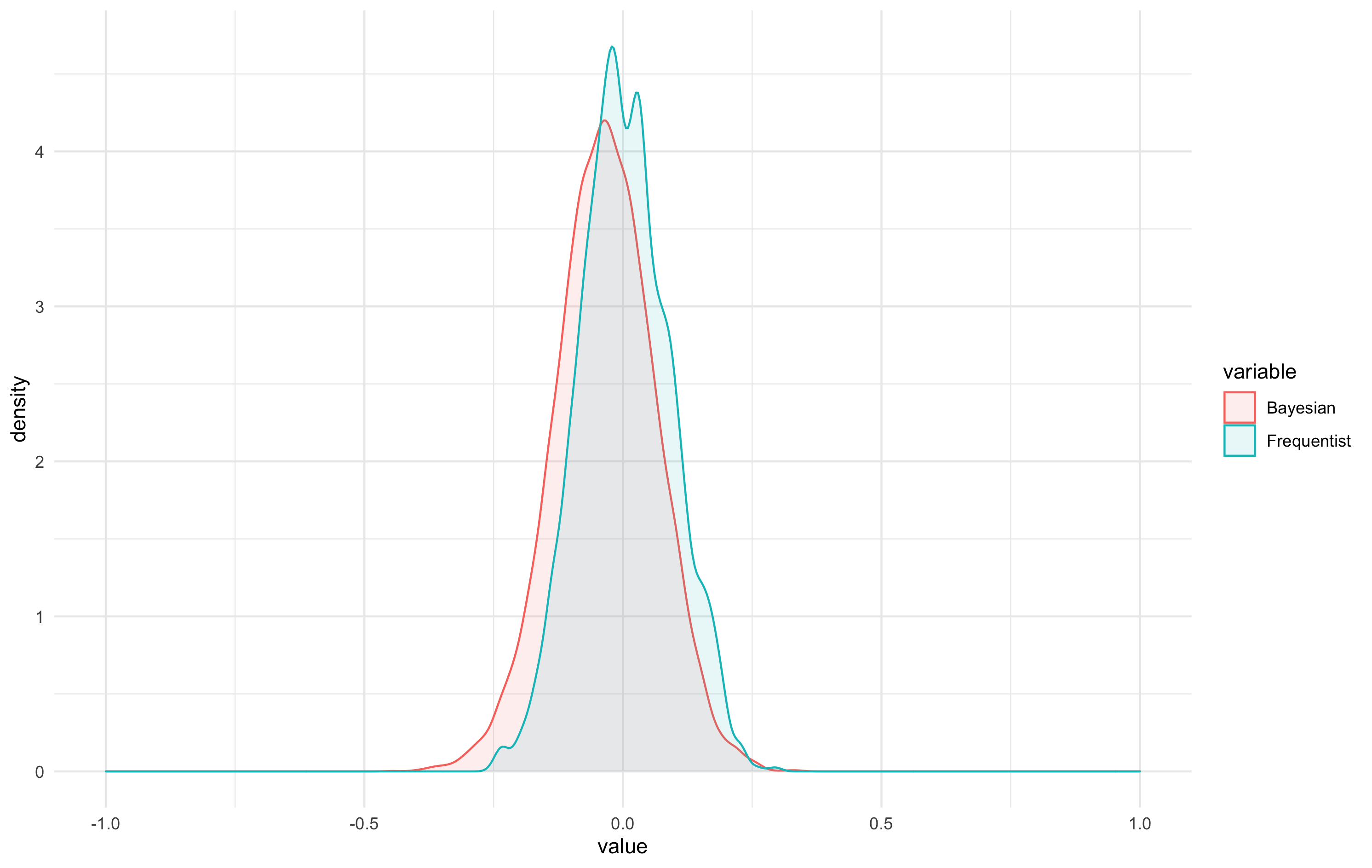}
  	\caption{$T = 70$}
  \end{subfigure}%
  \caption{Convergence of frequentist and Bayesian coverage as $T\to\infty$.}
    \begin{tablenotes}
		\small\item\textbf{Notes}: Kernel densities of the frequentist empirical distribution of the estimated treatment effect over 10000 draws and the Bayesian posterior distribution for one draw for different values of $T_0$. The potential outcomes are generated by the grouped factor model with 4 groups of 5 units with the same factor loadings and $\sigma = 0.25$.
   \end{tablenotes}
  \label{appendix_bvm_dense}
\end{figure}

\begin{figure}[ht!]
\centering
  \begin{subfigure}[b]{0.5\textwidth}
  \centering
  	\includegraphics[width=1\linewidth]{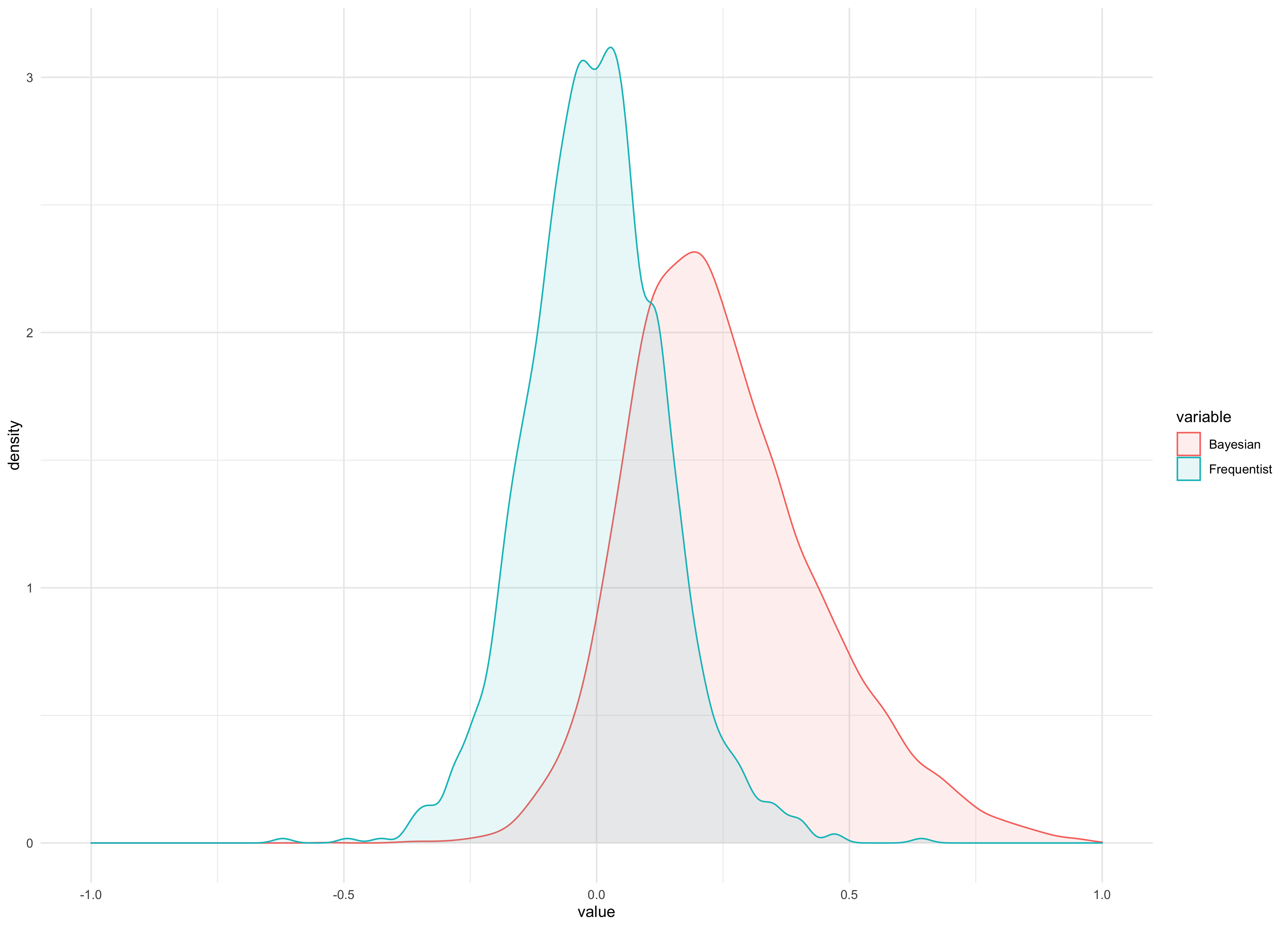}
  	\caption{$T = 30$}
  \end{subfigure}%
  \begin{subfigure}[b]{0.5\textwidth}
  \centering
  	\includegraphics[width=1\linewidth]{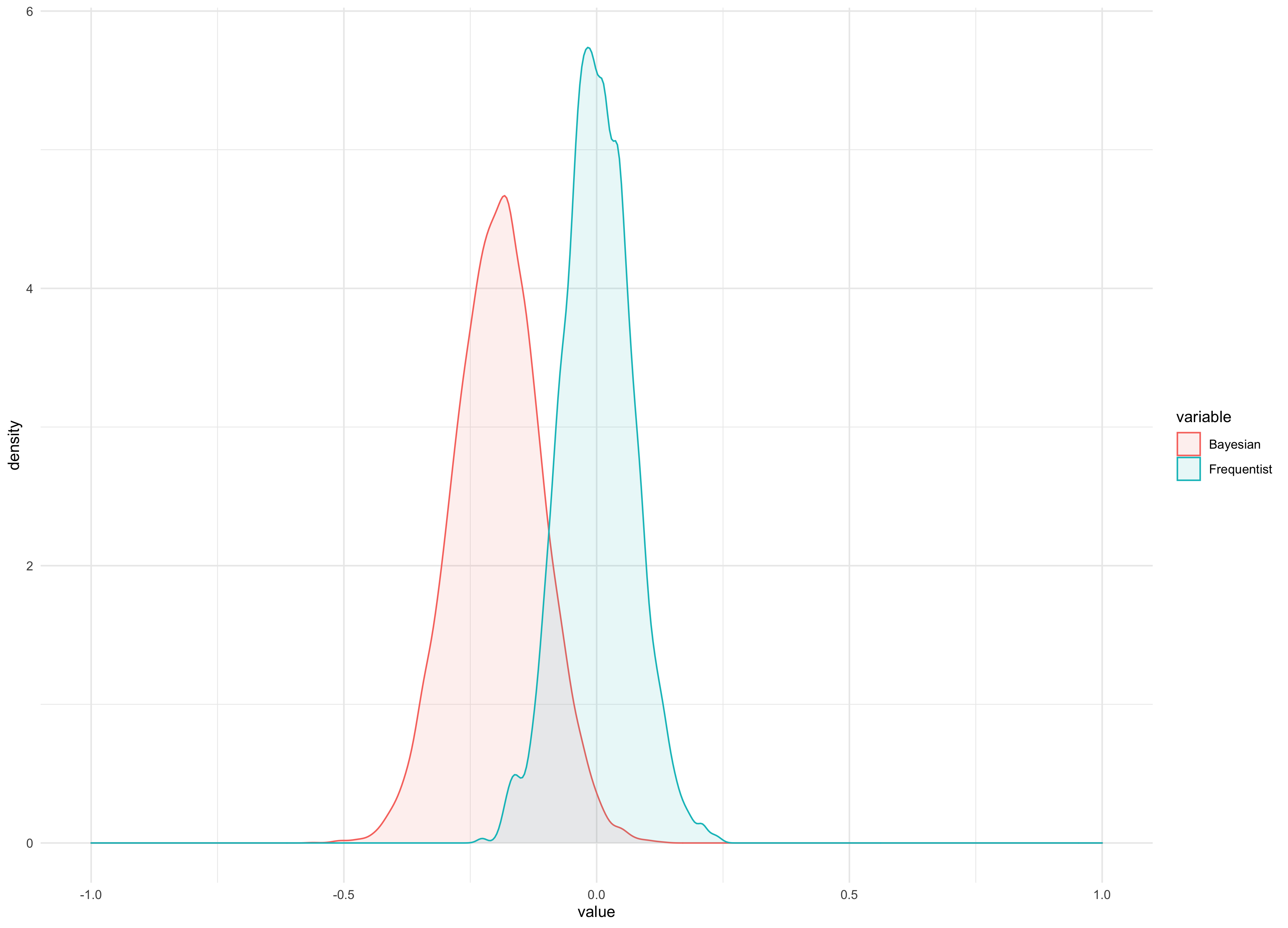}
  	\caption{$T = 100$}
  \end{subfigure}%
  
   \begin{subfigure}[b]{0.5\textwidth}
  \centering
  	\includegraphics[width=1\linewidth]{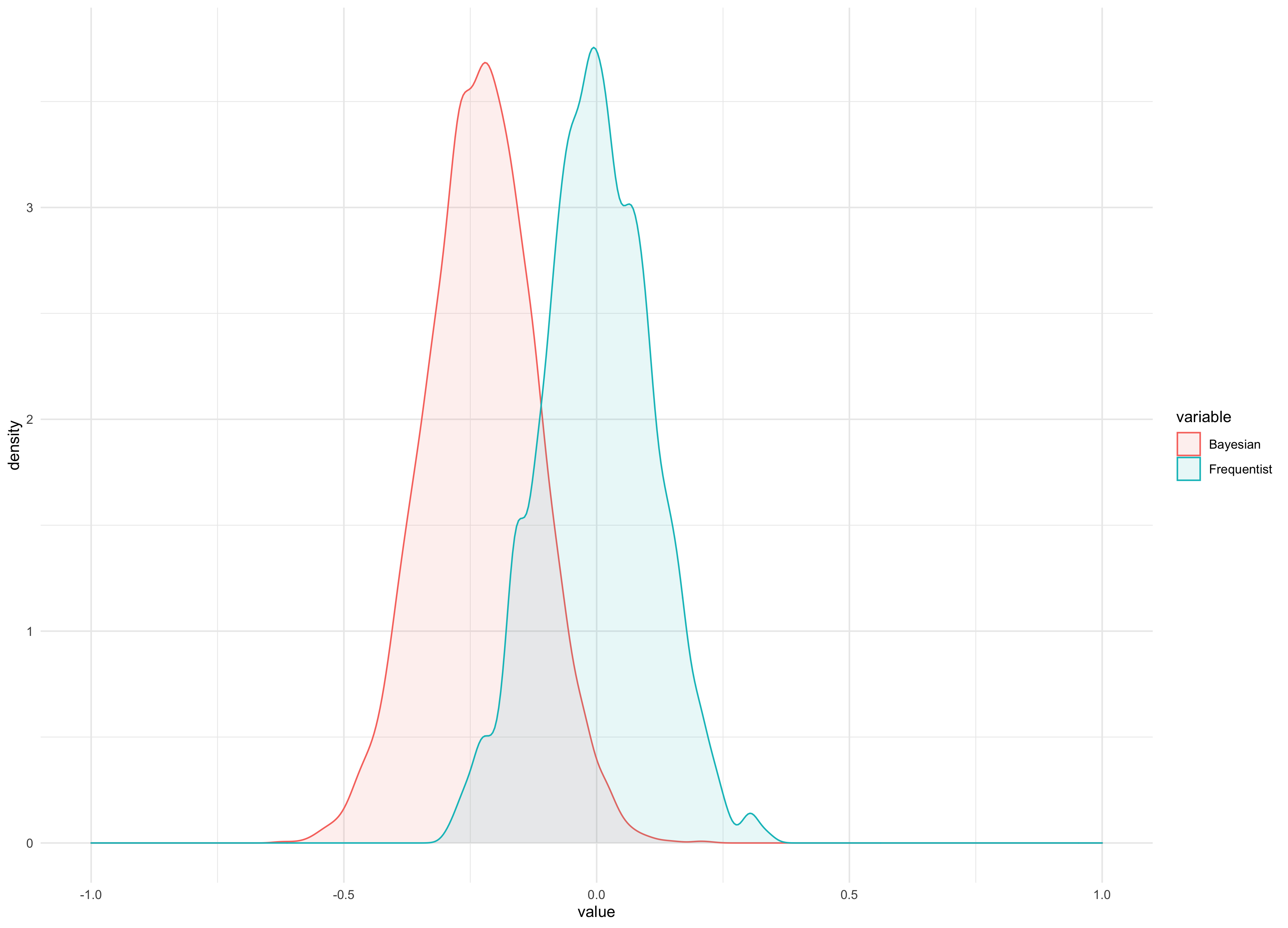}
  	\caption{$T = 500$}
  \end{subfigure}%
  \begin{subfigure}[b]{0.5\textwidth}
  \centering
  	\includegraphics[width=1\linewidth]{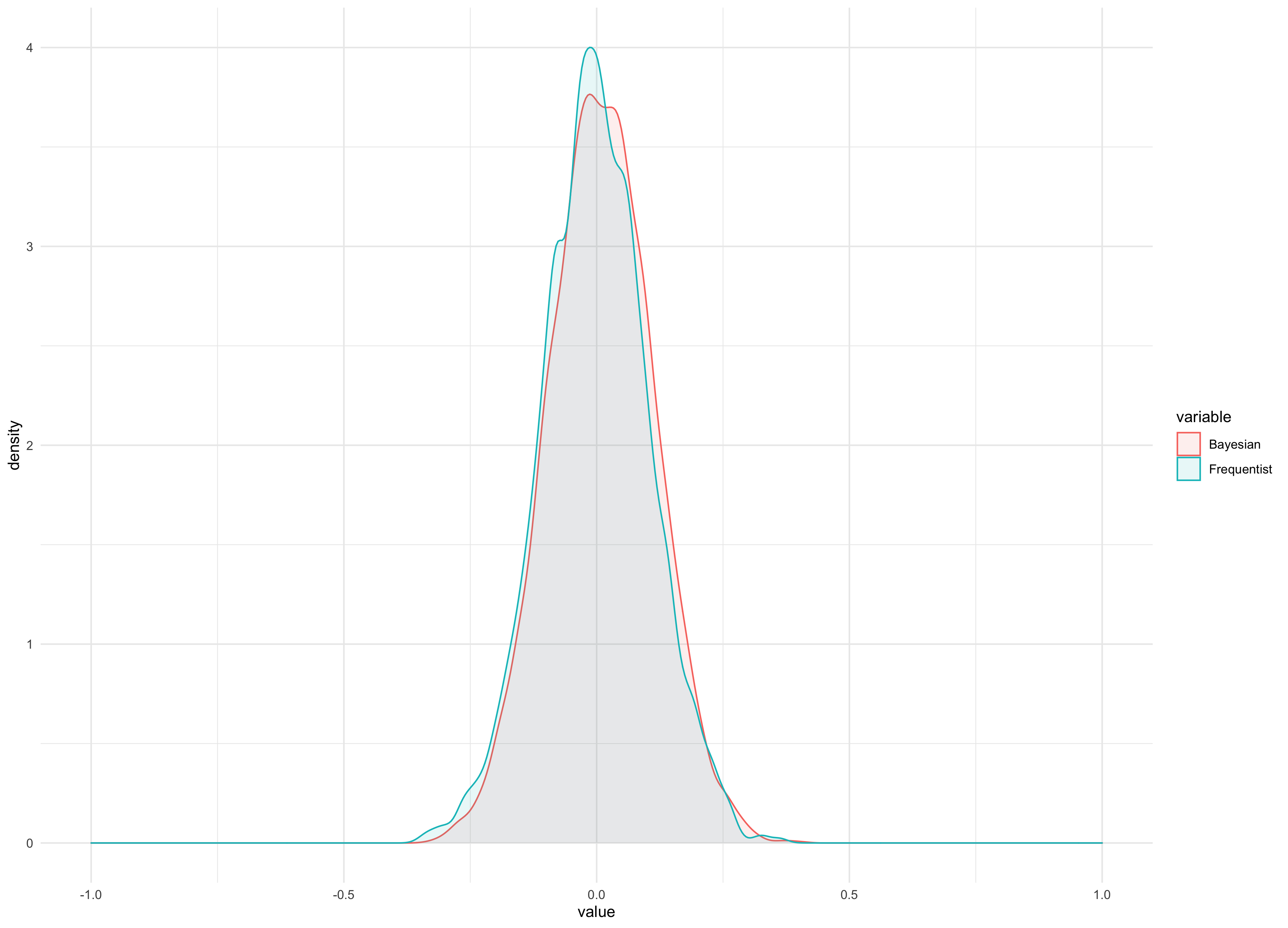}
  	\caption{$T = 1000$}
  \end{subfigure}%
  \caption{Convergence of frequentist and Bayesian coverage as $T\to\infty$.}
    \begin{tablenotes}
		\small\item\textbf{Notes}: Kernel densities of the frequentist empirical distribution of the estimated treatment effect over 10000 draws and the Bayesian posterior distribution for one draw for different values of $T_0$. The potential outcomes are generated by the grouped factor model (\ref{grouped_fm}) with 10 groups of 2 units with $\sigma = 0.25$.
   \end{tablenotes}
  \label{appendix_bvm}
\end{figure}

In this section we report some additional simulation results in which we compare the frequentist distribution of the treatment effect estimator over 10000 draws with the Bayesian posterior distribution over 1 draw. This simulations are meant to highlight that even with just 1 draw from the underlying DGP as $T_0\to \infty$ the Bayesian posterior distribution approximates the frequentist distribution. Figure \ref{appendix_bvm} highlights the BvM convergence in a sparse setting given by the grouped factor model \ref{grouped_fm} in which unit 2 is the only unit to have the same factor loading as unit 1. Figure \ref{appendix_bvm_dense} replicates Figure \ref{appendix_bvm} for a denser setting in which we have four groups of five units with equal factor loadings. Unit 1 has the same factor loading as units 2 to 5. The Figure shows that convergence is achieved earlier than in the sparse case, with good coverage around $T=70$ instead that when $T=1000$. This is in line with the theory that suggests that a requirement for convergence is that the weights are evenly distributed amongst many units. However, observe that the convergence rate is faster than expected as for 20 units the theory would suggest that at least $20^2 = 400$ time periods are necessary.

\section{German re-unification additional plots}

In this section we provide additional materials for the study of the German re-unification. In Figure \ref{fig_germany_more} we show the correlation between the implicit weight distributions and the distribution of an upper bound on the bias (as proposed by \citet{AbaDiaHai2010}) for each Bayesian draw.

\begin{figure}[ht!]
\centering
  \begin{subfigure}[b]{0.5\textwidth}
  \centering
  	\includegraphics[width=1\linewidth]{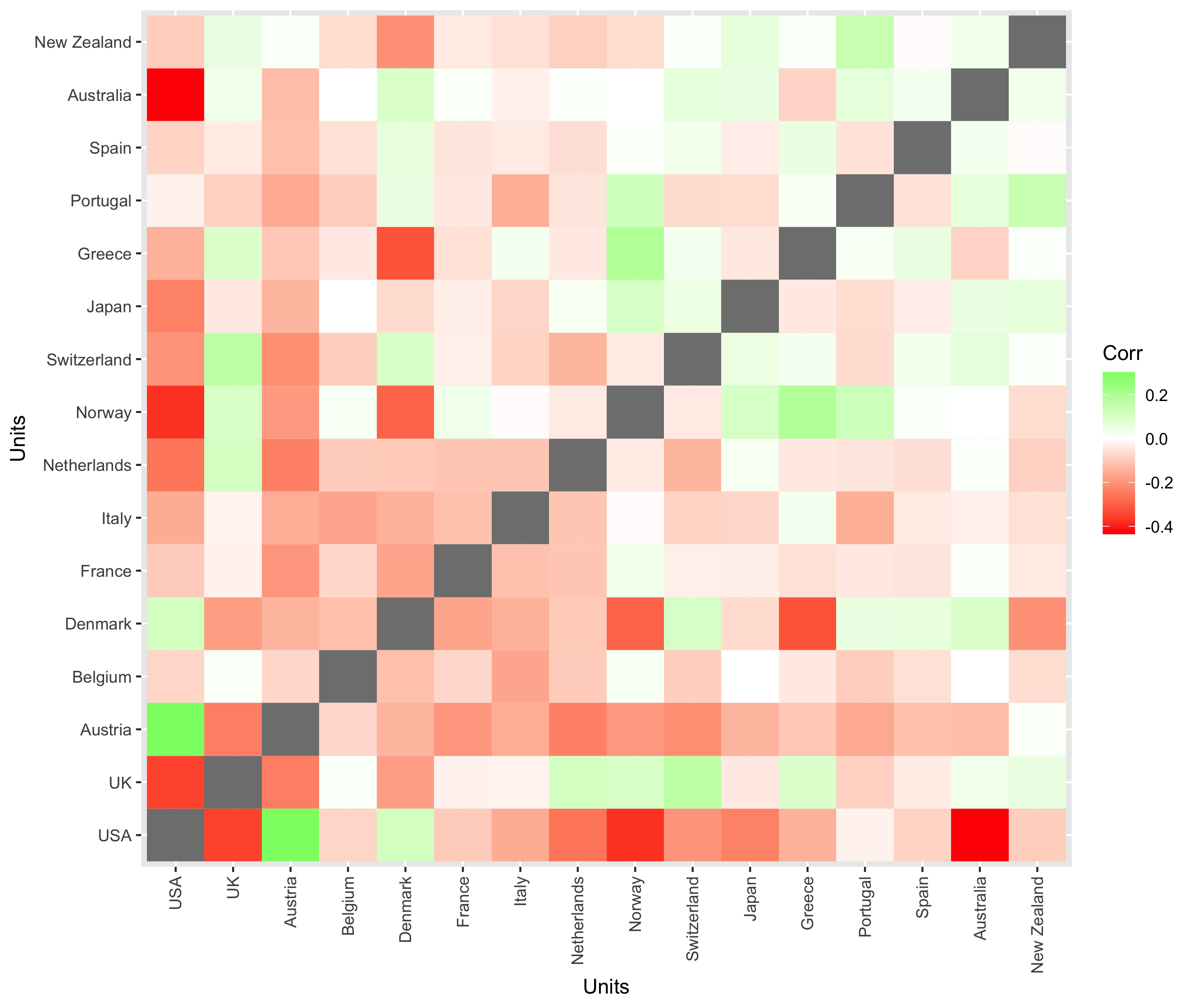}
  	\caption{Implicit weights correlations}
  \end{subfigure}%
  \begin{subfigure}[b]{0.5\textwidth}
  \centering
  	\includegraphics[width=1\linewidth]{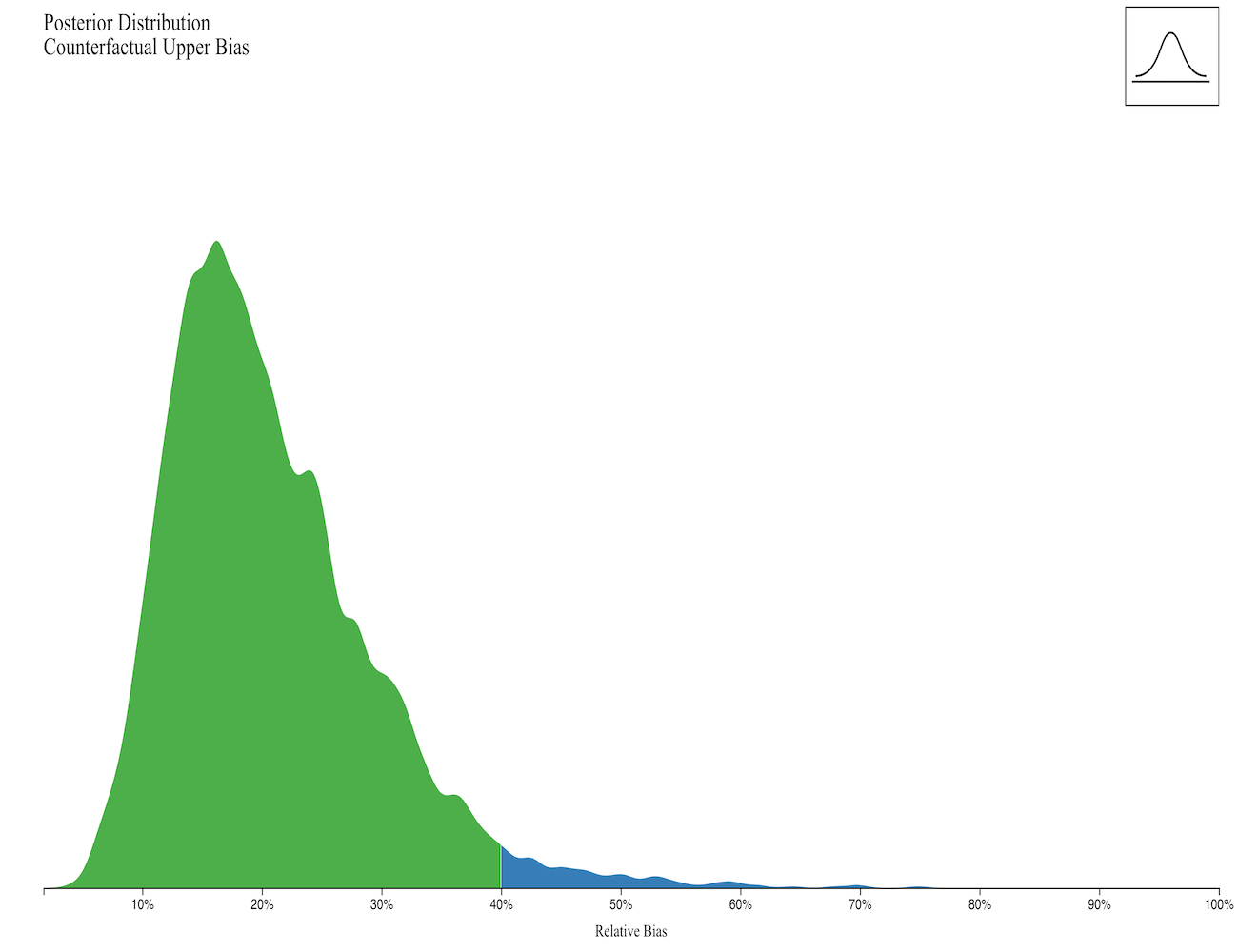}
  	\caption{Bias bound distribution}
  \end{subfigure}%
  \caption{Additional Plots}
      \begin{tablenotes}
		\small\item\textbf{Notes}: Panel (a) shows the correlations between implicit weights of the donor countries. Panel (b) shows the distribution of the bias bound term (computed as the sum of MADs of the pre-treatment outcomes) relative to the size of the mean treatment effect for each Bayesian draw. 
   \end{tablenotes}
  \label{fig_germany_more}
\end{figure}

\section{Catalan secession movement additional plots}

\begin{figure}[ht!]
\centering
  	\includegraphics[width=0.5\linewidth]{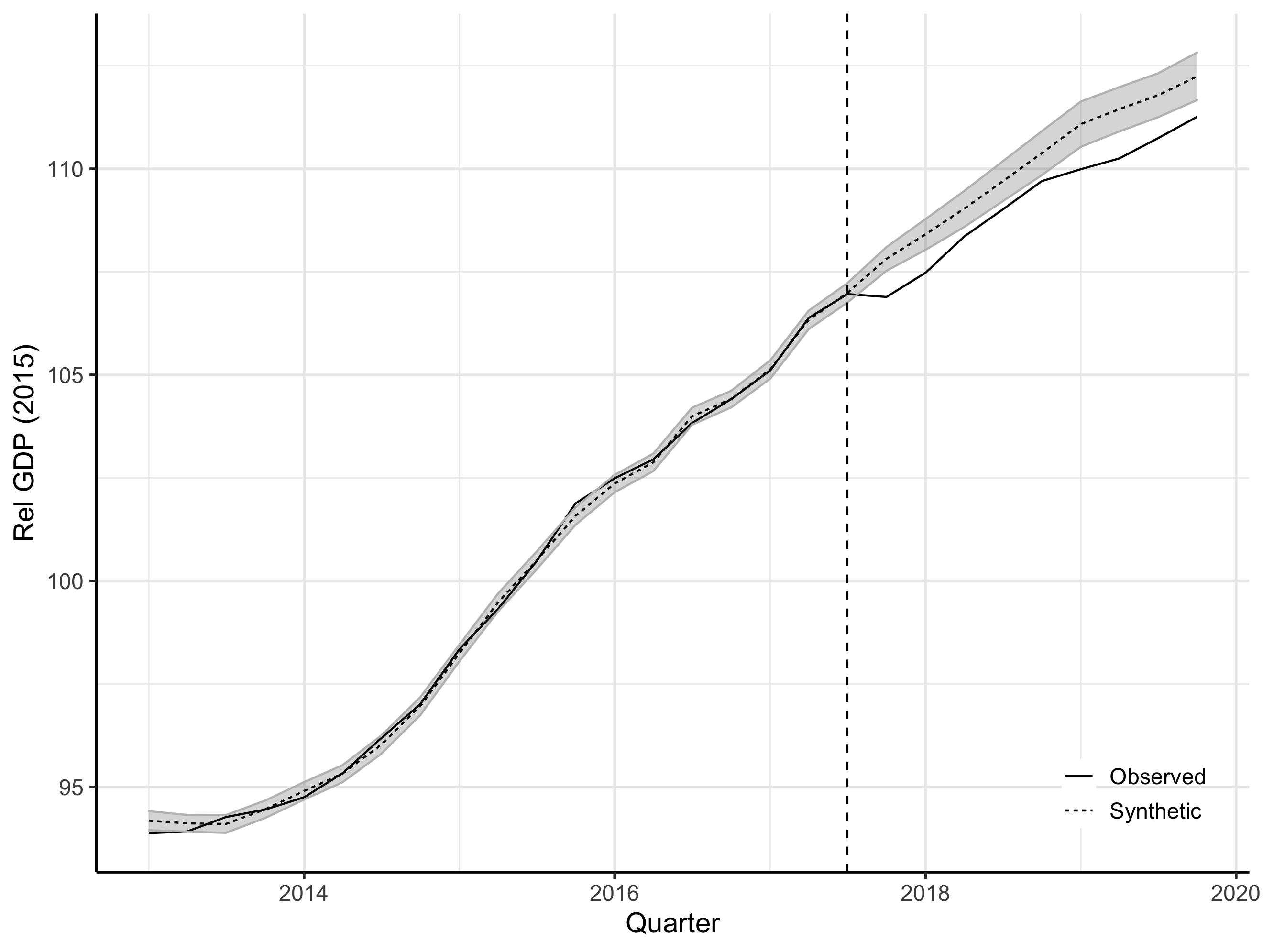}
  \caption{Additional Plot Synthetic Catalonia}
      \begin{tablenotes}
		\small\item\textbf{Notes}: Bayesian synthetic control for Catalonia excluding Madrid and Valencia from the donor pool. The average treatment effect relative to the third quarter of 2017 is -0.85\%.
   \end{tablenotes}
  \label{fig_germany_more}
\end{figure}

\end{document}